\documentclass[epj,final,twocolumn]{svjour}

\usepackage{graphicx}
\usepackage{url}
\usepackage{amssymb}

\twocolumn

\begin{document}

\title{Primer on ILC Physics and SiD Software Tools}
\author{Chris Potter\inst{1}\inst{2}}
\institute{Physics Department, University of Oregon, \email{ctp@uoregon.edu} \and Institute for Fundamental Science, University of Oregon }
\date{Received: \today / Revised version: \today}

\abstract{We first outline the Standard Model (SM) of particle physics, particle production and decay, and the expected signal and background at a Higgs factory like the International Linear Collider (ILC). We then introduce high energy colliders and collider detectors, and briefly detail the ILC and the Silicon Detector (SiD), one of the two detectors proposed for the ILC. Next we review the available software tools for ILC event generation, SiD detector simulation, and event reconstruction. Finally we suggest open avenues in research for detector optimization and physics analysis. The pedagogical level is suitable for advanced undergraduate and beginning graduate students in physics and research scientists in related fields.
}

\maketitle

\section{Introduction: physics goals}

\begin{figure*}[t]
\begin{center}
\framebox{\includegraphics[width=0.9\textwidth]{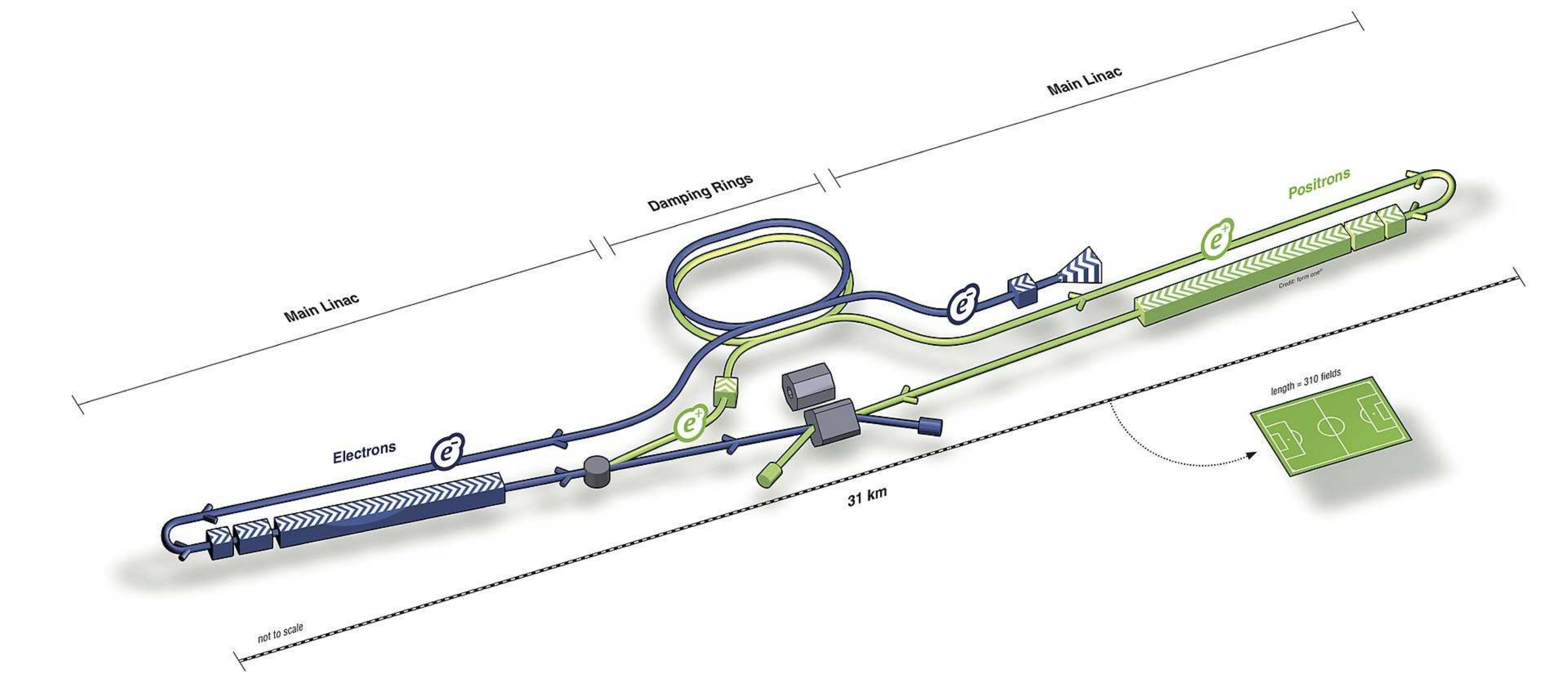}}
\caption{Schematic diagram (not to scale) of the main systems of the ILC assuming the nominal TDR design. Shown are the electron linac, the positron linac, the damping rings and two detectors at the collision point. Credit: ILC TDR \cite{Behnke:2013xla,Baer:2013cma,Phinney:2007gp,Behnke:2013lya}}
\label{fig:ilc}
\end{center}
\end{figure*}

\begin{figure*}[t]
\begin{center}
\vspace{-0.8in}
\resizebox{0.9\textwidth}{!}{\includegraphics*{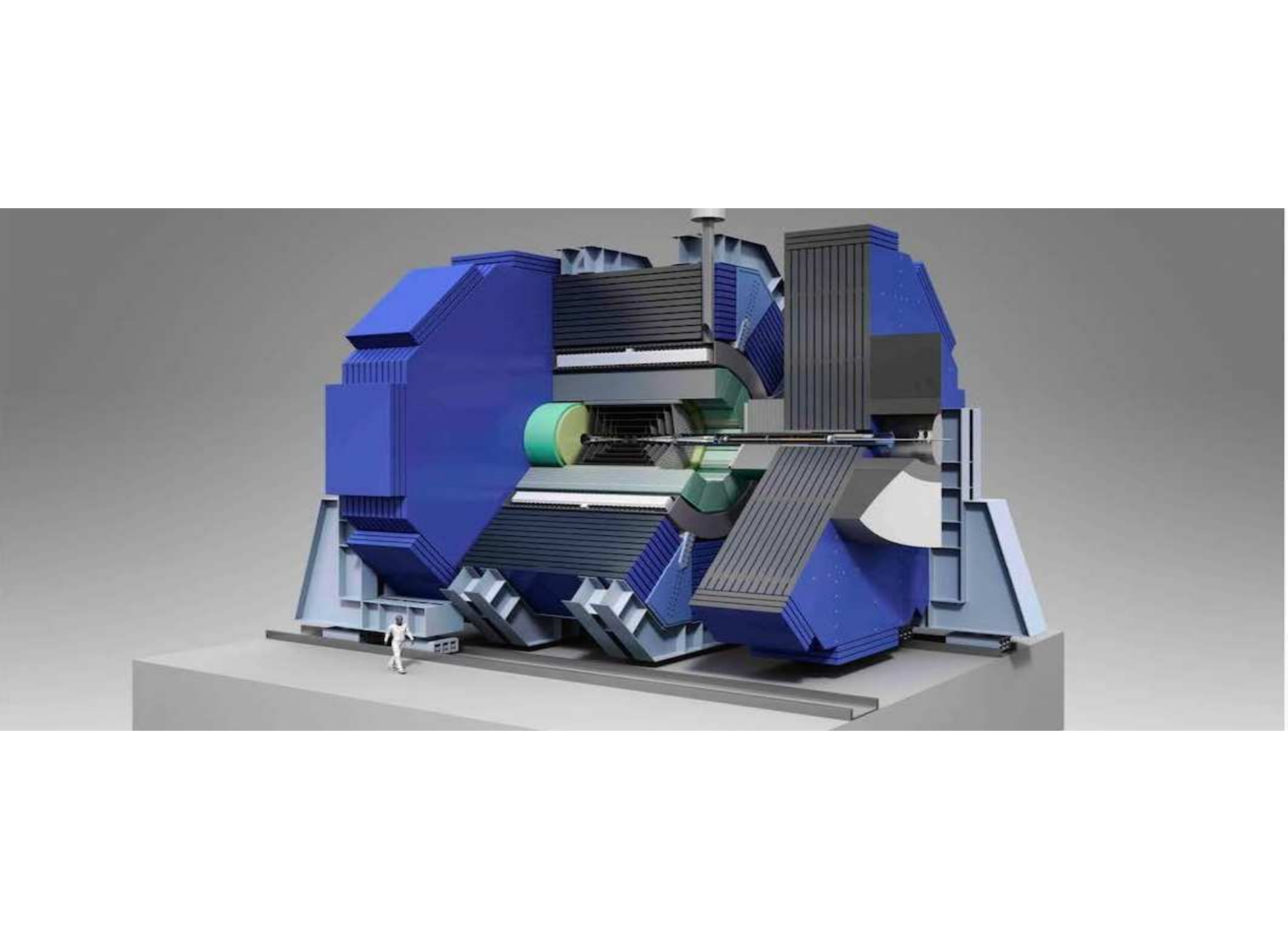}}
\vspace{-0.8in}
\caption{Rendering of SiD. The ILC electron and positron beams collide in the detector center. Credit: SiD Consortium.}
\label{fig:sid}
\end{center}
\end{figure*}

The International Linear Collider (ILC) is a mature proposal for the next major high energy accelerator after the Large Hadron Collider (LHC). The ILC Technical Design Report (TDR) \cite{Behnke:2013xla,Baer:2013cma,Phinney:2007gp,Behnke:2013lya} demonstrates that the accelerator project is technically feasible and construction ready. Moreover two detector designs detailed in the TDR, the Silicon Detector (SiD) and the International Large Detector (ILD), are prepared to enter the technical design phase. See figs. \ref{fig:ilc} and \ref{fig:sid} for renderings of the ILC and SiD. 

The primary motivation for the ILC is the precision study of the Higgs boson. The Higgs phenomenon was independently proposed in 1964 by Higgs \cite{Higgs:1964pj} and Englert and Brout \cite{Englert:1964et} as a possible explanation for how the $W$ and $Z$ bosons obtain their mass. In fact the Higgs mechanism can explain how every particle obtains its mass. The scalar particle $H$ associated with the Higgs field, which mediates the Higgs mechanism, was jointly discovered in 2012 at the CERN LHC by the ATLAS \cite{Aad:2012tfa} and CMS \cite{Chatrchyan:2012ufa} Collaborations. In 2013, 49 years after their papers were published in the same journal, Higgs and Englert were awarded the Nobel Prize in Physics. 

The ILC and its detectors are multipurpose, and address secondary physics motivations which elaborate their scientific merit. Undiscovered new particles and interactions postulated by various theoretical models can be discovered, constrained or ruled out with a full ILC program. In brief, the ILC goals outlined in the TDR, are as follows:

\begin{enumerate}
\item Measuring Higgs boson branching ratios and other properties with high precision
\item Searching for new particles, including dark matter and supersymmetric particles
\item Constraining new interactions by high precision measurements of the $W,Z$ and $t$ particles
\end{enumerate}

\noindent We quote at length from the executive summary in the TDR Volume 1 \cite{Behnke:2013xla}, which elaborates on these goals. First, precision study of the Higgs boson:

\begin{quote}
The initial program of the ILC for a 125 GeV Higgs boson will be centered at an energy of 250 GeV, which gives the peak cross section for the reaction $e^+ e^- \rightarrow Zh$.  In this reaction, the identification of a $Z$ boson at the energy appropriate to recoil against the Higgs boson tags the presence of the Higgs boson.  In this setting, it is possible to measure the rates for all decays of the Higgs boson - even decays to invisible or unusual final states — with high precision \dots

The study of the Higgs boson will continue, with additional essential elements, at higher energies. At 500 GeV, the full design energy of the ILC, measurement of the process $e^+ e^- \rightarrow \nu \bar{\nu} h$ will give the absolute normalization of the underlying Higgs coupling strengths, needed to determine the individual couplings to the percent level of accuracy.  Raising the energy further allows the ILC experiments to make precise measurements of the Higgs boson coupling to top quarks and to determine the strength of the Higgs boson’s nonlinear self interaction \dots
\end{quote}

\noindent Next, the search for new particles:

\begin{quote}
The ILC also will make important contributions to the search for new particles associated with the Higgs field, dark matter, and other questions of particle physics.  For many such particles with only  electroweak  interactions,  searches  at  the  LHC  will  be  limited  by  low  rates  relative  to  strong interaction  induced  processes,  and  by  large  backgrounds.   The  ILC  will  identify  or  exclude  these particles unambiguously up to masses at least as high as the ILC beam energy \dots
\end{quote}

\noindent Finally, constraining new interactions:

\begin{quote}
The  ILC  will  also  constrain  or  discover  new  interactions  at  higher  mass  scales  through  pair production of quarks and leptons, $W$ and $Z$ bosons, and top quarks.  Much of our detailed knowledge of the current Standard Model comes from the precision measurement of the properties of the $Z$ boson  at $e^+ e^-$ colliders.  The  ILC  will  extend  this  level  of  precision  to  the $W$ boson  and  the  top quark.  The ILC will measure the mass of the top quark in a direct way that is not possible at hadron colliders, fixing a crucial input to particle physics calculations \dots
\end{quote}

The TDR outlines several reasons why the ILC is the preferred tool for these goals. First, \emph{cleanliness}. At the LHC a large number of background events contaminate each collision event, constraining the detector design to improve radiation hardness and forcing some detector elements away from the collision point. At the ILC the number of background events from spurious collisions is much lower, so that detectors are not as limited by radiation hardness constraints and may be placed very near the collision point. Second, \emph{democracy}. ILC signal cross sections are not much smaller than background cross sections since all backgrounds are electroweak in origin. At the LHC background from strong interaction processes are very high compared to signal processes. Third, \emph{calculability}. ILC theoretical cross sections are calculated with much greater precision because the associated uncertainty on QCD calculations are large; in contrast, $e^+ e^-$ cross sections are calculated at very high precision so that experimental deviations from the SM are more readily apparent. Finally, \emph{detail}. Due to the clean event environment and the potential to polarize beams, the detailed spins of initial and final states can be reconstructed.

Realizing the physics goals of the ILC program will require knowledge of the theoretical and experimental techniques fundamental to high energy physics, as well as the software written to simulate the underlying physics at the ILC and its detectors. The target audience for this primer is advanced undergraduates and beginning graduate students who have not yet had the benefit of a course in particle physics and who may be starting research on the ILC and one of its detectors. The goal is not to introduce particle physics at the ILC with depth and rigor, but rather to provide a fairly complete story in one place, together with references and suggestions for further reading where more depth and rigor can be found. The exercises are not meant to be deeply challenging but rather to provide a good starting point and working knowledge of particle physics and the technology used to study it.

In the first section we focus on the Standard Model (SM) of particle physics, describing the particles and gauge fields which mediate their interactions. Gauge invariance and the Higgs phenomenon are described. Next we focus on quantum scattering, first the nonrelativistic version in the Born Approximation and then the relativistic version encoded in the Feynman Calculus. We describe the production and decay of particles with prescriptions for how to calculate cross sections and lifetimes, then turn to the Higgs signal and background processes expected at a Higgs factory like the ILC. Suggestions for further reading follow at the end of the section, and exercises can be found in Appendix \ref{appendix1}.

In the following section we first survey the historical development of particle physics and the evolution in size and complexity of the machines driving that development. We then describe the fundamentals of particle accelerators and colliders, as well as the detectors built to study the results of particle collisions. We then focus on the technical designs of the ILC and SiD. SiD was first described in detail in the Letter of Intent (LoI) \cite{Aihara:2009ad}. In the following section We switch from ILC physics to software meant to simulate that physics. We describe event generators, which produce particle four-vectors produced after collisions, and detector simulations, which simulate the response of a detector like SiD to the particles and their decay products. Techniques for the reconstruction of shortlived particles like the Higgs boson are elaborated. Suggestions for further reading follow at the end of both sections, and exercises can be found in Appendix \ref{appendix1}.

Most software in high energy particle physics runs on the Linux operating system. At the time of writing, CERN CentOS 7 (CC7), a version of CentOS, is the default distribution. Instructions on downloading and installing CC7 are available on the web. All of the simulation software discussed in this primer is freely downloadable on the web. Installation instructions can be found on the webpages easily located with a search engine. Familiarity with a shell like \emph{bash} or \emph{csh} is required for installing the software, but only a small subset of shell commands is required. Instructions for installing and using ILCsoft, the nominal software for the global ILC effort, can be found in Appendix \ref{appendix2}.

\section{Higgs factory physics}

\subsection{Standard Model \label{sec:sm}}

\begin{table*}
\begin{center}
\begin{tabular}{|c|c|c|c|c|c|c|c||c|c|c|c|} \hline
\multicolumn{4}{|c|}{Leptons} & \multicolumn{4}{|c||}{Quarks} & \multicolumn{4}{|c|}{Bosons} \\ \hline
 & Q & M & ID &  & Q & M & ID &  & Q & M  & ID \\ \hline \hline
$e^{\pm}$ & $\pm$1 & 0.0005 & $\mp$11 & $u$ & $+2/3$ & 0.002  & 2 & $g$ & 0 & 0  & 21\\
$\nu_e$ & 0 & 0 & 12 & $d$ & $-1/3$ & 0.005  & 1 & $\gamma$ & 0 & 0 & 22\\ \cline{1-8}
$\mu^{\pm}$ & $\pm$1 & 0.106 & $\mp$13 & $c$ & $+2/3$ & 1.28  & 4 & $Z$ & 0 & 91.2 & 23 \\
$\nu_{\mu}$ & 0 & 0 & 14 & $s$ & $-1/3$ & 0.095 & 3 & $W^{+}$ & $+1$ & 80.4  & 24 \\ \cline{1-8}
$\tau^{\pm}$ & $\pm$1 & 1.78  & $\mp$15 & $t$ & $+2/3$ & 173  & 6 & $W^{-}$ & $-1$ & 80.4  & -24  \\ \cline{9-12}
$\nu_{\tau}$ & 0 & 0  & 16 & $b$ & $-1/3$ & 4.18 & 5 & $H$ & 0 & 125.1 & 25 \\ \hline
\end{tabular}
\caption{Elementary fermions and bosons of the Standard Model (SM) with their electric charge (in $e$), mass (in GeV) and Particle Data Group (PDG) identification numbers. Masses are rounded and uncertainties are suppressed. For current mass precision, see the PDG \cite{Tanabashi:2018oca}.}
\label{tab:elementary}
\end{center}
\end{table*}

The Standard Model (SM) of particle physics comprises the elementary (noncomposite) particles and their strong, weak and electromagnetic interactions. The elementary spin 1/2 fermions are the quarks and leptons, while the elementary bosons are the spin 1 gauge bosons, which mediate interactions, and the spin 0 Higgs boson. See Table \ref{tab:elementary}. The SM also accounts for composite particles  in bound states of quarks $q$ like the mesons ($q\bar{q}$) and baryons ($qq^{\prime}q^{\prime \prime}$). See Table \ref{tab:mesons}.

The Lagrangian density $\mathcal{L}_{SM}=\mathcal{L}_{particles}+\mathcal{L}_{interactions}$ encodes the SM. If fields $\phi$ represent a scalar (the Higgs boson), $\psi$ a fermion (lepton or quark) and $A^{\mu}$ a vector boson, then

\begin{eqnarray}
\mathcal{L}_{particles} & = & \mathcal{L}_0 +\sum_{\psi} \mathcal{L}_{1/2}+ \sum_{A^{\mu}} \mathcal{L}_{1}
\end{eqnarray}

\noindent where $\mathcal{L}_0$, $\mathcal{L}_{1/2}$, $\mathcal{L}_1$, are the Lagrangians appropriate for spins 0, 1/2 and 1, namely

\begin{eqnarray}
\mathcal{L}_0  & = & \frac{1}{2}(\partial_{\mu} \phi)(\partial^{\mu}\phi)-\frac{1}{2} m_{\phi}^2 \phi^2 \\
\mathcal{L}_{1/2} & = &  \bar{\psi}(i\gamma^{\mu} \partial_{\mu}-m_{\psi}) \psi \\
\mathcal{L}_{1} & = & -\frac{1}{4} F_{\mu \nu}F^{\mu \nu} + \frac{1}{2} m_{A}^2 A^{\mu}A_{\mu} \label{eqn:l1}
\end{eqnarray}

\noindent Here $\gamma_{\mu}$ are the Gamma matrices and $F_{\mu \nu}$ is the field strength tensor. When the Euler-Lagrange equation is applied to these Lagrangians, they yield the Klein-Gordon, Dirac, and Maxwell equations. See Table \ref{tab:elementary} for the masses of the elementary fermions associated with $\psi$ and vector bosons associated with $A^{\mu}$ of the SM. 

\begin{figure*}[t]
\begin{center}
\includegraphics[width=0.9\textwidth]{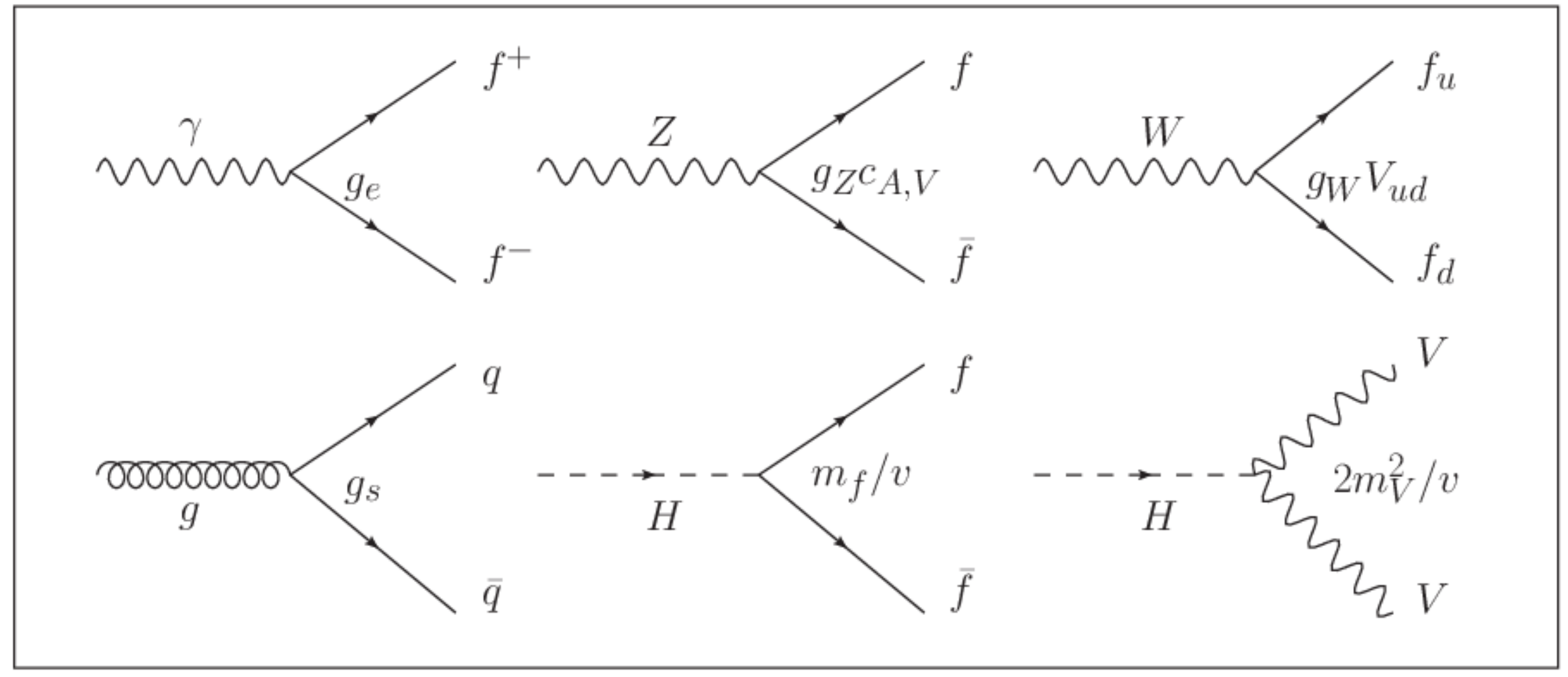}
\caption{Fundamental SM interaction vertices and their couplings for fermions (solid lines), gauge bosons (wavy or loopy lines) and the Higgs boson (dashed lines). For the $Z$ boson vertex, $c_{A}^{f},c_{V}^{f}$ are axial and vector factors, and for the $W$ vertex $V_{ud}$ is the Cabibbo, Kobayashi, Maskawa (CKM) matrix element. For the Higgs boson vertices, $v\approx 246$~GeV and $V=W,Z$.}
\label{fig:vertices}
\end{center}
\end{figure*}

The term $\mathcal{L}_{interactions}$ describes the electromagnetic, weak and strong interaction of particles in the SM, and their form is determined by demanding \emph{gauge invariance}. The field theories of the electromagnetic and strong interactions are referred to as Quantum Electrodynamics (QED) and Quantum Chromodynamics (QCD). The gauge groups are $U(1)$ for the electromagnetic interaction, $SU(2) \times U(1)$ for the electroweak interaction and $SU(3)$ for the strong interaction. See fig. \ref{fig:vertices} for diagrams of the SM $\gamma f^{+}f^{-}$, $Zf\bar{f}$, $Wf_u f_d$ and $gq\bar{q}$ interactions demanded by gauge invariance and the $Hf\bar{f}$ and $HVV$ interactions determined by the Higgs mechanism (see below). In addition to these interactions the SM includes the triple and quadruple boson interactions which do not involve fermions: $3g$, $4g$, $3H$, $4H$, $HHWW$, $HHZZ$, $ZWW$, $ZZWW$, $4W$, $\gamma WW$, $\gamma \gamma WW$, $\gamma Z WW$. The associated \emph{couplings} show the relative strength of each interaction with respect to the others. We have $g_s/g_e =\sqrt{\alpha_s/\alpha} \approx 4$, explaining why the strong interaction is considered strong (and why nuclei hold together). According to the \emph{electroweak unification condition} $g_e=g_W \sin \theta_W=g_Z \cos \theta_W$ where $\sin^2 \theta_W \approx 0.231$ is the \emph{weak mixing angle}, so in general weak interactions are not much weaker than electromagnetic ones. It is only for low energy weak phenomena $E/m_W \ll 1$, \emph{e.g.} nuclear beta decay, that the weak interaction is suppressed by $m_W^2$ relative to the electromagnetic interaction.

Not all vertices in fig. \ref{fig:vertices} have been observed in nature, in particular for vertices involving neutrinos. A particle which has nonzero spin may align its spin either with its direction of motion (righthanded, helicity +1) or against it (lefthanded, helicity -1). Processes in nature are \emph{spatially invariant}, or \emph{parity conserving}, if they occur with lefthanded or righthanded particles nonpreferentially. Most processes are parity conserving, but processes involving neutrinos violate parity conservation because righthanded neutrinos $\nu_{R}$ and lefthanded antineutrinos $\bar{\nu}_{L}$ are not observed in nature, and therefore excluded from the SM. Only $\nu_{L}$ and $\bar{\nu}_{R}$ exist in the SM. This has a critical consequence for weak interactions involving neutrinos. For example, the vertices $We^{-}_{L}\bar{\nu}_{R}$ and $We^{+}_{R}\nu_{L}$ exist in the SM, but $We^{-}_{R}\bar{\nu}_{L}$ and $We^{+}_{L}\nu_{R}$ do not.

The interaction vertices of fig. \ref{fig:vertices}, together with conservation laws, explain how the unstable particles of the SM decay. The electron and all neutrinos are stable against decay. Single quarks decay only within bound states, apart from the top quark through $t \rightarrow bW$ due to its short lifetime. Electron decay is ruled out by energy conservation, but muon decays can proceed through two connected diagrams: $\mu \rightarrow W^{\star} \nu_{\mu}$, where the $W$ is off mass shell, and $W^{\star} \rightarrow e\nu_e$. No other decay channel is available due to energy conservation. For the $\tau$ lepton, $\tau \rightarrow W^{\star} \nu_{\tau}$ and $W^{\star} \rightarrow e\nu_e, \mu \nu_{\mu}, q_u q_d$ are accessible if $q_u$, $q_d$ are first and second generation quarks. Onshell gauge bosons decay via $Z \rightarrow f\bar{f}$ and $W \rightarrow f_u f_d$ for all fermions except the top quark. The photon is stable. The Higgs boson decays through single vertices $H \rightarrow f\bar{f}$ and $H \rightarrow ZZ^{\star},WW^{\star}$ where one gauge boson is off mass shell, and also through multiple vertex processes.

The photon $\gamma$, stable with zero mass, is the exemplar of the gauge boson. Maxwell's field equations exhibit the canonical gauge invariance under $U(1)$ transformation, but only if the term $\frac{1}{2} m_{A}^2 A^{\mu}A_{\mu}$ in eq. \ref{eqn:l1} is zero. Similarly, the strong interaction exhibits invariance under $SU(3)$ transformation only if $m_{g}=0$. In contrast, the $Z$ and $W$ are \emph{massive}: the only particles more massive in the SM are the Higgs boson and the top quark. Early weak interaction theory predicted what their masses should be based on experimental data, but their nonzero mass was a puzzle. Electroweak $SU(2) \times U(1)$ gauge invariance is spoiled if the term in $\mathcal{L}_{1}$ (eq. \ref{eqn:l1}) quadratic in the $W$ and $Z$ fields is nonzero. How do mass terms for $W$ and $Z$ appear in $\mathcal{L}_{SM}$ if not through eq. \ref{eqn:l1}? The $W$ and $Z$ were discovered at the Super Proton Antiproton Synchrotron (S$p\bar{p}$S) collider at CERN in 1984 with $m_{W} \approx 80.4$~GeV and $m_{Z}\approx 91.2$~GeV, just where the experimental data pointed.

The explanation for how $W$ and $Z$ mass terms appear in $\mathcal{L}_{SM}$ is addressed by the \emph{Higgs mechanism}, which also predicts the Higgs boson $H$ and its couplings. We postulate a complex scalar field $\phi=(\phi_1 + i \phi_2)/\sqrt{2}$ in a potential $V(\phi)$. A generic potential is $V(\phi)=\sum_{n} c_{n} (\phi^{\star}\phi)^{n}$, but imposing theoretical constraints like renormalizability require  $c_{1}<0,c_{2}>0$, and $c_{n}=0$ for $n\neq 1,2$, so

\begin{eqnarray}
V(\phi) & = & \mu^2 \phi^{\star}\phi + \lambda (\phi^{\star} \phi)^2
\end{eqnarray}

\noindent where $\mu^2 < 0$ and $\lambda >0$. Then by construction we have symmetric ground states $\phi_0=\frac{v}{\sqrt{2}} e^{i \theta}$ ($v=\sqrt{-\mu^2/\lambda}$) for the Lagrangian  $\mathcal{L}_{\phi}=T-V=1/2( \partial_{\mu} \phi)^{\star} (\partial^{\mu} \phi) -V(\phi)$. If the symmetry is broken by choosing a particular $\theta$, \emph{e.g.} $\theta=0$, we have a ground state $\phi_0=\frac{v}{\sqrt{2}}$. For excitations $\kappa_1 + i \kappa_2$ near $\phi_0$, $\phi=\phi_0+\frac{1}{\sqrt{2}}(\kappa_1 + i \kappa_2)$, $\mathcal{L}_{\phi^{\prime}}$ yields a term quadratic in $\kappa_1$ with a coefficient $m_{\kappa_1}=\sqrt{-2 \mu^2}$. A mass term for a boson $\kappa_1$ has been generated by \emph{spontaneous symmetry breaking}. 

The SM fermions exhibit an interesting pattern: they fit into three \emph{generations} ordered by mass. Within each generation, fermions are paired together in $SU(2)$ \emph{electroweak doublets}. Each charged lepton $\ell$ is paired with its corresponding neutrino $\nu_{\ell}$ in the doublet $(\nu_{\ell} \ell)^{T}$, and each \emph{up-type} quark $q_u$ is paired with a \emph{down-type} quark $q_d$ in the doublet $(q_u q_d)^{T}$. Each fermion also has an anti-fermion of the same mass but opposite charge. The Large Electron Positron (LEP) collider established that there are exactly three generations, assuming $m_{\nu}<\frac{1}{2}m_{Z}$ for all neutrinos of generation four or higher, by measuring the cross section for $e^{+}e^{-} \rightarrow Z \rightarrow \sum_{\ell} \nu_{\ell} \bar{\nu}_{\ell}$ to to high precision.

The first generation, the least massive, contains the electron $e$ and its neutrino $\nu_e$ as well as the up quark $u$ and down quark $d$. Bound states of the $e$, $u$ and $d$ explain all ordinary matter bound up in atoms: the proton ($p=uud$) and the neutron ($n=udd$) are bound states of three first generation quarks. An atom with atomic number $Z$ and atomic weight $A$ contains $Z$ electrons bound to a nucleus with $Z$ protons and $A-Z$ neutrons. Quarks carry fractional charge: $q_{u}=+2/3$ and $q_{d}=-1/3$. Thus $u$ and $d$ quarks also explain the pions discovered in cosmic rays in 1937 ($\pi^+=u\bar{d},\pi^-=\bar{u}d$). 

The second generation contains the muon $\mu$ and its neutrino $\nu_{\mu}$, and the charm quark $c$ and strange quark $s$. The muon can be considered a heavy copy of the electron, with $m_{\mu}/m_{e} \approx 210$, but an unstable one since the muon can decay without violating energy conservation. The muon was discovered, like the pion, in cosmic rays. The second generation quarks are copies of the first generation quarks, with $m_{c}/m_{u} \approx 640$ and $m_s/m_d \approx 19$. Similar to the first generation, $q_{c}=+2/3$ and $q_{s}=-1/3$. While the second generation quarks do form bound states, the bound states are all unstable and decay to first generation free or bound fermions. The strange quark was discovered in 1947 in cosmic rays in the decay $K^0 \rightarrow \pi^+ \pi^-$ ($K^0=d\bar{s}$), while the the charm quark, or rather the bound state $J/\psi$ ($J/\psi=c\bar{c}$), was codiscovered in 1974 at the Stanford Positron Electron Accelerator Ring (SPEAR) and the Brookhaven Alternating Gradient Synchrotron (AGS) accelerator.

The third generation contains the $\tau$ and its neutrino $\nu_{\tau}$, and the top quark $t$ and bottom quark $b$. The $\tau$, the heaviest lepton, has $m_{\tau}/m_{e} \approx 3560$ and is unstable like the muon but has many more decay channels open. Like the $J/\psi$, the $\tau$ was discovered at SPEAR. The bottom quark $b$ has $m_{b}/m_{d}\approx 840$ and the top quark has a whopping $m_{t}/m_{u} \approx 8.65 \times 10^4$! Similar to the first generation, $q_{t}=+2/3$ and $q_{b}=-1/3$. The $b$ quark forms bound states with other quarks but the $t$, alone among quarks in this regard, decays well before it can form a bound state. The $b$ was discovered at Fermilab in 1977 in its bound state $\Upsilon$ ($\Upsilon=b\bar{b}$), while the $t$ had to wait until 1995 for discovery at the Fermilab Tevatron.

\begin{table*}[t]
\begin{center}
\begin{tabular}{|c|c|c|c|c|c|c|c|} \hline
\multicolumn{8}{|c|}{Mesons} \\ \hline
 & $q\bar{q}^{\prime}$ & Q & M [GeV]& $c\tau$ & ID & Decay1(BR) & Decay2(BR) \\ \hline
$\pi^{+}$ & $u \bar{d}$ & +1 & 0.140 & 7.80m & +211 & $\mu^{+} \nu_{\mu}$(1.000) & $e^{+} \nu_{e}$ (0.000) \\ 
$\pi^0$ & $u\bar{u}-d\bar{d}$ & 0 & 0.135 & 25.5nm & 111 & $\gamma \gamma$(0.988) & $e^+e^- \gamma$(0.012) \\ \hline
$K^+$ & $u\bar{s}$ & +1 & 0.494 & 3.71m & +321 & $\mu^+ \nu_{\mu}$(0.636) & $\pi^+ \pi^0$(0.207) \\ 
$K^{0}_{S}$ & $d\bar{s}$ & 0 & 0.498 & 2.68cm & 310 & $\pi^+ \pi^-$(0.692) & $\pi^0 \pi^0$(0.307) \\
$K^{0}_{L}$ & $d\bar{s}$ & 0 & 0.498 & 15.3m & 130 & $\pi^{\pm} e^{\mp} \nu_e$(0.406) & $\pi^{\pm} \mu^{\mp} \nu_{\mu}$(0.270) \\ 
$\phi$ & $s\bar{s}$ & 0 & 1.019 & 46.5fm  & 333 & $K^+ K^-$(0.492) & $K^{0}_{L}K^{0}_{S}$(0.340) \\ \hline
$D^+$ & $c\bar{d}$ & +1 & 1.870 & 312$\mu$m & +411 & $K^{0} \bar{K^{0}}X$(0.61) & $K^-X$(0.257) \\
$D^0$ & $c\bar{u}$ & 0 & 1.865 & 123$\mu$m & 421 & $K^- X$(0.547) & $K^{0} \bar{K^{0}}X$(0.47)  \\
$J/\psi$ & $c\bar{c}$ & 0 & 3.097 & 2.16pm & 443 & $ggg$(0.641) & $\ell^+ \ell^-$(0.119) \\ \hline
$B^+$ & $u\bar{b}$ & +1 & 5.279 & 491$\mu$m & +521 & $\bar{D^0}X$(0.79) & $D^- X$(0.099) \\
$B^0$ & $d\bar{b}$ & 0 & 5.280 & 455$\mu$m & 511 & $\bar{D^0}X$(0.474) & $D^- X$(0.369) \\
$\Upsilon$ & $b\bar{b}$ & 0 & 9.460 & 3.63pm & 553 & $ggg$(0.817) & $\ell^+ \ell^-$(0.075) \\ \hline
\multicolumn{8}{|c|}{Baryons} \\ \hline
& $qq^{\prime}q^{\prime \prime}$ & Q & M & $c\tau$ & ID & Decay1(BR) & Decay2(BR) \\ \hline
$p$ & $uud$ & +1 & 0.938 & $\infty$ & 2212 & - & - \\
$n$ & $udd$ & 0 & 0.940 & 264Gm &  2112 & $pe^- \bar{\nu_e}$(1.00) & - \\ \hline
$\Sigma^{+}$ & $uus$ & +1 & 1.189 & 2.40cm & 3222 & $p\pi^0$(0.516) & $n\pi^+$(0.483) \\
$\Sigma^0$ & $uds$ & 0 & 1.193 & 22.2pm & 3212 & $\Lambda \gamma$(1.00) & - \\
$\Sigma^{-}$ & $dds$ & -1 & 1.197 & 4.43cm & 3112 & $n\pi^-$(0.998) & $ne^- \bar{\nu}_e$(0.001) \\ \hline
\end{tabular}
\caption{Some common meson and baryons, their valence quark content, charge (in $e$), mass (in GeV), $c$ times lifetime, PDG identification number and two dominant decays with their branching ratios. Measured values are rounded and uncertainties are suppressed. For current precision, see the PDG \cite{Tanabashi:2018oca}.}
\label{tab:mesons}
\end{center}
\end{table*}

Because of a property of QCD known as \emph{confinement}, single quarks $u,d,s,c,b$ are not observed. Rather, when produced they form bound states with other quarks produced either in association or pulled from the vacuum. Bound states of quarks,  mesons ($q\bar{q}$) and baryons ($qq^{\prime}q^{\prime \prime}$), are \emph{colorless}. Color charge, the QCD analog of electric charge in QED, is either red, green or blue ($r,g,b$) and mesons carry color charge $r\bar{r}$, $g\bar{g}$ or $b\bar{b}$ while baryons carry color charge $rgb$ or $\bar{r} \bar{g}\bar{b}$. We have seen the first generation mesons ($\pi^+=u\bar{d}$), second generation mesons ($K^0=u\bar{s}$, $\phi=s\bar{s}$, $J/\psi=c\bar{c}$), as well as the third generation meson ($\Upsilon=b\bar{b}$), but there are many more. Similarly, the first generation baryons $p=uud$ and  $n=udd$ are only the tip of the iceberg. See Table \ref{tab:mesons} for a slightly larger tip of the iceberg and the PDG \cite{Tanabashi:2018oca} for the complete iceberg as it is presently known.

A meson, a bound state of two quarks, may have a variety of total spin and total angular momenta. Another degree of freedom, \emph{weak isospin}, is analogous to spin, and adds further to the variety. Thus mesons with the same valence quark content may nevertheless be distinct based on how their spin and isospin add. Distinct radial and orbital angular momentum quantum numbers can also yield distinct mesons. For example, see Table \ref{tab:g1nm} for the first generation mesons $\pi^0$,$\pi^{\pm}$,$\eta^0$,$\omega^0$,$\rho^0$ and  $\rho^{\pm}$, all of which have \emph{valence} quark content $u\bar{u},d\bar{d}$,$u\bar{d}$, $\bar{u}d$.

\begin{table}[t]
\begin{center}
\begin{tabular}{|c|c|c|c|c|c|c|} \hline
Meson & J & I & M [GeV]& $\Gamma$ [MeV]& ID \\ \hline
$\pi^{0}$ & 0 & 1 & 0.135 & $7.73 \times 10^{-6}$ & 111  \\
$\pi^{\pm}$ & 0 & 1 & 0.140 & $2.53 \times 10^{-14}$  & $\pm211$  \\
$\eta^0$ & 0 & 0 & 0.548 & $1.31 \times 10^{-3}$ & 221 \\ 
$\omega^0$ & 1 & 0 & 0.783 & 8.49 & 223 \\
$\rho^0$ & 1 & 1 & 0.775 & 149 & 113 \\ 
$\rho^{\pm}$ & 1 & 1 & 0.775 & 149 & $\pm213$ \\ \hline
\end{tabular}
\caption{Some first generation mesons with valence quark content $u\bar{u}$,$d\bar{d}$,$u\bar{d}$, $\bar{u}d$. They differ in their total angular momentum J and isospin I. Decays of the $\rho$, $\omega$ and  $\eta$ are to pions and photons. Measured values are rounded and uncertainties are suppressed. For current precision, see the PDG \cite{Tanabashi:2018oca}.}
\label{tab:g1nm}
\end{center}
\end{table}

\subsection{Quantums scattering \label{sec:scattering}}

\begin{figure*}[t]
\begin{center}
\framebox{\includegraphics[width=0.9\textwidth]{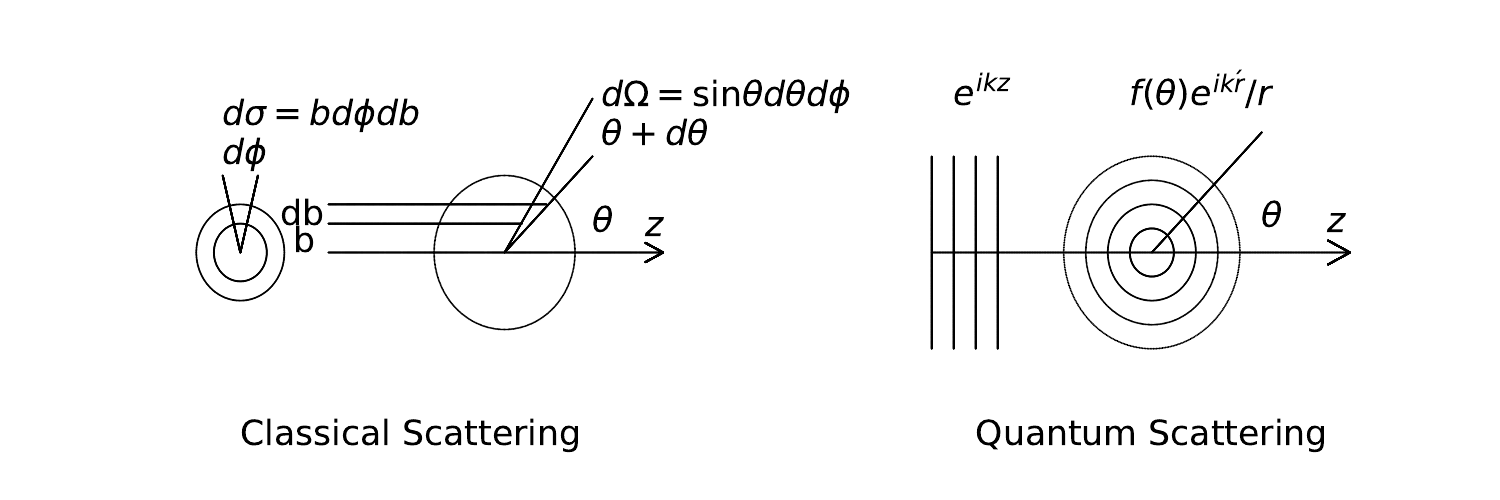}}
\caption{Diagrams for classical and nonrelativistic quantum elastic scattering. For classical scattering, the crucial functional relation is $\theta=\theta(b)$ determined by the force law. For nonrelativistic quantum scattering, it is the scattering amplitude $f(\theta)$ determined by the Schr{\"o}dinger equation.}
\label{fig:scattering}
\end{center}
\end{figure*}

The fundamental quantities which determine the number and kind of events produced in particle collisions are essentially geometric: the \emph{cross section} for the process and the \emph{luminosity} of particle production. The number of events $N$ produced in a process with cross section $\sigma$ and luminosity $\mathcal{L}$ is $N=\sigma \mathcal{L}$. The units of cross section are area, typically the \emph{barn} $b=10^{-28}$~m$^2$. Most processes of interest in modern particle physics have cross sections with a few femtobarns (fb) or picobarns (pb). Luminosity $\mathcal{L}$, also known as the \emph{integrated} or \emph{total} luminosity, therefore has units of inverse area, typically fb$^{-1}$ or pb$^{-1}$.

In classical scattering the incident and target particle are treated as Newtonian particles and the cross section can be calculated geometrically given a force law. For a central force the critical ingredient for the calculation of a cross section is the relation between the \emph{impact parameter} $b$ and the \emph{scattering angle} $\theta$. If the coordinate system is taken so that the origin is the target location and and the $z$ axis point in the direction of the incident particle, $b$ is the distance in the $xy$ plane from the incident particle to the $z$ axis, and $\theta$ is the polar angle from the $z$ axis. The target presents a cross-sectional area $\sigma$ to the incident particle. See fig. \ref{fig:scattering} (left).

Each small solid angle $d\Omega$ the incident particle scatters into contributes a quantity $d\sigma$ to the total cross section, the quantitative amount $d\sigma/d\Omega$ depending on the nature of the force law. This quantity $d\sigma/d\Omega$ is the \emph{differential cross section}. From fig. \ref{fig:scattering} it is clear that $d\sigma=bd\phi db$ and $d\Omega=\sin \theta d\theta d\phi$, so that

\begin{eqnarray}
\frac{d\sigma}{d\Omega} & = & \frac{b}{\sin \theta} \left\vert \frac{db}{d\theta} \right\vert
\end{eqnarray}

\noindent In the case of a pointlike particle scattering off a hard sphere with radius $R$, for example, the relation is $b=R\cos \frac{1}{2}\theta$. In this case $d\sigma/d\Omega=\frac{1}{4}R^2$ and $\sigma=\int d\sigma=\pi R^2$, the cross-sectional area of the sphere. For $b \leq R$ we have one collision, so the luminosity $\mathcal{L}=1/\pi R^2$. For $b > R$ we have no collision so $\mathcal{L}=0$. 

In accelerators, we generalize this notion of luminosity to include bunches of colliding particles, not just single particles, repeatedly colliding at fixed intervals of time. The \emph{instantaneous luminosity} is $L=d\mathcal{L}/dt$. See eq. \ref{eqn:lumi}. The total number of scattering events from beam collisions is 

\begin{eqnarray}
N & = & \int dt \frac{d\mathcal{L}}{dt} \int d\Omega \frac{d\sigma}{d\Omega}
\end{eqnarray}

\noindent where the integrations are over time and solid angle.

In nonrelativistic quantum scattering in a central potential $V(r)$, the incident particle is treated as a plane wave and the scattered particle is treated as a spherical wave. The ansatz is a superposition of these two,

\begin{eqnarray}
\Psi(r,\theta) & = & A \left( \exp (ikz)+\frac{f(\theta)}{r} \exp(ikr) \right)
\label{eqn:ansatz}
\end{eqnarray}

\noindent where $f(\theta)$ is the \emph{scattering amplitude} determined by solving the time-independent Schr{\"o}dinger equation. See fig. \ref{fig:scattering} (right). By equating the plane wave probability flowing into the scattering center with the spherical wave probability flowing out it can be shown that $d\sigma/d\Omega=\vert f(\theta) \vert^2$.

This ansatz follows naturally from the \emph{first Born approximation}. The time independent Schr{\"o}dinger equation can be cast in integral form with the aid of a Green's function,

\begin{eqnarray}
\Psi(r) & = & \Psi_0(r)+\int d^3r_0 g(r-r_0)V(r_0)\Psi(r_0) \\
g(r) & = & - \frac{m}{2 \pi \hbar^2}\left( \frac{\exp (i k r}{r} \right)
\end{eqnarray}

\noindent where $g$, which resembles the Green's function, is known as the \emph{propagator}. If $\Psi_0=A\exp(ikz)$ is the plane wave of the incident particle, the Born series iteratively bootstraps solutions 

\begin{eqnarray}
\Psi_1 & = & \Psi_0 + \int gV \Psi_0 \\
\Psi_n & = &  \Psi_{n-1} + \int g^nV^n \Psi_0
\label{eqn:born}
\end{eqnarray} 

\noindent where for clarity some notation has been omitted. Each successive term is a correction to the previous terms which, in principle, converges to the solution.

The first Born approximation is simply $\Psi_1$, the plane wave plus the Fourier transform of the potential. By comparing $\Psi_1$ with eq. \ref{eqn:ansatz}, the scattering amplitude can be extracted for a potential $V(r)$ which is localized at the scattering center and drops to zero elsewhere:

\begin{eqnarray}
f(\theta) & = & - \frac{2m}{\hbar^2 \kappa} \int dr r V(r) \sin (\kappa r) \label{eqn:scatamp}
\end{eqnarray}

\noindent where $\kappa=\vert k-k^{\prime} \vert$.

\begin{figure*}[t]
\begin{center}
\includegraphics[width=0.8\textwidth]{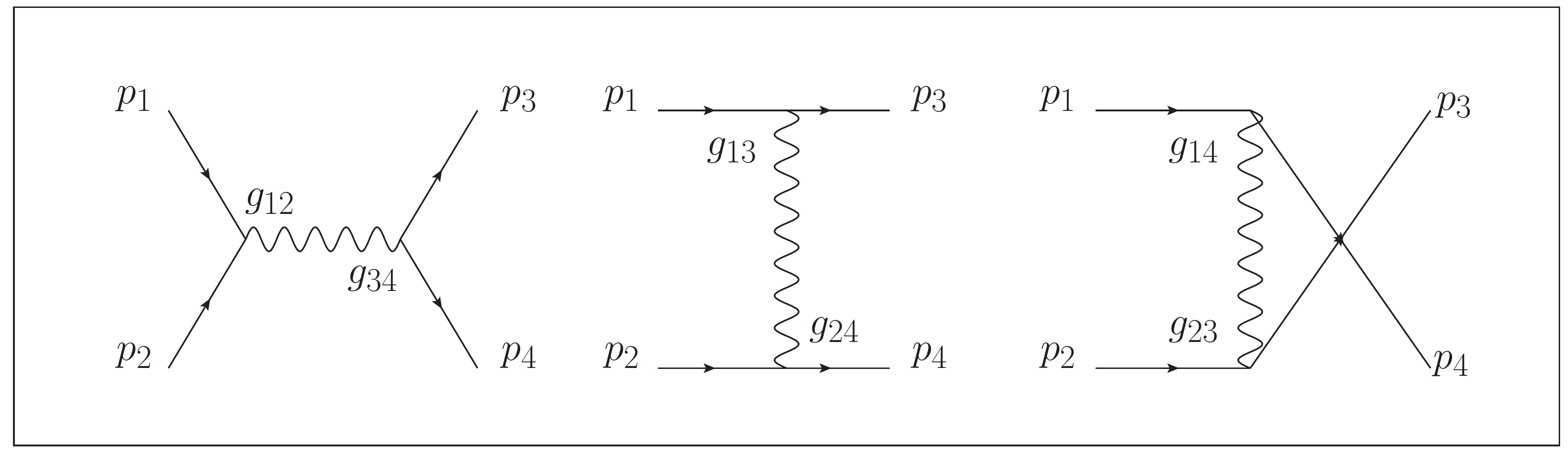}
\caption{Feynman diagrams for two-body scattering in the $s$-channel (left), the $t$-channel (middle) and $u$-channel (right). Amplitudes will have vertex factors $g_{12}g_{34}$, $g_{13}g_{24}$ and $g_{14}g_{23}$ respectively. Cross sections will depend on the Mandelstam variables $s$, $t$, and $u$ respectively.}
\label{fig:mandelstam}
\end{center}
\end{figure*}

In classical scattering and nonrelativistic Born scattering discussed above, the scattering is \emph{elastic}. We now generalize to relativistic scattering and consider \emph{inelastic} scattering, in which the interaction may produce new particles distinct from the incident particles and the concepts of luminosity and cross section generalize. We show how differential cross sections and lifetimes are calculated with the fully relativistic Feynman prescription.  The \emph{amplitude} $\mathcal{M}$ of the process is the key to calculating both cross sections and decay rates.

Fermi's Golden Rule states that the rate of a process from initial state $i$ to final state $f$ is the product of the phase space available in the final state $PS$ times the modulus of the amplitude squared, $\vert \mathcal{M} \vert ^2$:

\begin{eqnarray}
T_{i \rightarrow f} & = & \frac{2\pi}{\hbar} \vert \mathcal{M} \vert^2 \times PS
\end{eqnarray}

\noindent The amplitude $\mathcal{M}$ is calculated using Feynman rules described below. For each particle in the final state $f$ there is a contribution $\frac{c}{(2\pi)^3} \frac{d^3p}{2E}$ to $PS$ and an overall delta function to enforce energy conservation.

In the case of scattering the amplitude $\mathcal{M}$ is similar to the scattering amplitude $f(\theta)$ from Born scattering. The coupling of the scattered particles is a factor in the amplitude. In the limit of zero coupling or zero $PS$ in the final state, the transition rate is zero. In either case the process will not occur. For large couplings and $PS$, the transition rate is large. A large coupling can be counterbalanced by small $PS$, and \emph{vice versa}.

From Fermi's Golden Rule it can be shown that, after integrating phase space for two-body scattering $1+2 \rightarrow 3+4$ and two-body decay $1 \rightarrow 2+3$ of a particle with mass $m$, the differential cross section and decay rate are

\begin{eqnarray}
\frac{d\sigma}{d\Omega} & = &  S\left( \frac{\hbar c}{8\pi} \right)^2 \frac{\vert \vec{p}_f \vert}{\vert \vec{p}_i \vert} \frac{\vert \mathcal{M} |^2}{(E_1+E_2)^2}  \\
\Gamma & = &  \frac{S \vert \vec{p}_f \vert}{8\pi \hbar c}  \frac{\vert \mathcal{M} \vert^2}{m^2}
\label{eqn:twobody}
\end{eqnarray}

\noindent where $E_1+E_2$ is the sum of energies in the initial state, $\vec{p}_f$ is the momentum of either final state particle, and $\vec{p}_i$ is the momentum of either initial state particle. The statistical factor $S=1/2$ for identical final state particles and $S=1$ for distinct final state particles.

For a given scattering or decay amplitude, there is a corresponding \emph{Feynman diagram} which connects the initial state particles to the final state particles through any number of intermediate vertices. The Feynman prescription for calculating an amplitude $\mathcal{M}$ for a Feynman diagram is this:

\begin{enumerate}
\item \emph{Momenta}. Label external momenta $p_i$, and internal momenta $q_j$.
\item \emph{Vertex Factor}. For each vertex with coupling $g$, write a factor $-ig$.
\item \emph{Propagator}. For each internal momentum $q_j$, write a factor $\frac{i}{q_{j}^{2}-m_{j}^2 }$.
\item \emph{Energy Conservation}. For each vertex $k_1,k_2,k_3$, write a factor $(2\pi)^4 \delta(k_1+k_2+k_3)$ .
\item \emph{Integration}. For each internal momentum $q_j$, integrate $\frac{1}{(2\pi)^4} \int d^4 q_{j}$.
\end{enumerate}

\noindent What remains after this procedure is $-i\mathcal{M}$. In fact this simplified prescription applies to scalars, rather than fermions or vector bosons, but broadly the idea is the same. Only a little more complexity is required to describe strong and electroweak interactions of fermions and vector bosons.

\subsection{Particle production and decay\label{sec:production}}

Particle production cross sections and decay rates are related in that both are calculated with the Feynman prescription using Feynman diagrams. We consider the cases of two-body production and two-body decay.

\textbf{Production.} It should be evident that a total cross section must be a relativistic invariant. In two-body scattering $1+2 \rightarrow 3+4$, the cross section must depend on the four-vectors $p_{\mu}^{1}$,$p_{\mu}^{2}$,$p_{\mu}^{3}$,$p_{\mu}^{4}$, but if it is a relativistic invariant it can \emph{only} depend on functions of the four-vectors which are relativistic invariants like the contractions $p_{\mu}p^{\mu}$. 

\begin{figure*}[t!]
\begin{center}
\includegraphics[width=0.9\textwidth]{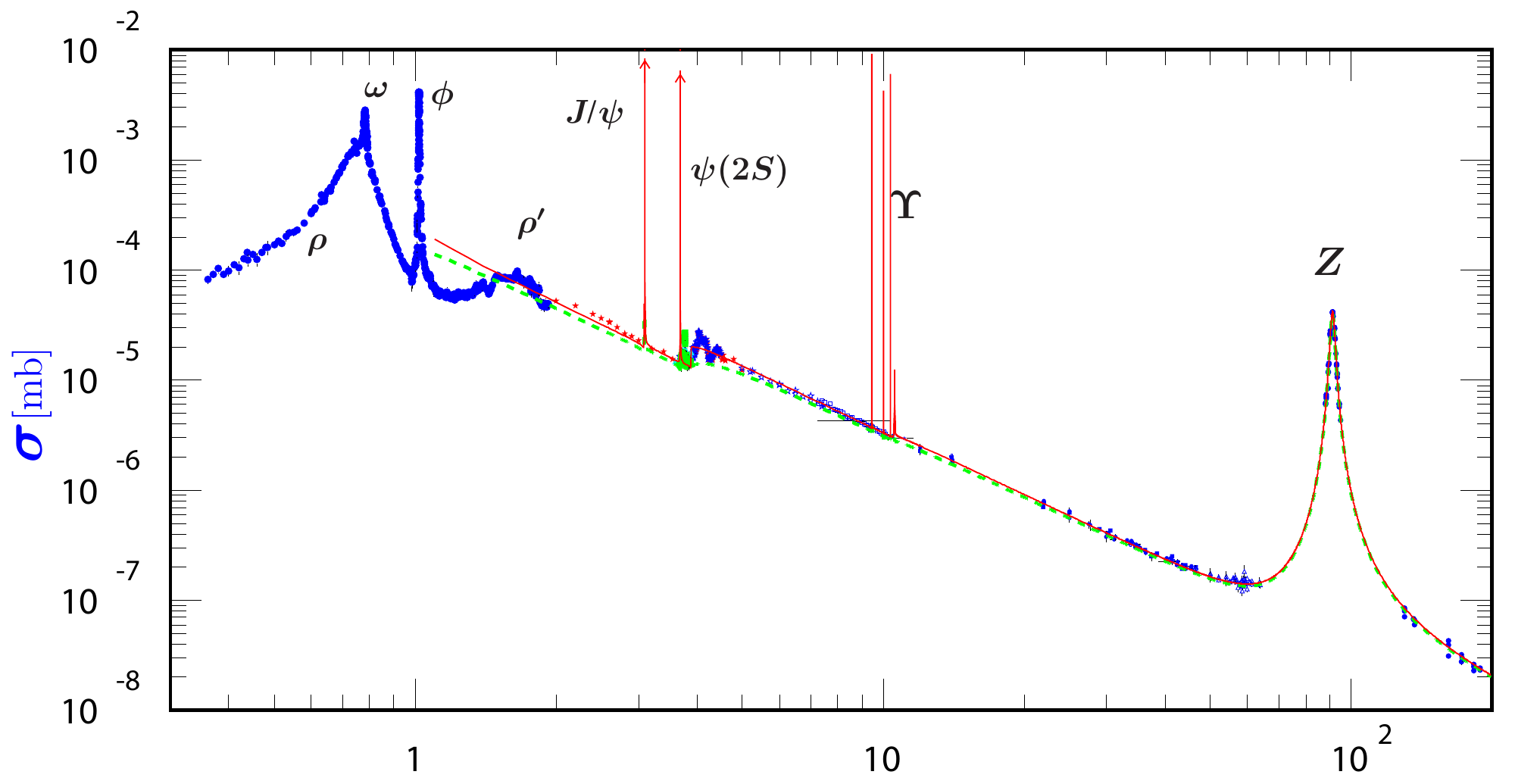}
\caption{Cross section for $e^+ e^- \rightarrow \sum_{q} q\bar{q}$ \emph{vs.} $\sqrt{s}$ (in GeV) with experimental data from various sources. Breit-Wigner meson resonances $u\bar{u},d\bar{d},s\bar{s},c\bar{c},b\bar{b}$ lie atop a nonresonant component with a $1/s$ dependence. At low $\sqrt{s}$ the $\gamma^{\star}$ process dominates, while at higher $\sqrt{s}$ the $Z^{\star}$ process dominates. The Breit-Wigner $Z$ resonance dominates at $\sqrt{s}=m_{Z}$. Credit: PDG \cite{Tanabashi:2018oca}.}
\label{fig:eetohad}
\end{center}
\end{figure*}

We define the \emph{Mandelstam variables} $s,t,u$, for two-body scattering $1+2 \rightarrow 3+4$ which are contractions and therefore relativistic invariants:

\begin{eqnarray}
s & = & (p_1+p_2)_{\mu}(p_1+p_2)^{\mu}  \\
t & = & (p_1-p_3)_{\mu}(p_1-p_3)^{\mu} \\
u & = & (p_2-p_3)_{\mu}(p_2-p_3)^{\mu} 
\end{eqnarray}

\noindent Note that in two-body scattering there are ten distinct contractions $p_{\mu}^{i} p_{j}^{\mu}$ and four conservation constraints $p_{\mu}^{1}+p_{\mu}^{2}=p_{\mu}^{3}+p_{\mu}^{4}$. There are seven $m_{i}^2$,s,t,u but it can be shown that $s-t-u=\sum_i m_{i}^2$. Therefore any two-body scattering cross section can be written as a combination of $s,t,u$ and three masses, or $s,t$ and all four masses. If the masses are negligible compared to $s,t,u$, then the latter are sufficient.

For each Mandelstam variable, there is a corresponding type of two-body scattering Feynman diagram or \emph{channel}: $s$-channel, $t$-channel, $u$-channel. See fig. \ref{fig:mandelstam}. When a propagator is combined with the delta function meant to enforce energy conservation at a vertex, the result is a simple function of a Mandelstam variable. For simplicity, we consider $m_i/E_i \ll 1$. For an $s$-channel Feynman diagram $\int d^4 q \delta(p_1+p_2-q)/q^2 \propto 1/s$. Thus the Mandelstam variables enter amplitudes naturally through the propagators. Moreover, the couplings $g_1$ and $g_2$ associated with the two vertices will contribute a factor $g_{1}g_{2}$ to the amplitude. If both $s$- and $t$-channel scattering are possible, then the total amplitude will be a sum $\mathcal{M}=\mathcal{M}_{s}+\mathcal{M}_{t}+\mathcal{M}_{u}$ and the diagrams will produce interference terms in $\vert \mathcal{M} \vert^2$.

We cite a few instructive inelastic scattering cross sections. First compare the cross sections for analogous processes, $e^+ e^- \rightarrow \gamma \gamma$ from QED and $q\bar{q} \rightarrow gg$ from QCD. Both have $t$- and $u$-channel diagrams, the former mediated by a virtual electron and the latter by a virtual quark:

\begin{eqnarray}
\left( \frac{d\sigma}{d\Omega}\right)_{e\bar{e} \rightarrow \gamma \gamma} & = & \frac{\alpha^2}{2s} \left( \frac{t^2+u^2}{ut} \right) \\
\left(\frac{d\sigma}{d\Omega}\right)_{q\bar{q} \rightarrow gg}  & = & \frac{8\alpha_{s}^2}{27s} \left( \frac{t^2+u^2}{ut} \right) \left(1-\frac{9tu}{4s^2} \right)
\end{eqnarray}

\noindent The vertices are from fig. \ref{fig:vertices}. Because $g_e=\sqrt{4 \pi \alpha}$ and $g_s=\sqrt{4 \pi \alpha_s}$, the processes have amplitudes proportional to $\alpha$ and $\alpha_s$ respectively, and cross sections proportional to $\alpha^2$ and $\alpha_{s}^2$. In the case of gluon pair production, however, there is also an $s$-channel process which interferes with the $t$-channel process due to the $3g$ vertex (there is no SM $3\gamma$ vertex).

Next consider quark pair production $e^+ e^- \rightarrow q\bar{q}$ in the $s$-channel mediated either by a virtual $\gamma$ or a virtual $Z$. For the virtual $\gamma$ process, there will be a vertex factor $g_e$ at the $e^+ e^-$ vertex and another vertex factor $g_e Q_{q}$ for the $q\bar{q}$ vertex, leading to an overall cross section factor of $\alpha^2 Q_{q}^2$, and the characteristic $1/s$ dependence. For the virtual $Z$ process, the vertex factors will yield an overall cross section factor of $X_q X_e g_{Z}^2$, where the shorthand $X_{f}\equiv (c_{V}^{f} )^2 + (c_{A}^{f} )^2$ is used. Because the $Z$ is unstable (whereas the $\gamma$ is not) the $Z$ propagator must also be modified, $m_{Z}^2 c^2 \rightarrow m_{Z}^{2}c^2 - i \hbar m_{Z} \Gamma_{Z}$, to account for the fact that the $Z$ decays with decay rate $\Gamma_{Z}>0$. 

The total cross sections are given by

\begin{eqnarray}
\sigma_{e\bar{e} \rightarrow \gamma^{\star} \rightarrow q\bar{q}} & = & 3 Q_{q}^2 \frac{4 \pi \alpha^2}{3s} \\
\sigma_{e\bar{e} \rightarrow Z^{\star}\rightarrow q\bar{q}} & = & X_{q} X_{e} \frac{g_{Z}^2 }{192 \pi} \frac{s}{(\sqrt{s}-m_{Z})^2 + m_{Z}^{2} \Gamma_{Z}^{2}}
\end{eqnarray}

\noindent For $\sqrt{s} \ll m_{Z}$, the total cross section is dominated by the virtual photon process, but closer to the $Z$ mass the virtual $Z$ diagram dominates. 

Note that the cross section diverges if the $Z$ is stable, \emph{i.e.} its decay width $\Gamma_Z=0$. Indeed all unstable particles have their propagators modified in this way. The $Z$ cross section is an example of a \emph{Breit-Wigner} cross section, characteristic of \emph{resonances} with very small lifetimes $1/\Gamma$. The full width at half maximum of a Breit-Wigner resonance is just the width $\Gamma$. See fig. \ref{fig:eetohad} for the total cross section for $e^+ e^- \rightarrow \sum_q q\bar{q}$ \emph{vs.} $\sqrt{s}$ together with experimental data. Breit-Wigner meson resonances $u\bar{u},d\bar{d},s\bar{s},c\bar{c},b\bar{b}$, along with the $Z$ resonance, lie atop a nonresonant component with a $1/s$ dependence.

\textbf{Decay.} Consider the decay rate $\Gamma$ of an unstable particle. Elementary particle decay is a purely statistical process, and occurs without regard for the history of the particle. For $N$ particles the small change in $dN$ in a small amount of time $dt$ is $dN=-N \Gamma dt$, from which it follows that $N(t)=N_0 \exp(-\Gamma t)$ and the mean \emph{lifetime} of a single particle is $\tau=1/\Gamma$. In general unstable particles may decay to a variety of final states, so the total decay rate is $\Gamma=\sum_f \Gamma_{f}$ where the sum is over all final states. The \emph{branching ratio} for an unstable particle to a particular final state $f$ is $BR(i \rightarrow f)=\Gamma_f/\Gamma$.

There is a natural connection between the decay rate $\Gamma$ of a particle and the uncertainty in its mass through the Heisenberg Uncertainty Principle, $\Delta E \Delta t \geq \hbar/2$. A shortlived particle with mass $m$ and mass uncertainty $\Delta E/c^2$ may exist for a short lifetime $\Delta t=1/\Gamma$, so at the minimum $\Delta E = \hbar \Gamma/2$. The mass will be in the interval $(m-\Delta E/c^2,m+\Delta E/c^2)$, so the \emph{natural width} of the particle is $2\Delta E/c^2=\hbar \Gamma$, or in natural units the width is $\Gamma$, the decay rate, with units of energy. Hereafter the terms decay width and decay rate are used interchangeably. 

The only leptons which decay are the $\mu$ and the $\tau$, both by virtual $W$ emission, the former via $\mu \rightarrow e \nu_e \nu_{\mu}$  with branching ratio near unity, the latter in a plethora of final states. The top quark decays via $t \rightarrow bW$ before hadronization can occur,with branching ratio near unity. All hadrons except the proton decay, also \emph{via} virtual $W$ emission from a quark within the hadron. The reader is referred to the PDG \cite{Tanabashi:2018oca} for $\tau$ and hadron partial decay widths and branching ratios. We consider here partial widths and branching ratios of the bosons $W,Z,H$.

For the $W$, the decay is either to leptons $\ell \nu_{\ell}$ or quark pairs $q_i \bar{q}_j$. For the $Z$, the decay is either to lepton pairs $\ell^+ \ell^-,\nu_{\ell},\bar{\nu}_{\ell}$ or quark pairs $q\bar{q}$. Applying the Feynman rules with vertex factors from fig. \ref{fig:vertices} yields

\begin{eqnarray}
\Gamma_{W \rightarrow \ell \bar{\nu},q_{i} \bar{q}_{j}} & = & \frac{\sqrt{2} G_{F} m_{W}^3}{12\pi} \times \left(1,  3 \vert V_{ij}\vert^2 \right)  \\
\Gamma_{Z \rightarrow \ell^+ \ell^-, \nu \bar{\nu}, q\bar{q}} & = &  \frac{\sqrt{2} G_F m_{Z}^3}{6 \pi} \times \left( X_{\ell}, X_\nu, 3X_q \right)
\end{eqnarray}

\noindent where $G_F\equiv \sqrt{2}g_{W}^{2}/8m_{W}^2$ is Fermi's constant, which absorbs the vertex factors. The extra factors $3$ appear for quarks because they carry an extra three degrees of freedom: strong color $r,g,$ or $b$. Leptons $\ell_i \bar{\nu}_{i}$ carry no color charge. See Table \ref{tab:wzbr} for the measured decay rates and branching ratios for the $W$ and $Z$.

\begin{table}[t]
\begin{center}
\begin{tabular}{|c|c|c|} \hline
Decay & $\Gamma_{f}$ (GeV) & BR (\%)\\ \hline
$W \rightarrow \ell \nu_{\ell}$ & 0.226 & $10.86\pm 0.09$  \\
$W \rightarrow$ hadrons & 1.41 & $67.41 \pm 0.27$ \\ \hline
$Z \rightarrow \ell^+ \ell^-$ & 0.08398 & $3.3658 \pm 0.0023$\\
$Z \rightarrow$ invisible & 0.4990 & $20.000 \pm 0.055$\\
$Z \rightarrow$ hadrons & 1.744 & $69.911 \pm 0.056$ \\ \hline
\end{tabular}
\caption{Measured $W,Z$ boson partial widths and branching ratios. Partial width uncertainties are suppressed. Total $\Gamma_W=2.085\pm0.042$~GeV, $\Gamma_Z=2.4952\pm0.0023$~GeV. From the PDG \cite{Tanabashi:2018oca}.}
\label{tab:wzbr}
\end{center}
\end{table}

\begin{table}[t]
\begin{center}
\begin{tabular}{|c|c|c|c|} \hline
Decay & $\Gamma_{f}$/MeV & BR/\% & $\mu/\mu_{SM}$\\ \hline
$H \rightarrow b\bar{b}$ & 2.35 & $57.7^{+3.21}_{-3.27}$ & $1.02^{+0.15}_{-0.15}$ \\
$H \rightarrow WW^{\star}$ & 0.875 & $21.5^{+4.26}_{-4.20}$ &  $ 1.08^{+0.18}_{-0.16} $\\
$H \rightarrow gg$ & 0.349 & $8.57^{+10.22}_{-9.98}$ & -\\
$H \rightarrow \tau^+ \tau^-$ & 0.257 & $6.32^{+5.71}_{-5.67}$ & $1.11^{+0.17}_{-0.17}$\\
$H \rightarrow c\bar{c}$ & 0.118 & $2.91^{+12.17}_{-12.21}$  & -\\
$H \rightarrow ZZ^{\star}$ & 0.107 & $2.64^{+4.28}_{-4.21}$ & $1.19^{+0.12}_{-0.11} $\\ 
$H \rightarrow \gamma \gamma$ & 0.00928 & $0.228^{+4.98}_{-4.89}$ & $1.10^{+0.10}_{-0.09}$ \\
$H \rightarrow \mu^+ \mu^-$ & 0.000891 & $0.0219^{+6.01}_{-5.86}$ & $0.6^{+0.8}_{-0.8}$ \\ \hline
Combined  & 4.07 & 100.0 & $1.10 \pm 0.11$ \\ \hline
\end{tabular}
\caption{Theoretical Higgs boson partial widths (uncertainties suppressed) and branching ratios for $m_{H}=125$~GeV at highest current order from the LHC Higgs Cross Section Working Group \cite{Dittmaier:2011ti,Dittmaier:2012vm,Heinemeyer:2013tqa}, and the measured signal strength relative to the SM $\mu/\mu_{SM}$ from the PDG \cite{Tanabashi:2018oca}. }
\label{tab:hbr}
\end{center}
\end{table}

\begin{figure*}[t]
\begin{center}
\includegraphics[width=\textwidth]{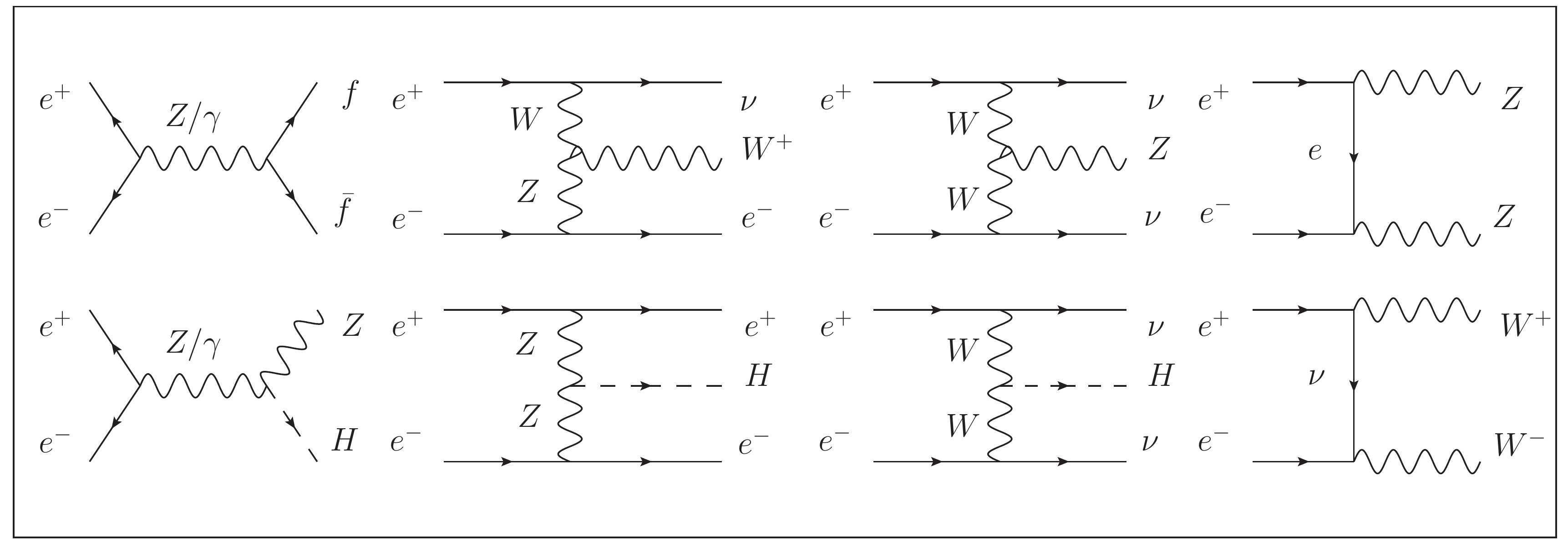}
\caption{Feynman diagrams for some main signals and backgrounds at the ILC. At far left, the $s$-channel diagrams for fermion pair production (top) and Higgstrahlung (bottom). The remaining $t$-channel diagrams, from left to right,  show $WZ$ fusion production of a single $W$ (top), $ZZ$ fusion production of a single $H$ (bottom), $WW$ fusion production of a single $Z$ (top) and $H$ (bottom), and diboson production of $ZZ$ (top) and $WW$ (bottom).}
\label{fig:ilcdiagrams}
\end{center}
\end{figure*}

Now  consider decay rates of the Higgs boson to fermion pairs $f\bar{f}$ and gauge boson pairs $ZZ,W^+W^-,gg,\gamma \gamma$.  For fermion pairs the Feynman diagram has a single vertex and the amplitude at first order is simply the vertex factor since there are no internal momenta.  The results at first order are:

\begin{eqnarray}
\Gamma_{H \rightarrow f\bar{f}} & = & n_{c} \frac{G_F m_{f}^2 m_{H}}{4 \pi \sqrt{2}} \left[ 1 - \frac{4m_{f}^2}{m_{H}^2} \right] ^{3/2} 
\end{eqnarray}

\noindent where $n_c=1$ for leptons and $n_c=3$ for quarks.

For the $ZZ$ and $W^+W^-$ diagrams the amplitudes are $\mathcal{M} \propto -2i m_{Z}^2/v,-2i m_{W}^2/v$. The results for $V=Z,W$ at first order are

\begin{eqnarray}
\Gamma_{H \rightarrow VV} & = & S \frac{G_F}{8\pi \sqrt{2}} m_{H}^3 (1-4 \lambda_{V})^{1/2} (12 \lambda_{V}^2-4 \lambda_V +1)
\label{eqn:offshell}
\end{eqnarray}

\noindent where $\lambda_V = (m_V/m_H)^2$ and $S=1$ for $V=W$ and $S=1/2$ for $V=Z$. Decay rates to offshell gauge bosons $ZZ^{\star}$ and $WW^{\star}$ are complicated due to the fact that one gauge boson is virtual since $m_{H}<2m_{V}$, and the decay rates in eqn. \ref{eqn:offshell} must be adjusted by phase space factors before comparison to experiment. 

Finally, decays to $gg$ and $\gamma \gamma$ do not occur at first order, and require integrations over internal momenta in top quark loops. The vertex factors are $g_{s}^2 m_{t}/v$ and $g_{e}^2 m_{t}/v$:

\begin{eqnarray}
\Gamma_{H \rightarrow gg,\gamma \gamma} & = & \frac{G_F m_{H}^3}{36 \sqrt{2} \pi} \left[ \frac{\alpha_s}{\pi} \right]^2 \vert I \vert^2, \frac{G_F m_{H}^3}{8 \sqrt{2} \pi} \left[ \frac{\alpha}{\pi} \right]^2 \vert I \vert^2
\end{eqnarray}

\noindent where $\vert I \vert^2 \approx 1$ contains the $m_t$ dependence.

As noted, these contributions to decay rates only represent the first order at which they occur. Higher order corrections can be large. See Table \ref{tab:hbr} for the calculated Higgs boson decay rates for $m_{H}=125$~GeV at the current highest order together with the currently measured signal strengths.

\subsection{ILC signal and background}

All processes at the ILC can be classified according to the number of fermions $f$ in their final state after boson decay. Thus $e^+e^- \rightarrow f\bar{f}$ is a 2f process, while $e^+e^- \rightarrow ZZ,WW$ are 4f processes. If the beam electron or positron splits $e \rightarrow \gamma e$, or if both split, then the initial state may contain one or two photons. Thus $\gamma e \rightarrow \gamma e$ is a 1f process, while $\gamma e \rightarrow e Z, \nu W$ are 3f processes. Processes 2f,4f also arise from $\gamma \gamma$ initial states: $\gamma \gamma \rightarrow f\bar{f},WW$. See fig. \ref{fig:ilcdiagrams} for the Feynman diagrams of some of the main 2f,4f,6f signals and backgrounds at the ILC.

\begin{figure*}[t]
\begin{center}
\includegraphics[width=0.9\textwidth]{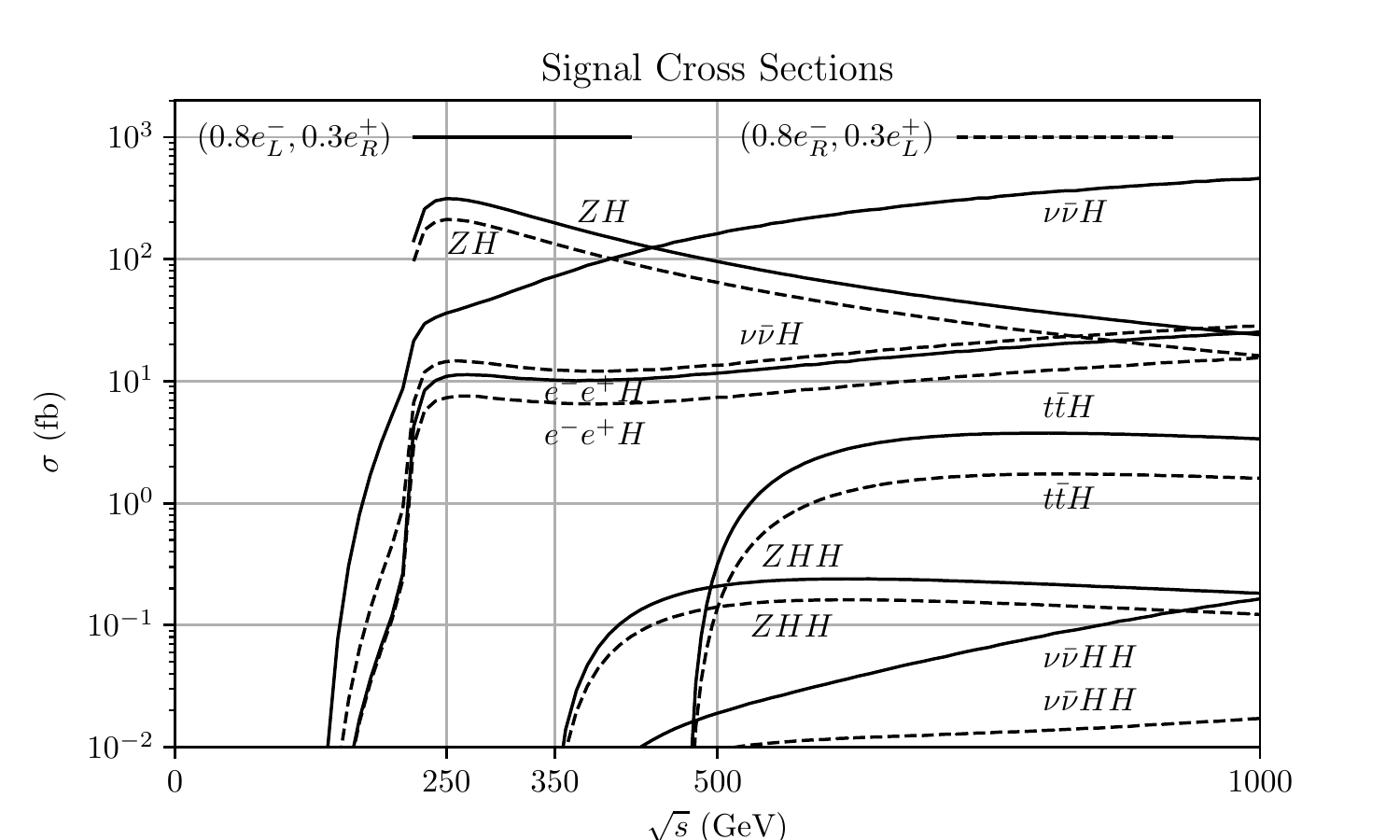}
\caption{Cross sections for Higgstrahlung, $WW$ fusion, $ZZ$ fusion, $t\bar{t}$ associated production and double $H$ production with $m_{H}=125$~GeV. ISR is included. Solid lines indicate polarized beams with 80\% lefthanded electrons and 30\% righthanded positrons. Dashed lines indicate 80\% righthanded electrons and 30\% lefthanded positrons. Obtained with Whizard 2.6.4 \cite{Kilian:2007gr}.}
\label{fig:hxsec}
\end{center}
\end{figure*}

\textbf{Signal.} The Higgs boson can be produced singly at the ILC in four ways: $e^+ e^- \rightarrow ZH$ (\emph{Higgstrahlung}), $e^+ e^- \rightarrow \nu \bar{\nu} H$ (\emph{WW fusion}), $e^+ e^- \rightarrow e^+ e^- H$ (\emph{ZZ fusion}) and $e^+ e^- \rightarrow t\bar{t}H$ (\emph{$t\bar{t}$} associated). Higgs bosons may also be produced doubly in more rare processes, $e^+ e^- \rightarrow ZHH$ (\emph{double Higgstrahlung}) and $e^+ e^- \rightarrow \nu \bar{\nu} HH$ (\emph{double WW fusion}).  In double Higgstrahlung the triple Higgs coupling $HHH$ is accessible, while in double $WW$ fusion the $HHWW$ coupling is accessible. Associated $t\bar{t}$ production and double Higgs production are only available at or above $\sqrt{s}=500$~GeV.

Higgstrahlung is an $s$-channel process in which the $H$ is radiated from a $Z$. Higgstrahlung turns on near threshold at $\sqrt{s} \approx m_{Z}+m_{H}$ with the $e_{R}^{+}e_{L}^{-}$ cross section rapidly reaching a maximum of $\sigma_{ZH} \approx 300$~fb near $\sqrt{s} \approx 250$~GeV. Thereafter it decreases with the characteristic $1/s$ dependence of an $s$-channel process, reaching $\sigma_{ZH} \approx 200$~fb (100~fb) near $\sqrt{s}=350$~GeV (500~GeV). The $e_{L}^{+}e_{R}^{-}$ cross section is approximately 2/3 of the $e_{R}^{+}e_{L}^{-}$ cross section.

Vector boson ($ZZ$ or $WW$) fusion production is a $t$-channel process in which a $Z$ or a $W$ is exchanged and the large $ZZH$ and $WWH$ couplings produce a Higgs boson. $WW$ fusion turns on at threshold and the $e_{R}^{+}e_{L}^{-}$ cross section rises to $\sigma_{\nu \nu H} \approx$ 37~fb (72~fb,162~fb) at $\sqrt{s} \approx 250$~GeV (350~GeV,500~GeV). $ZZ$ fusion $e_{R}^{+}e_{L}^{-}$ cross section rises to $\sigma_{\nu \nu H} \approx$ 11~fb (10~fb,12~fb) at $\sqrt{s} \approx 250$~GeV (350~GeV,500~GeV). The $e_{L}^{+}e_{R}^{-}$ cross section is approximately 2/3 of the $e_{R}^{+}e_{L}^{-}$ cross section for processes involving a $Z$ boson but considerably smaller for processes involving a $W$ boson.

See fig. \ref{fig:hxsec} for signal cross sections \emph{vs.} $\sqrt{s}$, and Table \ref{tab:ilcxsec} for signal cross sections at $\sqrt{s}=250,350,500$~GeV in the Higgstrahlung and vector boson fusion production channels assuming the nominal ILC design beam polarizations.

\begin{figure*}[p]
\begin{center}
\vspace{-1in}
\includegraphics[width=1.\textwidth]{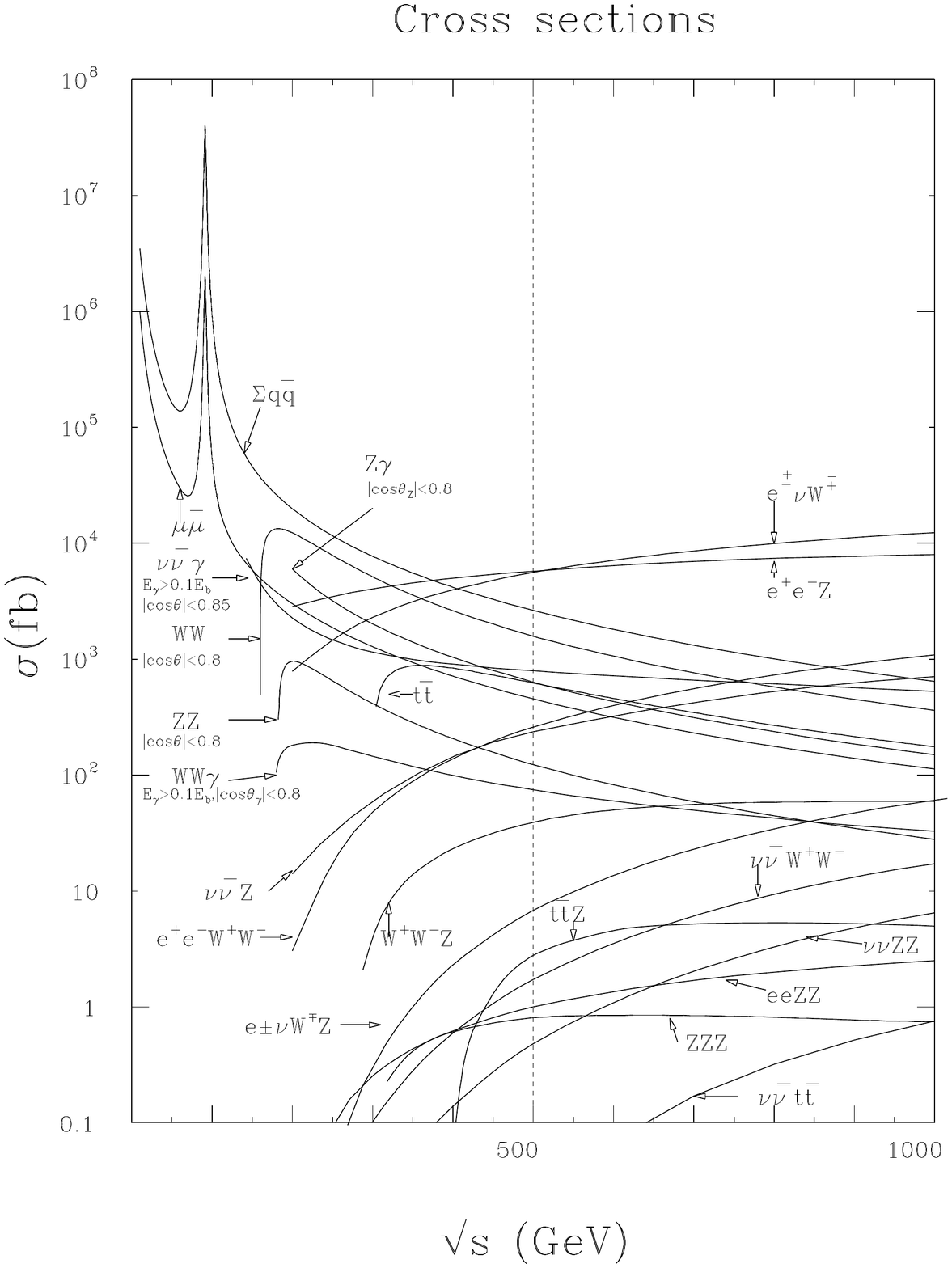}
\vspace{-0.5in}
\caption{Total cross sections for $e^+ e^-$ to various SM background final states \emph{vs.} $\sqrt{s}$ with unpolarized beams and without ISR or beamstrahlung.  The cross section for $\sum q\bar{q}$ below $\sqrt{s} \approx 100$~GeV is the same as fig. \ref{fig:eetohad}. Credit: ref. \cite{Murayama:1996ec}.}
\label{fig:ilcxsec}
\end{center}
\end{figure*}

\textbf{Background.} See fig. \ref{fig:ilcxsec} for the 2f,4f,6f bckground cross sections \emph{vs.} $\sqrt{s}$ assuming unpolarized beams. Since the main Higgs boson production processes are 4f or 6f, depending on its decay, and 2f backgrounds are fairly straightforwardly suppressed, the 4f and 6f backgrounds are the most important to consider here.

Beams at the ILC will be polarized. By polarizing the electrons and positrons $e_{L}^{+}e_{R}^{-}$ or $e_{R}^{+}e_{L}^{-}$ a process involving a $W$ boson can be turned on or off: if the process requires the $W$ to couple $\nu_R$ or $\bar{\nu}_{L}$ then it does not occur. Because it is not possible to polarize 100\% of electrons or positrons, there will be some fraction of the beams which do not contain the desired polarization. Hereafter we quote cross sections for 30\% polarized positron beams and 80\% polarized electron beams, the ILC design goal. For both signal and background, cross sections are higher for $e_{R}^{+}e_{L}^{-}$ than for $e_{L}^{+}e_{R}^{-}$ with the nominal polarization fractions, but in the case of background processes the difference is more dramatic.

See Table \ref{tab:ilcxsec} for 2f, 4f, 6f background cross sections for polarized $e^+e^-$ beams at $\sqrt{s}=$250,350,500 GeV calculated with Whizard 2.6.4 \cite{Kilian:2007gr}. No requirements have been imposed, except for the process $\gamma \gamma \rightarrow W^+ W^-$, for which a minimum $t$-channel momentum transfer requirement $q^2>1$~MeV and $e \rightarrow e\gamma$ splitting function $x>0.001$ are imposed in order to prevent divergence. Systematic uncertainties reported by Whizard are typically below 1\% but are in a few cases of order 10\%. More information on ILC backgrounds, including those with initial states $e\gamma$ and $\gamma \gamma$, can be found in \cite{Potter:2017rlo}, where the generator MG5 aMC@NLO \cite{Alwall:2014hca} has been used instead of Whizard.

\begin{table*}[t]
\begin{center}
\begin{tabular}{|l|c|c|c|c|c|c|c|c|c|} \hline
 Process & \multicolumn{2}{|c|}{$\sqrt{s}=250$~GeV} & \multicolumn{2}{|c|}{$\sqrt{s}^{\star}=250$~GeV} & \multicolumn{2}{|c|}{$\sqrt{s}=350$~GeV} & \multicolumn{2}{|c|}{$\sqrt{s}=500$~GeV} \\
         & $+,-$ & $-,+$ & $+,-$ & $-,+$ & $+,-$ & $-,+$ & $+,-$ & $-,+$  \\ \hline \hline
 $e^+ e^- \rightarrow ZH$ & 0.313 & 0.211 & 0.297 & 0.200 & 0.198 & 0.134 & 0.096 & 0.064 \\
 $e^+ e^- \rightarrow \nu \bar{\nu} H$ & 0.037 & 0.015 & 0.034 & 0.014 & 0.072 & 0.012 & 0.162 & 0.014\\
 $e^+ e^- \rightarrow e^+ e^- H$ & 0.011 & 0.007 & 0.010 & 0.007 & 0.010 & 0.006 & 0.012 & 0.007\\ \hline

$e^+ e^- \rightarrow b\bar{b}$ &  15.4 & 8.87 & 16.3 &  9.44 & 7.52  & 4.34 & 3.72 & 2.14 \\ 
$e^+ e^- \rightarrow c\bar{c}$ &  15.5 & 9.64 & 16.5 & 10.7 & 7.76 & 5.03 & 3.97 & 2.42\\ 
$e^+ e^- \rightarrow u\bar{u},d\bar{d},s\bar{s}$ & 47.0 & 28.0 & 49.5 & 29.9 & 23.1 & 13.6 & 11.3 & 6.92\\ 
$e^+ e^- \rightarrow \tau^+ \tau^-$ & 6.10 & 4.74 & 6.36 & 5.02 & 3.02 & 2.43 & 1.57 & 1.21\\ 
$e^+ e^- \rightarrow \mu^+ \mu^-$ & 6.19 & 4.63 & 6.43 & 5.15 & 3.00 & 2.48 & 1.50 & 1.20\\ \hline

$e^+ e^- \rightarrow WW$ & 37.5 & 2.58 & 37.9  & 2.62 & 27.1 & 1.79 & 17.9 & 1.15\\ 
$e^+ e^- \rightarrow e^{\pm} \nu W^{\mp}$ & 10.2 & 0.109 &  10.4 & 0.108 & 10.1 & 0.134 & 10.9 & 0.215\\  
 $e^+ e^- \rightarrow e^+ e^- Z$ & 2.51 & 2.63 & 2.38 & 2.13 & 2.64 & 2.23 & 2.64 & 3.04\\ 
 $e^+ e^- \rightarrow ZZ$ & 1.80 & 0.827 & 1.82 & 0.837 & 1.20 & 0.552 & 0.761 & 0.348 \\ 
$e^+ e^- \rightarrow \nu \bar{\nu} Z$ & 0.354 & 0.117  & 0.347 & 0.117 & 0.470 & 0.092 & 0.780 & 0.088 \\ \hline
 $e^+ e^- \rightarrow t\bar{t}$ & 0 & 0 & 0 & 0 & 0.267 & 0.117 & 0.890 & 0.421 \\ 
$e^+ e^- \rightarrow WWZ$ & 0 & 0 & 0 & 0 & 0.024 & 0.002 & 0.083 & 0.006 \\ 
 $e^+ e^- \rightarrow ZZZ$ & 0 & 0 & 0 & 0 & 0.001 & 0.000 & 0.002 & 0.001 \\ \hline
\end{tabular}
\caption{Cross sections (in pb) for some signal and background processes at the ILC for $\sqrt{s}=$250, 350, 500~GeV. ISR is included. For $\sqrt{s}^{\star}$ beamstrahlung is included, otherwise not. Beam polarization $+,-$ indicates 30\% righthanded positrons, 80\% lefthanded electrons while $-,+$ indicate 30\% lefthanded positrons and 80\% righthanded electrons. Obtained with Whizard 2.6.4 \cite{Kilian:2007gr}. }
\label{tab:ilcxsec}
\end{center}
\end{table*}

The cross sections in Table \ref{tab:ilcxsec} include an important effect: \emph{initial state radiation} (ISR). In a Feynman diagram with initial state $e^+e^-$ a photon may attach to either electron or positron. The photon carries away energy, effectively lowering the center of mass energy of the $e^+e^-$ system, subjecting the interacting particles to a cross section for a lower $\sqrt{s}$ than the nominal $\sqrt{s}$ of the beams. Thus ISR effectively increases the cross section for a process with a decreasing cross section \emph{vs.} $\sqrt{s}$, and decreases it for a process with an increasing cross section \emph{vs.} $\sqrt{s}$. The probability for ISR to occur and the resulting change in cross section is folded into the cross sections reported by Whizard. The effect can be dramatic: in \emph{radiative return} to the $Z$, including ISR at $\sqrt{s}=250$~GeV increases the $e^+ e^- \rightarrow q\bar{q}$ cross section fivefold.

As will be discussed in the next section, beam particles are bunched together and the \emph{bunches} are spaced discretely. One side effect is \emph{beamstrahlung}, photon radiation from electron or positron in the $e^+e^-$ system induced by the field of an oncoming bunch. The effect is similar to ISR: the effective $\sqrt{s}$ of the $e^+e^-$ system is lowered somewhat. For $\sqrt{s}=250$~GeV in Table \ref{tab:ilcxsec}, cross sections are reported both with beamstrahlung and without. Beamstrahlung is sensitive to the details of the beam parameters, and for the case shown in Table \ref{tab:ilcxsec} the parameters for the staged ILC250 \cite{Evans:2017rvt} are assumed.

Another side effect of bunching beam particles is \emph{pileup}. For signal processes and even many background processes, cross sections are low enough such that the probability of two overlaid events per bunch crossing is very low. However some background processes, like $t$-channel $\gamma \gamma \rightarrow e^+ e^-,q\bar{q}$, have high enough cross sections that they can contaminate the nominal event. The effect is to overlay the nominal event with one or more $e^+e^-$ or $q\bar{q}$ pairs. If pileup events are reconstructed correctly these 2f pairs are easily suppressed, but pileup events introduce problematic ambiguities in event reconstruction. These interactions are described in \cite{Schulte:1997nga}.

\subsection{Further reading and exercises}

For the SM, see \emph{Introduction to the Standard Model of Particles Physics} (Cottingham and Greenwood) \cite{Cottingham:396082} for a concise and elegant introduction. The chapter on nonrelativistic quantum scattering in \emph{Introduction to Quantum Mechanics} (Griffiths) \cite{Griffiths2004Introduction} is very good. 

For more depth and rigor, see \emph{Introduction to Elementary Particles} (Griffiths) \cite{Griffiths:2008zz}, \emph{Gauge Theories of the Strong, Weak, and Electromagnetic Interactions} (Quigg) \cite{Quigg:1983gw}, \emph{Quarks and Leptons} (Halzen and Martin) \cite{Halzen:1984mc} and \emph{Collider Physics} (Barger) \cite{Barger:1987nn}. For quantum field theory, see \emph{An Introduction to Quantum Field Theory} (Peskin and Schroder) \cite{Peskin:1995ev}.

The Particle Data Group review articles \emph{The Standard Model and Related Topics}, \emph{Kinematics, Cross Section Formulae and Plots} and \emph{Particle Properties} \cite{Tanabashi:2018oca} are invaluable. Particle physics continues to evolve, and the most recent and precise measurements can be found under the PDG \emph{Summary Tables} \cite{Tanabashi:2018oca}.

Exercises for this section can be found in sect. \ref{sec:higgs} of Appendix \ref{appendix1}.

\section{ILC: accelerators and detectors}

\subsection{Historical perspective}

\begin{table*}
\begin{center}
\begin{tabular}{|l|c|c|} \hline
Year & Recipient & Reason Given By Nobel Committee\\ \hline \hline
1984 & Carlo Rubbia* &  “discovery of the field particles W and Z” \\
1979 & Steven Weinberg* & theory of the unified weak and electromagnetic interaction \\
1976 & Burton Richter* & “discovery of a heavy elementary particle of a new kind” \\
1969 & Murray Gell-Mann & “classification of elementary particles and their interactions” \\
1968 & Luis Alvarez  & “technique of using hydrogen bubble chamber and data analysis” \\
1965 & Richard Feynman*  & “fundamental work in quantum electrodynamics” \\
1961 & Robert Hofstadter  &  “discoveries concerning the structure of the nucleons” \\
1958  &  Donald Glaser  &  “for the invention of the bubble chamber” \\
1945 & Wolfgang Pauli & “for the discovery of the Exclusion Principle” \\ \hline
1939  & E.O. Lawrence  &  “for the invention and development of the cyclotron”\\
1938 & Enrico Fermi & “demonstrations of the existence of new radioactive elements” \\
1936  &  Carl Anderson  &  “for his discovery of the positron” \\
1935  &  James Chadwick    &  “for the discovery of the neutron” \\
1931  & Paul Dirac*  &  “for the discovery of new productive forms of atomic theory” \\
1925  &  Charles Wilson  &  “making the paths of electrically charged particles visible”\\
1920  &  Niels Bohr  &  “structure of atoms and of the radiation emanating from them”\\
1921  & Albert Einstein  &  “discovery of the law of the photoelectric effect”\\
1906  &  J.J. Thomson  &  “conduction of electricity by gases”\\ \hline
\end{tabular}
\caption{Nobel Prizes in Physics awarded to physicists discussed in this section. Adapted from the Nobel Prize webpage \cite{nobelpage}. Asterisked names had co-recipients.}
\label{tab:nobel}
\end{center}
\end{table*}

In the previous section, the particles and interactions of the Standard Model (SM) were presented as an ahistorical \emph{fait accompli}, apart from mentioning where and when some particles were discovered. Such a presentation belies the metaphorical - and sometimes real - blood, sweat and tears of many physicists, both experimental and theoretical, over many decades, as well as the considerable cost of designing, building and operating the technology which provides the experimental foundation for the SM. See Table \ref{tab:nobel} for the physicists discussed in this section who were awarded the Nobel Prize in Physics.

The danger of this approach is in underestimating the magnitude of both the cost and the socio-technological challenge of building the ILC and its detectors. Before turning to the fundamentals of accelerators and detectors, we briefly remedy this shortcoming. The history of particle physics in the 20th century is a steady progression to higher energies required for resolving smaller particles. Shorter de Broglie wavelengths $\lambda=h/p$ are necessary for resolving smaller particles, requiring probes with ever increasing energy. At the beginning of the 20th century, probes from cathodes or radioactive nuclear decay were sufficient for tabletop discoveries, but as the century progressed more complex and expensive technology was required.

For the first generation of SM fermions, tabletop experiments carried out by one experimentalist, aided by a few assistants, were sufficient for major discoveries. J.J. Thomson discovered the electron in 1898 with a cathode ray tube, a simple handheld evacuated glass tube with low voltages for electron emission, acceleration and deflection. The detector was the glass tube itself. The experiment of Geiger and Marsden which led Ernest Rutherford to the discovery of the atomic nucleus in 1911 was a simple setup of a Radium source of incident alpha particles, a lead collimator, Gold foil for providing heavy nuclei targets and a phosphorescent screen of Zinc Sulfide for a detector.

Similarly, the discovery of the photon as the gauge boson which mediates the electromagnetic interaction occurred with considerable theoretical energy on the part of James Clerk Maxwell, Max Planck and Albert Einstein but, by today's standards, negligible cost and simple experimental technology. The photoelectric effect, blackbody spectrum, Compton scattering and Franck-Hertz experiments are easily demonstrated in beginning undergraduate physics courses. By the time the results of inexpensive spectroscopic experiments were being used by Niels Bohr and others to work out how the electron, nucleus and photon form the nonrelativistic quantum atom, the energies and event rates of the tabletop experiments were becoming insufficient for new discoveries. The tabletop experiments of nuclear beta decay, from which Wolfgang Pauli inferred the existence of the neutrino in 1930, and of James Chadwick, used to discover the neutron in 1932, were some of the last.

The first step away from the tabletop experiment came when physicists looked not to the earth for electrons traversing a voltage difference or nuclear fragments escaping a disintegrating nucleus, but to the heavens for a new source of energetic particles: secondary showers of particles created by collisions of highly energetic cosmic rays (protons or atomic nuclei) with atoms in the atmosphere. Like the tabletop experiments using radioactive nuclei, cosmic ray experiments could not provide a uniform energy or intensity, but the energies could be orders of magnitude larger than in the tabletop experiments and the event rates were large enough for new discoveries by sufficiently patient physicists.

The year the neutron was discovered, 1932, was also the year the positron $e^+$ was discovered among cosmic secondaries by Carl Anderson in a detector known as a cloud chamber, invented by Charles Wilson in 1911. The positron is the antimatter version of the electron $e^-$ first predicted by Paul Dirac with his fully relativistic quantum mechanics in 1931. The cloud chamber is an enclosed device filled with supersaturated water or alcohol which, when traversed by a charged particle, exhibits a visible track due to condensation centers made by ions created from the traversing charged particle. If a magnetic field is applied, the momentum can be inferred from the radius of curvature of the track. A few years after the positron was discovered, the charged pions $\pi^{\pm}$ and muons $\mu^{\pm}$ were discovered in photographic emulsions exposed to cosmic rays in the Bolivian Andes.

The neutral and charged kaons $K^0,K^{\pm}$ were also discovered in cosmic secondaries in 1947 in cloud chambers. The kaons were inferred from their visible decays $K^0 \rightarrow \pi^+ \pi^-$ and $K^+ \rightarrow \pi^+ \pi^- \pi^+$, unlike any known particle, and dubbed \emph{strange} particles. Shortly thereafter the meson discoveries were confirmed in accelerator experiments at Berkeley and Brookhaven (see below). Thereafter few major SM discoveries took place without accelerators, which provide the experimentalist with control over both the energies and event rates of their experiments. 

\begin{figure*}[t]
\begin{center}
\framebox{\includegraphics[width=0.3\textwidth, height=2in]{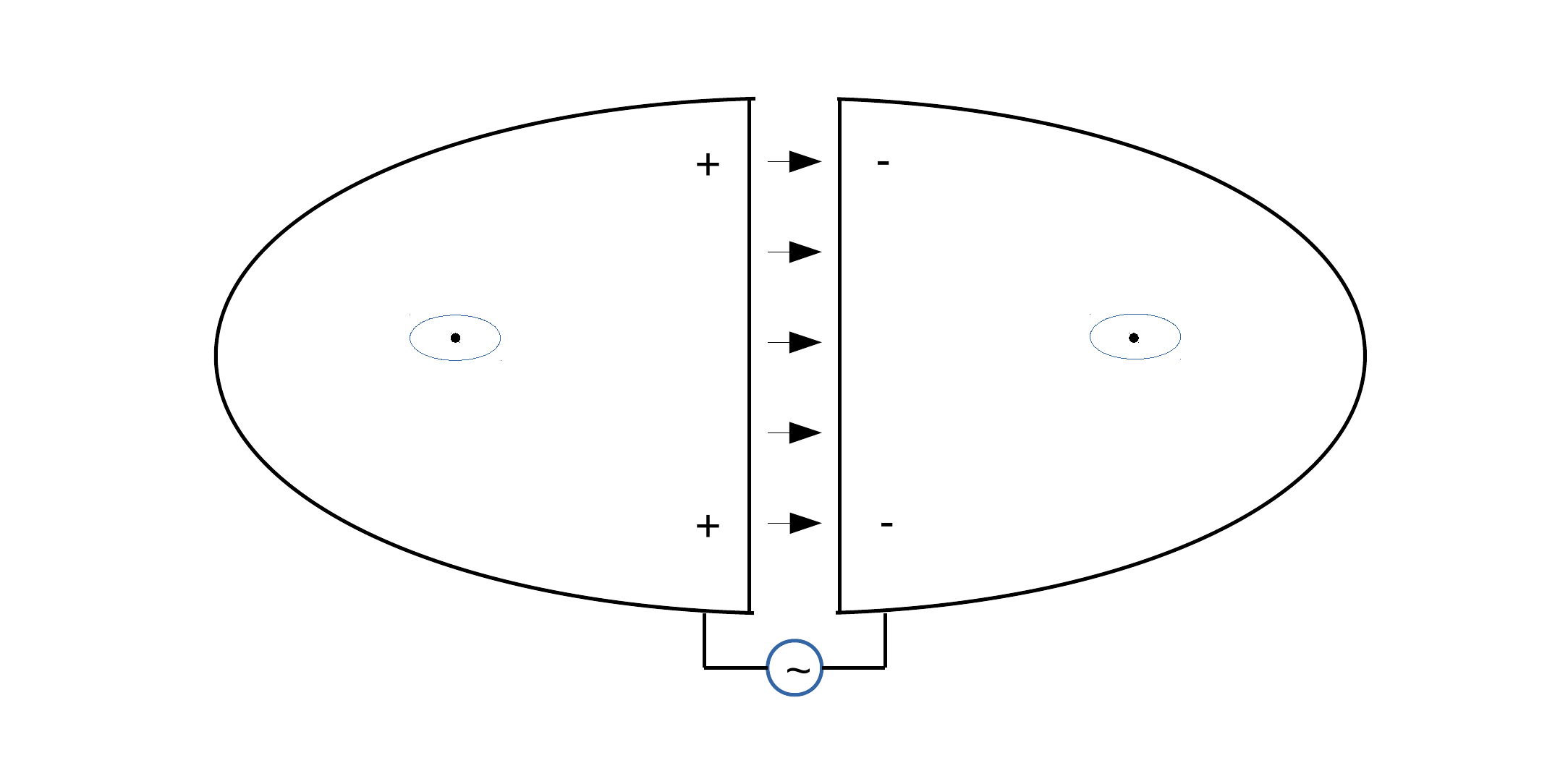}
\includegraphics[width=0.64\textwidth, height=2in]{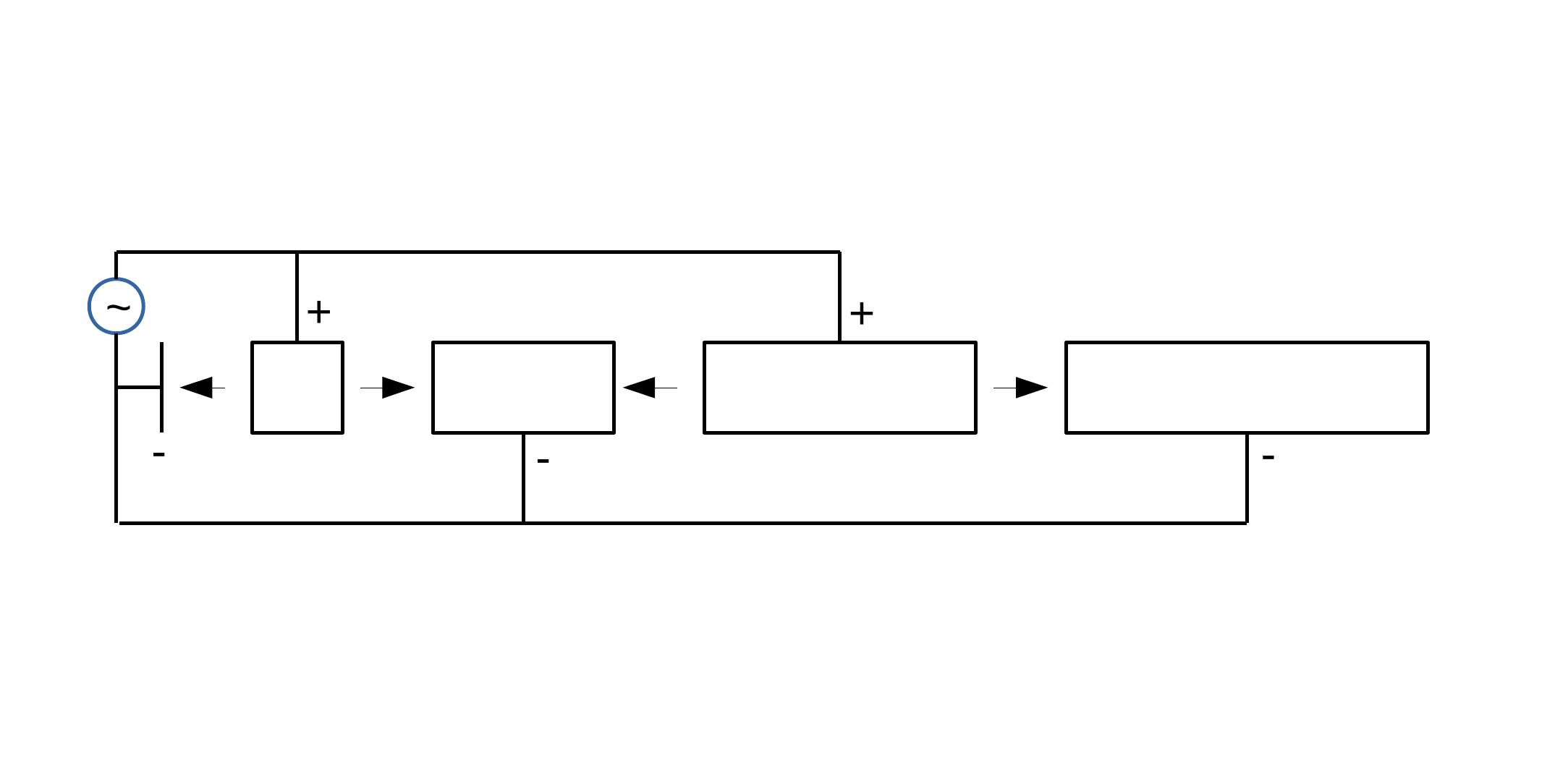}}
\caption{Schematic diagrams of a cyclotron (left) and linear accelerator (right). Encircled dots indicate constant magnetic fields perpendicular to the page in the cyclotron dees. Arrows indicate electric fields which oscillate in direction and magnitude due to the applied alternating current. Successively longer drift tubes after each linac voltage kick correspond to successively longer semicircles in the dees after each cyclotron voltage kick.}
\label{fig:linac}
\end{center}
\end{figure*}

But with experimental control comes the cost of the technology required for it, as well a new scale of scientific cooperation on a single experiment. It became clear that no single physicist, and no single university, could provide the funding  or personnel required for building and running the accelerator experiments. Only national governments could and, in the wake of World War II and the first use of nuclear weapons, many were willing to do so. Soon after the war many major national and international laboratories were formed to build and operate accelerators and their detectors at increasingly higher energies and luminosities.

In Europe, nations devastated by the war came together in 1954 to form a major new laboratory, the Conseil Europ{\'e}en pour la Recherche Nucl{\'e}aire (CERN) in Geneva. Shortly afterward in the Soviet Union, the Joint Institute for Nuclear Research (JINR) was established in Dubna in 1956. In West Germany, the Deutsches Elektronen Synchrotron (DESY) laboratory was formed in 1960 in Hamburg. In China, the Institute for High Energy Physics (IHEP) was established in 1973. In Japan, the K\={o} Enerug\={i} Kasokuki Kenky\={u} Kik\={o} (KEK, High Energy Accelerator Research Organization) was formed in 1997 in Tsukuba.

In the eastern US, a university consortium come together in 1947 to partner with the government to form a national laboratory at Brookhaven on Long Island, and in the western US the nuclei of later national laboratories were formed at Stanford University and the University of California at Berkeley. These later became the Brookhaven National Lab (BNL), the Stanford Linear Accelerator Center (SLAC) and the Lawrence Berkeley National Lab (LBNL). The Fermi National Accelerator Laboratory (FNAL), also known as Fermilab, came into being just outside Chicago in 1967. 

The center of accelerator research in postwar US was Berkeley under the leadership of Ernest O. Lawrence and later Luis Alvarez. Lawrence invented and developed the \emph{cyclotron}, a circular accelerator. The first version of the cyclotron was of tabletop dimensions, but subsequent versions were much larger. The cyclotron comprises two 'D' shaped magnets (\emph{dees}) placed back to back with a small gap, providing a uniform magnetic field $B$ pointing perpendicular to the faces of the dees. An alternating currently between the dees provides acceleration each time a charged particle traverses the gap, the polarity switching between gap crossings, and the magnetic field of the dees keeps the the particle in a circular orbit with increasing radius on each gap traversal. See fig. \ref{fig:linac} (left).

The resulting trajectory is an outward spiral. Including relativistic effects, the orbital frequency of a particle with mass $m$, charge $q$ and velocity $v$ is

\begin{eqnarray}
\nu & = & \frac{qB}{2\pi m} \sqrt{1-\frac{v^2}{c^2}}
\end{eqnarray}

\noindent where $\nu_0=qB/2\pi m$ is the \emph{cyclotron frequency} and $1/\gamma=\sqrt{1-v^2/c^2}$ is the relativistic correction factor. For a low energy particle like a 25 MeV proton, with $1/\gamma \approx 1$, an alternating current with fixed frequency will stay synchronized (within tolerance) with the particle, but for relativistic particles some correction must be applied to maintain synchronicity. 

In a \emph{synchrocyclotron}, the alternating current frequency is ramped to stay in sync with the particle. In a \emph{synchrotron} the magnetic field is ramped such that $B/\gamma$ is constant and the alternating current frequency can remain constant. In both cases accelerated particles must be bunched together at the same radius. Due to \emph{phase stability}, perturbations from the common radius are corrected by restoring forces. Both the cyclotron and the synchrocyclotron are limited by the amount of iron required for the dees, a severe cost constraint. With the synchrotron, a fixed orbital radius using magnets placed around a ring became possible and the dees were no longer necessary. One of the last functioning synchrocyclotrons built at Berkeley had a radius of 184 in and reached 720 MeV, the practical limit. The Berkeley Bevatron, a 6.5 GeV synchrotron, enabled the discovery of the antiproton in 1956 and the antineutron shortly afterward. All modern circular accelerators are synchrotrons.

By this time detectors had also advanced considerably over the cloud chamber. Donald Glaser invented the bubble chamber in 1952. In contrast to the cloud chamber, where tracks are formed from condensed liquid, the tracks in a bubble chamber are formed by vapor created by small energy deposits left by the traversing charged particle. The bubble chamber is filled with liquid gas just below the boiling point, then brought to expand with a piston into a supersaturated state which allows small vapor bubbles to form near the charged particle. Bubble chambers were used in experiments at the Brookhaven and  Berkeley machines, and thus the $\rho, \omega$ and $\eta$ mesons were observed in 1961. Synchronizing the bubble chamber piston with accelerator bunch timing, and using computers to analyze pictures of tracks in the bubble chamber, brought the state of affairs very close to modern accelerators and detectors. With the development of spark, streamer and drift chambers we have nearly arrived at modern trackers.

Meanwhile, at Stanford, the potential of the linear accelerator was being developed first under the leadership of Robert Hofstadter and later Pief Panofsky, SLAC Director from 1961 to 1984. See fig. \ref{fig:linac} (right). In 1954 Hofstadter discovered the finite size of the proton in an experiment using 188 MeV electrons from a linear accelerator, suggesting it was not a point particle but rather a composite particle. By measuring the scattering amplitude $f(\theta)$ of the electrons on a proton target, Hofstadter showed that it was not consistent with an amplitude predicted for a potential $V(r)$ from a pointlike proton. The proton had structure. A subsequent version of the linear accelerator first proposed by Hofstadter stretched 2 miles long and came into operation in 1966 under Panofsky with an energy of 17 GeV. Deep inelastic scattering experiments using electrons from the linear accelerator established the protons are noncomposite, with constituent \emph{partons}, thus helping establish the 1964 quark model of Gell-Mann. The theory of partons was developed by Richard Feynman, already famous for his work in QED. In a foray into circular machines, the Stanford Positron Electron Accelerator Ring (SPEAR) was built based on a design by Burton Richter, and SPEAR quickly discovered the $J/\psi$ and $\tau$.  Richter served as Director of SLAC from 1984 to 1999. When a positron linac beam was established and brought into collision with the electron linac beam in 1987, the first high energy linear collider was born as the Stanford Linear Collider (SLC).

At Fermilab, built and operated first under the directorship of Robert Wilson starting in 1967 and later Leon Lederman, the last major fixed target experiments were used to discover the $\Upsilon$ meson, the bound state of a $b$ quark and its antiquark. The Tevatron, a 1km radius $p\bar{p}$ synchrotron built to reach $\sqrt{s}$=1 TeV, came into operation in 1983. The top quark was discovered there in 1995. The threshold for Higgs boson production was reached by the Tevatron, but the luminosity was not sufficient for separation of signal from background.

In Europe, CERN had been aggressively pursuing large scale circular colliders and detectors. The Intersecting Storage Rings (ISR), which operated from 1971 until 1984, was the first hadron collider and reached energies up to $\sqrt{s}=$64 GeV. Under Herwig Schopper, Director General from 1981 to 1988, CERN not only carried out the experiment which led to the discovery of the $W$ and $Z$ bosons at the $\sqrt{s}=$540 GeV S$p\bar{p}$S synchrotron, but also proposed and began construction on the Large Electron Positron (LEP) collider, an 4.2km radius synchrotron which reached up to $\sqrt{s}=$200 GeV. Steven Weinberg and others had predicted the $W$ and $Z$ based on their theory of a unified electroweak interaction. Carlo Rubbia led the UA1 collaboration, which built the detector which discovered the $W$ and $Z$ in 1983, and served as CERN Director General from 1989 to 1993. LEP came online in 1989 and operated until 2000, when it had to make way for the Large Hadron Collider (LHC) in its tunnels. The LHC has reached $\sqrt{s}=$13 TeV.

\subsection{Accelerators and the ILC}

\subsubsection{Fixed target \emph{vs.} collider}

In a generic particle accelerator experiment, the particles in collision may have different momenta and a nonzero \emph{crossing angle}. In a \emph{fixed target} accelerator, one particle is stationary while the other is boosted. In a \emph{collider}, both particles are boosted. In a \emph{symmetric collider}, the colliding particles have equal but opposite momenta in the lab frame, while in an \emph{asymmetric collider} momenta are unequal in the lab frame. Early accelerator experiments were exclusively fixed target experiments but as the energy necessary to discover new particles increased, the collider came to dominate.

The reason is as follows. Consider two particles with four momenta $(E_1,\vec{p}_1)$ and $(E_2,\vec{p}_2)$ colliding to create a new particle. In a fixed target experiment $E_2=m_2$ and $\vec{p}_2=0$, so the sum is $(E_1+m_2,\vec{p}_1)$,  and

\begin{eqnarray}
m^2 & = & (E_1+m_2)^2-\vec{p}_{1}\cdot \vec{p}_1 \\
& = & m_{1}^{2}+m_{2}^{2}+2m_2 E_1
\end{eqnarray}

\noindent so that the mass reach is $m \propto \sqrt{E_1}$ for $m_1,m_2 \ll E_1$. But in a symmetric collider with no crossing angle colliding particles of equal mass, $E_1=E_2$ and $\vec{p}_1+\vec{p}_2=0$, so the sum is $(E_1+E_2,0)$ and

\begin{eqnarray}
m^2 & = & (E_1+E_2)^2-0^{2} \\
& = & 4E_{1}^2
\end{eqnarray}

\noindent so that the mass reach is $m \propto E_1$. 

Thus a collider with beam energies $\frac{1}{2}E$ can produce new particles of much higher mass than a fixed target accelerator with beam energy $E$. This is because in a fixed target accelerator most of the energy of the incident particle is used in conserving momentum and cannot go into creating a new particle. In a symmetric collider with no crossing angle all beam energy is available for new particle creation. Asymmetric colliders, colliders with crossing angles and colliders colliding particles with different mass are intermediate cases between fixed target and symmetric collider.

\subsubsection{Luminosity}

In sect. \ref{sec:scattering} we defined total luminosity $\mathcal{L}=1/A$ for a single collision of one incident particle with one target particle of cross sectional area $A$. Maximizing the rate of interesting events at a collider means maximizing the number of particles brought into collision per unit time, and  in accelerators particles are grouped and accelerated in \emph{bunches} of multiple particles.

\begin{table*}[t]
\begin{center}
\begin{tabular}{|c||c|c|c|c|c|} \hline
Collider & SPEAR $(e^+e^-)$ & S$p\bar{p}$S $(p\bar{p})$ & LEP $(e^+ e^-)$ & Tevatron $(p\bar{p})$ & LHC $(pp)$ \\ \hline \hline
$\sqrt{s}$ [GeV] & 8 & 630 & 209 & 1960 & 13000 \\
$C$ [m] & 234 & 6911 & 26659 & 6280 & 26659\\
$L$ [cm$^{-2}$s$^{-1}$] & $1 \times 10^{31}$ & $6 \times 10^{30}$ & $1 \times 10^{32}$ & $4.3 \times 10^{32}$ & $2 \times 10^{34}$ \\
Years & 1972-1990 & 1981-1990 & 1989-2000 & 1987-2011 & 2009-? \\
Laboratory & SLAC & CERN & CERN & Fermilab & CERN \\
Discoveries & $c,\tau$ & $Z,W$ & $n_{gen}=3$ & $t$ & $H$ \\ \hline
\end{tabular}
\caption{Historic circular colliders, their center of mass energies, circumference, peak luminosities, operational years, host labs and main discoveries. All are synchrotrons. Adapted from the PDG \cite{Tanabashi:2018oca}.}
\label{tab:circular}
\end{center}
\end{table*}

\begin{table*}[t]
\begin{center}
\begin{tabular}{|c||c|c|c||c|c|c|} \hline
$e^+ e^-$ & \multicolumn{3}{|c||}{Linear} & \multicolumn{3}{|c|}{Circular} \\ \cline{2-7}
Collider & SLC & ILC & CLIC & LEP & CEPC & FCCee \\ \hline
$\sqrt{s}$ [GeV] & 100 & 250,500 & 380,3000 & 209 & 240 & 240,366 \\ 
$D$ or $C$ [km] & $2 \times 1.5$ & $2 \times 20.5,31$ & $2 \times 11,50$ & 27 & 100 & 98 \\
$L$ [cm$^{-2}$s$^{-1}$] & $2.5 \times 10^{30}$ & $1.8 \times 10^{34}$ & $6 \times 10^{34}$ & $1 \times 10^{32}$ & $3.2 \times 10^{34}$ & $ 2.3 \times 10^{36}$ \\
Years & 1989-1998 & - & - & 1989-2000 & - & - \\ 
Laboratory & SLAC & KEK? & CERN? & CERN & IHEP? & CERN? \\ \hline
\end{tabular}
\caption{Parameters of possible future $e^+ e^-$ colliders and their direct antecedents, the Stanford Linear Collider (SLC) and LEP.  Adapted from the PDG \cite{Tanabashi:2018oca}.}
\label{tab:lepton}
\end{center}
\end{table*}

If $n_1$ particles in a bunch are incident on $n_2$ targets in a colliding bunch, and bunches are brought into collision at frequency $f$, then the time rate of particle-particle interactions is $f n_1 n_2$. Then we generalize the integrated luminosity $\mathcal{L}=1/A$ to the \emph{instantaneous luminosity},

\begin{eqnarray}
  L & = & f \frac{n_1 n_2}{A}
\label{eqn:lumi}
\end{eqnarray}

\noindent where $f$ is the bunch frequency and $n_1,n_2$ are the bunch populations. Thus maximizing luminosity means maximizing $f,n_1$ and $n_2$ while minimizing $A$ within accelerator constraints. Note that only $f$ and $A$ have dimensions so the units of $L$ are cm$^{-2}$s$^{-1}$. \emph{Integrated or total luminosity} is time integrated $\mathcal{L}=\int dt L$, and has units of cm$^{-2}$ (fb$^{-1}$, pb$^{-1}$, \emph{etc.}).

For bunches with Gaussian populations of horizontal width $\sigma_x$ and vertical width $\sigma_y$ at the interaction point, the bunch cross section is elliptical with area $A=4 \pi \sigma_x \sigma_y$ assuming axes of length $2\sigma_x$ and $2\sigma_y$. In many cases bunches are collected into \emph{pulses}, so $f=f_{r} n_{b}$ where $f_{r}$ is the pulse repetition rate and $n_{b}$ is the number of bunches per pulse. Assuming these expressions, eq. \ref{eqn:lumi} becomes

\begin{eqnarray}
  L & = & f_{r} n_{b} \frac{n_1 n_2}{4 \pi \sigma_{x}^{\star} \sigma_{y}^{\star}} \epsilon
\label{eqn:lumi2}
\end{eqnarray}

\noindent where the star indicates evaluation at the interaction point. The $\epsilon$ factor has been introduced to account for luminosity reductions due to crossing angle and other small accelerator effects, with $\epsilon \approx 1$ but $\epsilon  < 1$.

The cross sectional area of the bunches is not constant. As the bunches move toward the interaction point the $\sigma_x$ and $\sigma_y$ exhibit harmonic oscillation due to electromagnetic fields with an amplitude determined by $\beta_x$ and $\beta_y$, the \emph{amplitude functions}. To reach maximum luminosity, the accelerator is thus \emph{tuned} so that at the interaction point the amplitude functions, $\beta_x^{\star}$ and $\beta_y^{\star}$ are minimal. Finally, the horizontal and vertical \emph{emittance} are defined to be $\epsilon_{x,y} \equiv \sigma_{x,y}^{2}/\beta_{x,y}$ and so that eq. \ref{eqn:lumi2} is often written using $\sigma_{x,y}=\sqrt{\epsilon_{x,y} \beta_{x,y}}$

\subsubsection{Circular \emph{vs.} linear colliders}

If the earliest accelerator experiments were fixed target experiments, they were also \emph{linear} accelerator experiments. At one end of the line was the source for beam particles, while at the other end was the fixed target. But physicists soon realized that if the accelerator could be bent back around upon itself in a circle, the final collision energy could be greatly enhanced by multiple transits of the same accelerator, as with Lawrence's cyclotron.

See Table \ref{tab:circular} for the parameters of five historically important circular colliders: SPEAR is the Stanford Positron Electron Accelerating Ring,  S$p\bar{p}$S is the Super Proton Antiproton Synchrotron, LEP is the Large Electron Positron collider and LHC is the Large Hadron Collider.

Strong bending dipole magnets are required for keeping the beams in circular orbits. By equating the Lorentz force on a particle with charge $e$ and transverse momentum $p$ passing through a magnetic field $B$ with the centripetal force necessary for a circular orbit of radius $R$, one obtains

\begin{eqnarray}
p & = & aBR 
\label{eqn:pisqbr}
\end{eqnarray}

\noindent where $a \approx 0.3$~GeV/mT. This result holds for relativistic particles as it does for nonrelativistic ones. 

In a circular collider, counterrotating beams can be brought into collision at an \emph{interaction point} by specialized dipole magnets. Detectors are placed around the collision point (or points) to study the results of the collisions. Since beams made of bunches of identical charged particles will become unfocused over time due to electromagnetic repulsion, focusing quadrupole magnets are necessary to bring the bunches back into focus. Focusing quadrupoles usually alternate with bending dipoles in a circular collider.

\begin{figure*}[t]
\begin{center}
\vspace{-0.8in}
\resizebox{0.9\textwidth}{!}{\includegraphics*{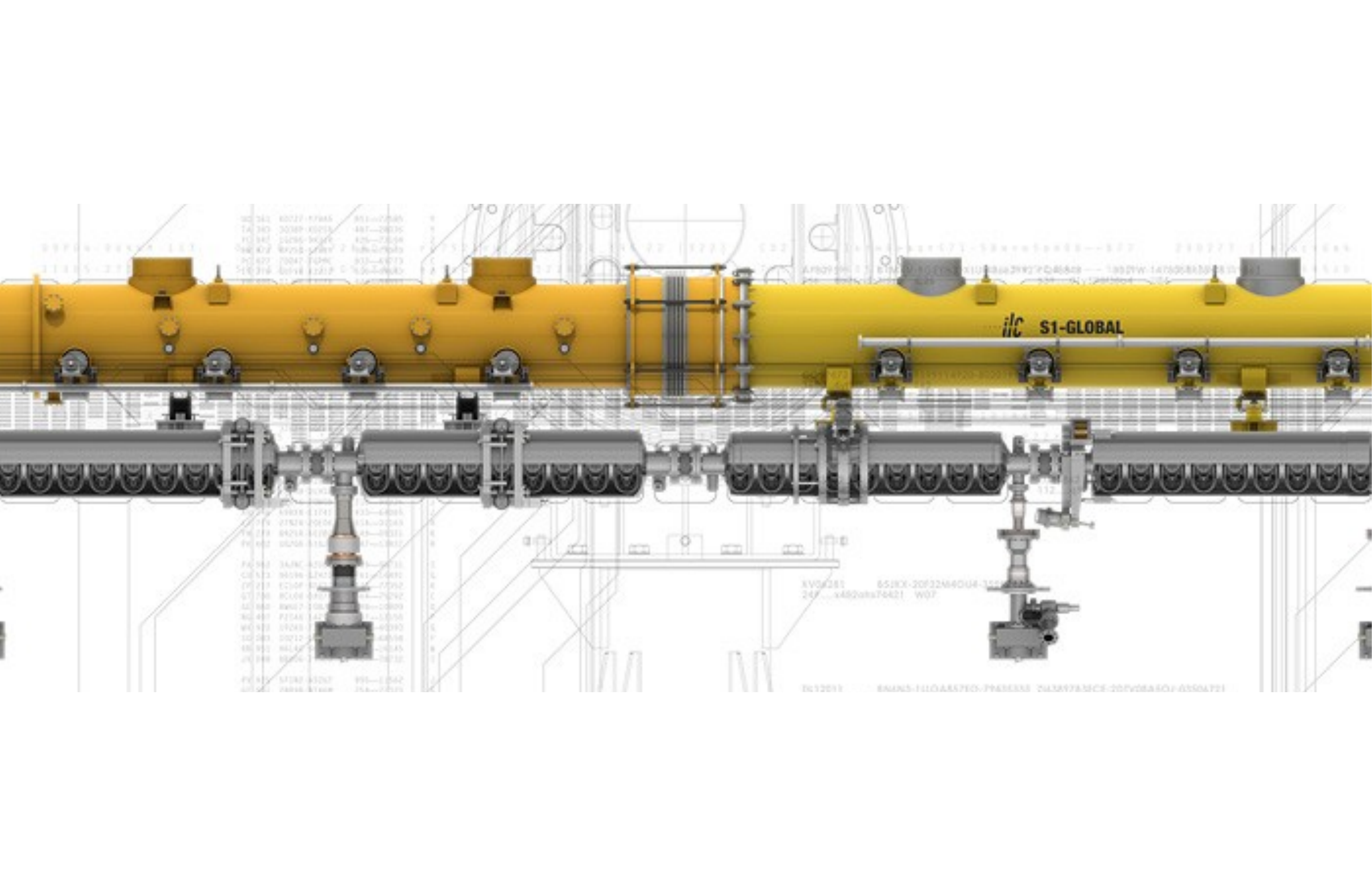}}
\vspace{-0.8in}
\caption{ILC cryogenic modules. Each module contains 9 cavities (Type A) or 8 cavities plus a quadrupole magnet (Type B), Helium tanks and support and extraction tubes. Approximately 1800 such modules are necessary for the nominal TDR ILC. Credit: $\textcopyright$ Rey Hori and KEK.}
\label{fig:module}
\end{center}
\end{figure*}

One important drawback for a circular collider is \emph{synchrotron radiation}. Charged particles in circular orbits radiate photons. For relativistic particles with charge $q$, energy $E$ and mass $m$,

\begin{eqnarray}
\Delta E & = & \frac{4\pi}{3 \epsilon_0} \frac{q^2}{R} \left( \frac{E}{m} \right)^4
\end{eqnarray}

\noindent for each orbit. Because $\Delta E \propto m^{-4}$, light particles are particularly susceptible to synchrotron radiation. Comparing  the two most commonly accelerated particles in a circular collider, $\Delta E_{e}/\Delta E_{p} = (m_{p}/m_{e})^4 \approx 10^{13}$, making electron losses considerably more severe than proton losses. Since energy lost to synchrotron radiation must be injected back into the beams in order to maintain fixed $\sqrt{s}$, the power required for $e^+ e^-$ colliders can be prohibitive. Since losses $\Delta E \propto E^4$, as higher center of mass energies are required to probe new physics the technological challenge of circular $e^+ e^-$ colliders will deepen.

For linear colliders there is no synchrotron radiation. In a simple linear accelerator, or \emph{linac}, drift tubes of successively greater length guide the beam particles while oscillating electric fields parallel to the beamline provide acceleration in the gaps between the drift tubes. Particles are bunched so they only experience the electric field when it accelerates them toward the target, and are within the drift tubes otherwise. A linear $e^+ e^-$ collider is made when a $e^+$ linac beam is brought into collision with a $e^-$ linac beam. A detector is then placed around the collision point. See Table \ref{tab:lepton} for the parameters of proposed linear and circular $e^+ e^-$ colliders ILC, the Compact LInear Collider (CLIC), the Circular Electron Positron Collider (CEPC) and the Future Circular Collider $e^+ e^-$ (FCCee), together with their direct historical antecedents, the Stanford Linear Collider (SLC) and LEP.

\subsubsection{International Linear Collider}

We note that for \emph{hadron colliders} like the LHC, both the luminosity and the center of mass energy can be misleading because they refer to the colliding hadrons, which are composite, not the underlying constituents of the hadron which undergo interactions during collision. By contrast, in a \emph{lepton collider} like the ILC the luminosity and center of mass energy refer directly to the elementary particles undergoing the interaction of interest.  

For processes at a proton collider like the LHC, the elementary particles interacting are gluons and quarks, and the share of the proton's energy carried by each gluon or quark described by a \emph{parton density function} is necessarily smaller than that of the proton. For processes at a lepton collider like the ILC, the elementary interacting particles are leptons where the parton density function is identically unity. Furthermore, in a lepton collider the initial state of an interaction is known on an event-by-event basis, whereas in a hadron collider it is not. In particular, momentum conservation along the beamline can be exploited at a lepton collider like the ILC but not at a hadron collider like the LHC.

The ILC design represents an international convergence of several decades of research and development. The design described in the TDR \cite{Behnke:2013xla,Baer:2013cma,Phinney:2007gp,Behnke:2013lya} calls for $\sqrt{s}=500$~GeV upgradeable to $\sqrt{s}=1000$~GeV, with 11 km linacs and a total footprint of 31 km including 6 km for damping rings and the beam delivery system, and another 3 km for the rings to the main linacs. In the ILC Machine Staging Report \cite{Evans:2017rvt}, the goal reverts to $\sqrt{s}=250$~GeV with possible upgrade to $\sqrt{s}=500$~GeV.

\begin{table*}[t]
\begin{center}
\begin{tabular}{|c|c|c|c|c|} \hline
Parameter & Staged $\sqrt{s}=250$~GeV & $\sqrt{s}=250$~GeV & $\sqrt{s}=350$~GeV & $\sqrt{s}=500$~GeV  \\ \hline \hline
$n_{1,2}$ & $2.0 \times 10^{10}$ & $2.0 \times 10^{10}$ &  $2.0 \times 10^{10}$ &  $2.0 \times 10^{10}$ \\
$f_{r}$ & 5Hz & 5Hz & 5Hz & 5Hz  \\
$n_{b}$ & 1312 & 1312 & 1312 & 1312 \\
$\sigma_{x}^{\star}$ & 516nm & 729nm & 684nm & 474nm \\
$\sigma_{y}^{\star}$ & 7.8nm & 7.7nm & 5.9nm & 5.9nm \\ \hline
$\int dt \mathcal{L}$ & 2000fb$^{-1}$ & 2000fb$^{-1}$ & 200fb$^{-1}$ & 4000fb$^{-1}$ \\ 
$-,+/+,-$ & 900/900fb$^{-1}$ & 1350/450fb$^{-1}$ & 135/45fb$^{-1}$ & 1600/1600fb$^{-1}$ \\ \hline
\end{tabular}
\caption{ILC beam parameters for various $\sqrt{s}$ from the TDR \cite{Phinney:2007gp} and the staging report \cite{Evans:2017rvt}, with projected integrated luminosities and luminosity sharing between 30\% negatively polarized positrons and 80\% positively polarized electrons $-,+$, and \emph{vice versa} $+,-$, from scenario H-20 in the operating scenarios report \cite{barklow2015ilc}.}
\label{tab:beams}
\end{center}
\end{table*}

Polarized electrons are produced by photoproduction with a polarized laser. Positrons are produced in pair conversion $\gamma \rightarrow e^+ e^-$, where the energetic photon is produced by a high energy electron beam passing through a superconducting undulator. Positron polarization is a significantly greater technical challenge than electron polarization, and the nominal design calls for 80\% polarized electrons and 30\% polarized positrons. 

Once produced, the electrons and positrons are injected into the main tunnel, where they are accelerated to 5 GeV and injected into the damping rings, storage rings with radius 1~km. See fig. \ref{fig:ilc} for reference. In the damping rings the beams are brought to the small cross sectional area necessary for high luminosity. They are then extracted and sent by transport lines for injection into the main linacs through bending rings. In the process the beams are accelerated to 15 GeV from 5 GeV and the bunches are compressed to their nominal bunch sizes.

The main linacs themselves consist of superconducting Niobium RF cavities cooled to 2K with supercooled He II. Each cavity is 1m long and consists of nine elliptical cells, which serve functions analogous to the drift tubes in the simple linac. Nine such cavities fit inside one cryogenic module of Type A. Eight such cavities, together with one focusing quadrupole magnet, fit inside a cryogenic module of Type B. Both modules A and B are 12.65 m in length and are assembled together in the pattern AABAAB to provide acceleration and beam focus. See fig. \ref{fig:module}. RF power is provided to the cavities by klystrons, yielding nominal nominal TDR accelerating gradients of 31.5 MV/m. 

At the end of the linacs, a beam delivery system collimates the beams, administers a final focus with quadrupole magnets and delivers the accelerated electrons and positrons to the interaction point at a 40 mrad crossing angle. See Table \ref{tab:beams} for ILC beam parameters for several $\sqrt{s}$. These parameters determine the ILC luminosity in eq. \ref{eqn:lumi2}. Table \ref{tab:beams} also shows projected ILC integrated luminosity and sharing between beam polarizations $(-,+)$ and $(+,-)$, that is 30\% negatively polarized positrons, 80\% positively polarized electrons and 30\% positively polarized positrons, 80\% negatively polarized electrons for scenario H-20 described in the operating scenarios report \cite{barklow2015ilc}.

In the ILC interaction region space is made for two  detectors in the \emph{push-pull} scheme, wherein one detector may be easily swapped into the interaction region as the other is swapped out. The two nominal ILC detectors are SiD and ILD. The advantage of using two detectors is scientific reproducability of results by two independent teams using distinct detector designs.

\subsection{Detectors and the SiD}

\begin{table*}
\begin{center}
\begin{tabular}{|c|c|c|c|c|c|c|} \hline 
Element & Z & A & $X_0$ [cm] & $\lambda$ [cm] & $\langle \frac{dE}{dx} \rangle$ [MeV/cm] & $\rho$ [g/cm$^3$] \\ \hline \hline
H$_2$ & 1 & 1.0 & 888.0 & 732.4 & 0.3 & 0.071 \\
C & 6 & 12.0 & 19.3 & 38.8 & 3.8 & 2.2 \\
Si & 14 & 28.1 & 9.4 & 46.5 & 3.9 & 2.3 \\
Fe & 26 & 55.8 & 1.8 & 16.8 & 11.4 & 7.9 \\
W & 74 & 183.8 & 0.4 & 9.9 & 22.1 & 19.3 \\
U & 92 & 238.0 & 0.3 & 11.0 & 20.5 & 19.0 \\ \hline
\end{tabular}
\caption{Atomic number, atomic mass, radiation lengths, nuclear absorption lengths, minimum mean ionization energy loss and density for several elements. Low Z materials are typically used in trackers to minimize $dE/dx$ and maximize $X_0$ and $\lambda$, while high Z materials are used to minimize $X_0$ ($\lambda$) in electromagnetic (hadronic) calorimeters.}
\label{tab:properties}
\end{center}
\end{table*}

\begin{table*}[t]
\begin{center}
\begin{tabular}{|c|c|c|c|c|c|} \hline
Parameter & SLD \cite{Rowson:2001cd} & OPAL \cite{1991275}  & ATLAS \cite{Collaboration_2008} & SiD \cite{Behnke:2013lya} \\ \hline \hline
Track $\Delta p_t/p_T$ & 0.010,0.0024 & -,0.0015 & 0.36,0.013 & 0.002,0.00002\\
ECal $\Delta E/E$ & -,0.08 & -,0.05  & 0.4, 0.10 & 0.01, 0.17 \\
HCal $\Delta E/E$ & -, 0.6 & -, 1.2 & 0.15,0.80 & 0.094,0.56 \\ \hline
\end{tabular}
\caption{Tracker and calorimeter performance parameters $a,b$ for several historically important collider detectors and one proposed collider detector. Parameters are obtained from data for the former, from simulated data for the latter. }
\label{tab:performance}
\end{center}
\end{table*}

\subsubsection{Collider detectors}

The quantitative signature of stable or quasistable particles traversing a collider detector is measured by energy transfers from the particle to the detector material mediated by electromagnetic or nuclear interactions. 

A particle's phenomenological signature in a collider detector can be classified as a \emph{shortlived} particle ($Z$, $W$, $t$, $\pi^0$, $\rho^{\pm}$, \emph{etc.}) with lifetimes to short to observe directly, a \emph{displaced vertex} ($B$, $D$, $\tau$, \emph{etc.}) with $\tau \approx 10^{-12}$~s, a \emph{quasistable} particle ($\mu$, $K$, $n$, \emph{etc.}) with lifetimes $\tau \approx 10^{-8}$~s or \emph{stable} ($e$ or $p$). Thus the ranges for relativistic particles are effectively of order $c\tau \approx$ 0, 1mm, 1m, $\infty$, respectively. For macroscopic detectors a few meters deep, only quasistable and stable particles are directly detected, but shortlived particles and displaced vertices can be reconstructed from their quasistable or stable decay products by four-vector addition. Neutrinos, because they only interact weakly, escape undetected.

For electrically charged particles, energy loss occurs through ionization, Coulomb scattering, bremsstrahlung induced by detector nuclei, and nuclear scattering or absorption if the particle is a hadron. For electrically neutral particles, energy loss occurs through photoelectric absorption, Compton scattering and pair production (for photons) or nuclear scattering and absorption (for hadrons).

For an example of energy loss, consider the mean ionization energy loss per unit length in a material given by the Bethe-Block equation,

\begin{eqnarray}
 \langle \frac{dE}{dx} \rangle & = & \frac{b}{\beta^2} \left( \frac{Z}{\rho A} \right) \left( \ln \left[ \frac{2m_e \beta^2}{I(1-\beta^2)} \right] -\beta^2 \right)
\end{eqnarray}

\noindent where $b$ is a constant, $Z$ is the atomic number, $A$ is the atomic weight, $\rho$ is the density, $I$ is the mean ionization potential and $\beta=v/c$. Hence the material dependence comes entirely in the factor $Z/\rho A$ and $I$, and the only remaining dependence is on $\beta$. After a $1/\beta^2$ fall at low $\beta$, the mean loss passes through a minimum near $\beta^2 \approx 0.9$ and begins a relativistic, logarithmic rise. See Table \ref{tab:properties} for the mean ionization energy loss for a minimum ionizing particle for several elements.

The modern collider detector is a complex, integrated system of interdependent subdetectors coordinated by fast electronics. It combines subdetectors like \emph{trackers}, which measure the spatial position and, if a magnetic field is applied, momentum of traversing charged particles, with \emph{calorimeters}, which trap charged and neutral particles to measure their spatial position and energy, and a variety of other specialized subdetectors. 

The earliest trackers were the photographic emulsions and cloud chambers used to study cosmic rays, which left visible tracks of chemical grains or condensation. With the advent of high energy colliders, new detector techniques were developed. Gaseous tracking detectors convert ionization electron avalanches from traversing charged particles to electric signals collected on wire cathodes. Modern trackers also employ semiconductors made of Silicon or Germanium, for example, in which the electron-ion pair in the gaseous tracker is replaced by an electron-hole pair in the valence and conduction bands of the semiconductor.

Whatever the tracker technology, the spatial \emph{hits} left in the tracker are mathematically fitted to reconstruct the trajectory of the traversing charged particle. If the active tracking region is subjected to a uniform magnetic field, the parameters of a charged particle's helical trajectory can be extracted from the fitted track and, from these parameters, the momentum is determined with eq. \ref{eqn:pisqbr}.  The vertex detector is a specialized tracker designed for precision tracking to resolve displaced vertices near the interaction point. Good spatial resolution in a tracker yields both precise spatial vertexing and precise momentum determination. 

While trackers are designed to induce minimal energy loss in traversing particles, calorimeters are designed to induce maximal energy loss. In the most common calorimeter configuration, a \emph{sampling calorimeter}, layers of absorbing material meant to induce showers alternate with layers of sensitive material to sample the energy deposition. The \emph{segmentation} of a calorimeter, the size of its sensitive elements, greatly impacts its energy resolution.

The electromagnetic calorimeter traps electrons and photons by inducing electromagnetic showers. In the presence of matter, the electron undergoes bremsstrahlung, $e \rightarrow \gamma e$, and a photon undergoes conversion $\gamma \rightarrow e^+ e^-$. Thus an incident electron $e \rightarrow \gamma e \rightarrow 3e \gamma \dots$ and an incident photon $\gamma \rightarrow e^+ e^- \rightarrow 2\gamma 2 e \dots$, producing a binary tree of cascading electrons and photons with successively lower energy until all electrons and photons are captured. 

For an incident electron (photon) with initial energy $E_0$, the energy at depth $x$ is described by $E_0 \exp( -x/X_0)$ ($E_0 \exp( -7x/9X_0)$), where $X_0$ is a characteristic of the traversed material called the \emph{radiation length}. For an electron, if the cross section for bremsstrahlung is $\sigma_{brem}$ and the radiation length is $X_{0}$, then the effective volume of an atom is $\sigma_{brem}X_0$. The effective volume is also $m_{atom}/\rho$, the atomic mass divided by the material density, or $1/n$, where $n$ is the number of atoms per unit volume. Therefore $X_0=1/n\sigma_{brem}$. Since the cross section for pair production $\gamma \rightarrow e^+ e^-$ is approximately $\sigma_{pair}=\frac{7}{9} \sigma_{brem}$, the effective pair production length is $\frac{9}{7} X_0$. See Table \ref{tab:properties} for the radiation lengths for several elements.

\begin{figure*}[t]
\begin{center}
\vspace{-0.5in}
\includegraphics[height=0.9\textwidth, angle=90]{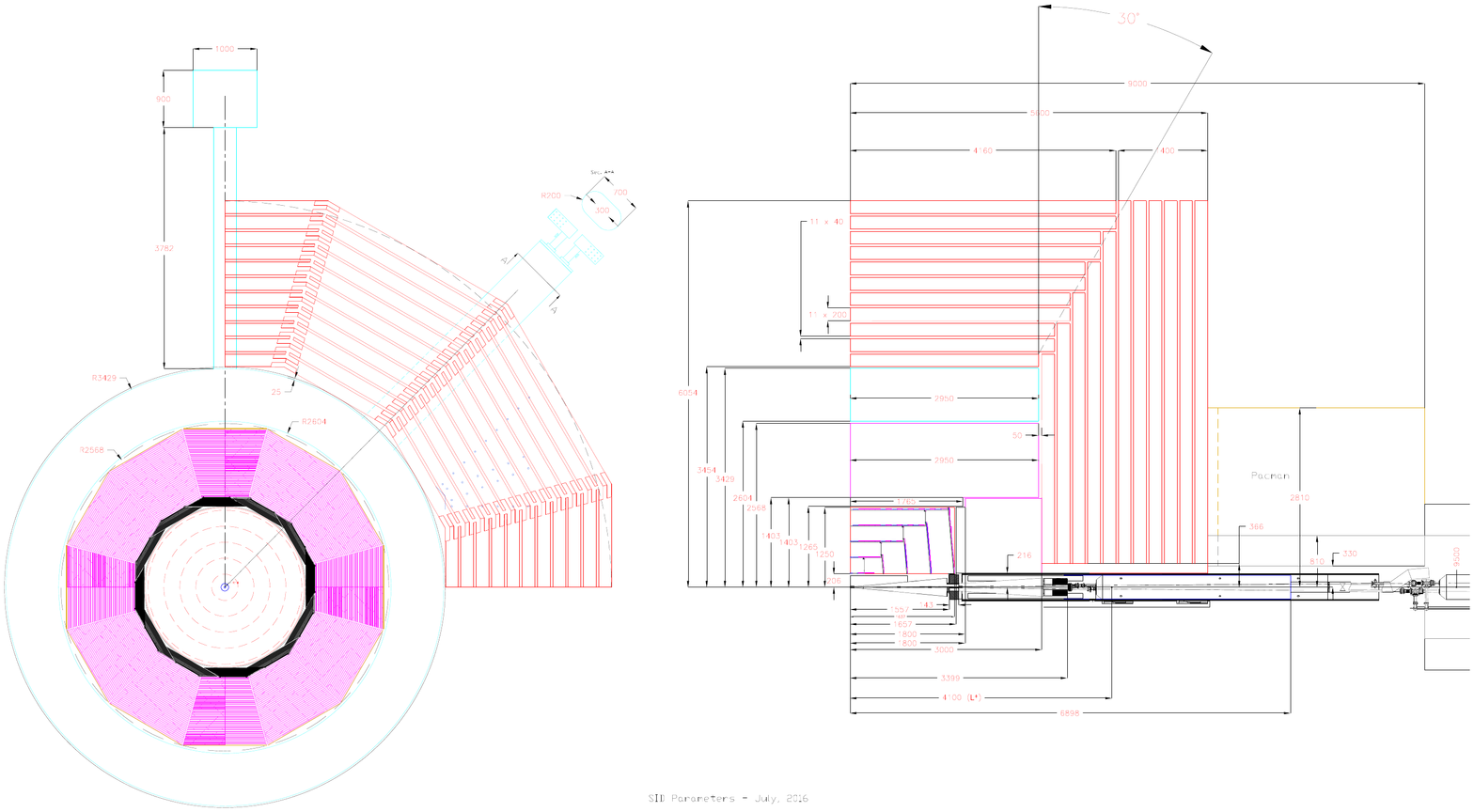}
\vspace{-0.5in}
\caption{Technical drawing of the SiD detector, barrel view (left) and quadrant view (right). Shown are the vertex detector (blue), the tracker (red), the ECal (black), the HCal (magenta), the solenoid (white) and the muon detector (orange).  Credit: SiD Consortium and SLAC.}
\label{figs.id}
\end{center}
\end{figure*}

The hadronic calorimeter traps charged and neutral hadrons by inducing hadronic showers in which the incident hadron and secondaries successively lose energy to nuclear collisions until complete absorption. Because the hadronic calorimeter is placed at a macroscopic distance from the interaction point, where unstable hadrons decay to stable or quasistable hadrons, the hadrons captured in a hadronic calorimeter are almost exclusively charged pions, kaons, protons and neutrons. In a hadronic calorimeter, unlike an electromagnetic calorimeter, not all energy from the incident hadron is seen due to (sometimes large) losses to nuclear binding energy, making it inherently less precise than an electromagnetic calorimeter. 

For an incident hadron of energy $E_0$, the energy at depth $x$ is $E_0 \exp(-x/\lambda)$, where $\lambda$ is the \emph{nuclear absorption length}, a characteristic of the traversed material. The relation between nuclear absorption length and inelastic nuclear scattering cross section $\sigma_{nuc}$ is straightforward. If we consider the effective volume of one nucleus of the traversed material, this is $\lambda \sigma_{nuc}$. The effective volume is also the nuclear mass $m_{nuc}$ divided by the density $\rho$, or $m_{nuc}/\rho=1/n$, where $n$ is the number of nuclei per unit volume. Therefore $\lambda=1/n\sigma_{nuc}$. See Table \ref{tab:properties} for the nuclear interaction lengths for several elements.

A calorimeter is meant to contain and measure all of the energy from an incident particle, and at $x=nX_0$ ($x=n\lambda$) the containment fraction is $\exp(-n)$ in an electromagnetic (hadronic) calorimeter. Thus at $x=3X_0$ an electromagnetic shower is 95\% contained on average, and at $x=5X_0$ is is 99\% contained on average, and similarly for a hadronic shower. Calorimeter showers are statistical processes, however, so shower penetration depth varies from shower to shower. 

The only particles which exit the tracker are quasistable or stable, almost exclusively electrons, muons, photons, pions, kaons, neutrons and protons. Since electrons and photons are absorbed by the electromagnetic calorimeter, while pions, kaons and nucleons are absorbed by the hadronic calorimeter, in principle only muons (and undetectable neutrinos) penetrate the hadronic calorimeter. In practice some hadronic (electromagnetic) showers do penetrate the hadronic (electromagnetic) calorimeter in \emph{leakage}, and individual particles can \emph{punch through}.

Muons, which are too heavy to undergo bremsstrahlung sufficient for absorption in the calorimetry and do not participate in nuclear interactions, are therefore easily identified in the muon detector, a tracker placed outside the hadronic calorimetry. Tracks reconstructed in the muon detector can be matched to tracks reconstructed in the main tracker, which typically measures momentum much more precisely.

The momentum resolution of a tracking detector can be parametrized with constants by the transverse momentum $p_T$ and the polar angle $\theta$ with respect to the beamline. The curvature $\Omega=R^{-1}$, so by eq. \ref{eqn:pisqbr} $d\Omega/dp=-qB/p^2$ and $\Delta p/p \propto p \Delta \Omega$. Similarly, the energy resolution of a calorimeter can be parametrized with constants by the energy $E$. Showers in calorimeters are statistical processes which deposit energy $E \propto N$, where $N$ is the number of shower particles, and $\Delta E \propto \sqrt{N}$. Thus $\Delta E/E \propto 1/\sqrt{E}$. 

Thus the tracking and calorimeter performance can be parametrized by the following expressions:

\begin{eqnarray}
\frac{\Delta p_T}{p_{T}^{2}} & = & a \oplus \frac{b}{\sin \theta} \label{eqn:parametrize1}\\
\frac{\Delta E}{E} & = & a \oplus \frac{b}{\sqrt{E}}
\label{eqn:parametrize2}
\end{eqnarray}

\noindent where $x\oplus y \equiv \sqrt{x^2+y^2}$ is addition in quadrature. See Table \ref{tab:performance} for a comparison of tracking and calorimeter performance for several historically important detectors: SLD is the SLC Detector \cite{Rowson:2001cd}, OPAL is the OmniPurpose Apparatus at LEP \cite{1991275}, ATLAS is A Toroidal LHC ApparatuS \cite{Collaboration_2008}. SiD is the ILC Silicon Detector \cite{Behnke:2013lya}.

\subsubsection{Silicon Detector (SiD)}

\begin{table*}[t]
\begin{center}
\begin{tabular}{|c|c|c|c|c|c|c|} \hline
Subdetector & Technology & $n_{layer}$ & Thickness  & $r_{in}$ [cm] & $r_{out}$ [cm] & $\pm z_{max}$ [cm] \\ \hline \hline
Vertex Detector & Si Pixels & 5 & 0.015$X_0$& 1.4 & 6.0 & 6.25 \\
Tracker & Si Strips & 5 & 0.1$X_0$ & 21.7 & 122.1 & 152.2 \\
ECal & W-Si Pixels & 20+10 & 26$X_0\approx 1 \lambda$ & 126.5 & 140.9 & 176.5 \\
HCal & RPC-Steel & 40 & 4.5$\lambda$ & 141.7 & 249.3 & 301.8 \\
Solenoid & 5T SC & - & - & 259.1 & 339.2 & 298.3 \\
Muon Detector & Sc-Steel & 10 & - & 340.2 & 604.2 & 303.3 \\ \hline
\end{tabular}
\caption{Parameters of the baseline SiD barrel design, adapted from \cite{Behnke:2013lya}: technology, number of layers, thickness in $X_0$ or $\lambda$, inner radius, outer radius and $z$ extent. In the ECal there 20 thin Tungsten layers and 10 thick Tungsten layers. In the HCal the sensitive elements are Resistive Plate Chambers (RPC). In the muon detector the sensitive elements are scintillators (Sc).}
\label{tab:sid}
\end{center}
\end{table*}

SiD was designed to meet the physics goals of the ILC \cite{Behnke:2013lya}. SiD comprises a precise vertex detector, a main tracker, a sampling electromagnetic calorimeter (ECal) with Tungsten absorber, a sampling hadronic calorimeter (HCal) with Iron absorber, a 5T solenoid, and a muon detector instrumented in the solenoid flux return. See fig. \ref{figs.id} for a technical drawing of the SiD detector, and Table \ref{tab:sid} for a summary of the important SiD barrel parameters.

Throughout, SiD was designed to enable the \emph{particle flow} reconstruction technique in which charged particle trajectories are extrapolated from the tracker to the calorimeters and matched either to an ECal energy deposit, an HCal energy deposit, or a muon tracker track. Remaining calorimeter deposits which are unmatched to tracker tracks are designated neutral, either photons in the ECal or neutral hadrons in the HCal. This enables the substitution of the far more accurate tracking momentum measurement for the calorimeter energy measurement in the case of charged particles.

The vertex detector is made of five barrel layers at radii from $r=1.4$~cm to $r=6.0$~cm, centered on the beamline and capped by four endcap disks perpendicular to the beamline. The barrel layers and endcap disks are instrumented with $20 \times 20$~ $\mu$m$^2$ Silicon pixels. The primary goals of the vertex detector are 5 $\mu$m hit resolution, less than 0.3\% $X_0$ per layer, low power consumption and single bunch timing resolution. Achieving these goals enables precise vertexing with minimal energy loss in the active volume and rejection of backgrounds out of time with the bunch crossings.

The main tracker comprises five barrel layers ranging from $r=22$~cm to $r=122$~cm, capped by four slightly conical disks instrumented with Silicon microstrips. The performance goals of the main tracker include hermetic coverage for polar angles $10^{\circ}< \theta < 170^{\circ}$, momentum resolution $\delta (1/p_T) \approx 2 \times 10^{-5}$/GeV, less than $0.1X_0$ in the central region, and greater than 99\% hit efficiency.

The ECal barrel ranges from $r=127$~cm to $r=141$~cm with 20 thin layers of Tungsten and 10 thick layers of Tungsten, each absorbing layer alternating with sensitive layers with 13 mm$^2$ Silicon pixels. ECal endcaps close the barrel at $z=\pm 177$~cm. The ECal performance goal is energy resolution $\Delta E/E =0.01 \oplus 0.17/\sqrt{E}$. 

The HCal barrel ranges from $r=142$~cm to $r=249$~cm with 40 Steel absorber layers alternating with gas Resistive Plate Chambers (RPC) sensitive layers. RPCs alternate highly resistive layers at high voltage with gas layers. When traversing charged particles ionize in the gas, the ionization electron induces an avalanche of secondary ionization electrons which are then read out by sensitive strips. Endcaps close the HCal barrel at $z=\pm 302$~cm. The HCal performance goal is energy resolution $\Delta E/E =0.094 \oplus 0.56/\sqrt{E}$. 

A solenoid occupying $r=259$~cm to $r=339$~cm provides the 5T magnetic field necessary for measuring momentum in the tracker and particle flow in the calorimetry. Finally, a steel flux return for the solenoid occupies $r=340$~cm to $r=604$~cm and is instrumented with scintillators for the muon tracker.

\subsection{Further reading and exercises}

\emph{The Experimental Foundations of Particle Physics} (Cahn and Goldhaber) \cite{cahn_goldhaber_2009} reprints key papers in the experimental development of the SM and presents bridging commentary and exercises. For accounts of the historical development of the SM and the colorful characters involved, see \emph{Inward Bound} (Pais) \cite{Pais:1988} and \emph{From X-Rays to Quarks} (Segr\`e) \cite{Segrè:100978}, both written by physicists deeply involved in the story.

For brief pedagogical introductions to accelerators and detectors, the chapters on experimental methods in \emph{Particle Physics} (Martin and Shaw) \cite{Martin:2211679} and \emph{Introduction to High Energy Physics} (Perkins) \cite{Perkins:396126} are good. For textbooks on accelerators see \emph{An Introduction to the Physics of High Energy Accelerators} (Edwards and Syphers) \cite{Edwards:1993qe} and \emph{RF Linear Accelerators} (Wangler) \cite{Wangler:368392}. \emph{Experimental Techniques in High Energy Nuclear and Particle Physics} (Ferbel) \cite{Ferbel:238481} reprints several good review papers on tracking, calorimetry and particle identification. The Particle Data Group reviews on \emph{Accelerator Physics of Colliders}, \emph{High Energy Collider Parameters}, \emph{Passage of Particles Through Matter} and \emph{Particle Detectors at Accelerators} \cite{Tanabashi:2018oca} are invaluable lifelong, regularly updated references. 

The SiD LoI \cite{Aihara:2009ad} is the most complete technical documentation of SiD. The ILC Technical Design Report is essential reading. There are four volumes, \emph{Volume 1: Executive Summary} \cite{Behnke:2013xla}, Volume 2: Physics \cite{Baer:2013cma}, Volume 3: Accelerator \cite{Phinney:2007gp} and Volume 4: Detectors \cite{Behnke:2013lya}. For a more recent overview see \cite{Bambade:2019fyw}. Research and development of SiD technologies has continued since the TDR as reported in the Linear Collider Collaboration Detectors R\&D Liaison Report \cite{jan_strube_2020_3766518}.

Exercises for this section can be found in sect. \ref{sec:ilc} of Appendix \ref{appendix1}.

\section{SiD: simulation and reconstruction}

\subsection{Generation of ILC events}

\subsubsection{Monte Carlo integration}

Two-body scattering and decay yield straightforward expressions for differential cross sections and decay widths (eq.s \ref{eqn:twobody}), and for tree level Feynman diagrams with only one internal momentum $q$ there is just one four-vector integration $\int d^4 q$ prescribed by the Feynman rules. But for $n$-body processes with arbitrary $m$ internal momenta, calculations can quickly become intractable analytically, and very challenging numerically.

The \emph{Monte Carlo} integration technique is best suited for such calculations because it promises faster convergence for arbitrary $n$ and $m$ compared to other numerical techniques. The idea is to randomly sample the integrand $f$ over the sampling volume $V^{\prime}$ and form a mean $\bar{f}$. Then the integral is $\int_{V} dV f \approx V^{\prime} \bar{f}$, with convergence at a rate $\propto 1/\sqrt{N}$ where $N$ is the number of samples. This converges for any reasonably well behaved $f$ and boundary. But by performing all possible integrations by hand before using the Monte Carlo technique, the convergence improves considerably. 

For example, consider the integral $\int_{\mathbb{S}^2} dA f(x,y)$, where $\mathbb{S}^2$ is the unit circle in $\mathbb{R}^2$ and $f(x,y)=1$. This integral is easily calculated analytically, $\int_{\mathbb{S}^2} dA=\pi$. With the Monte Carlo technique we sample at $N$ points $(x,y)\in [-1,1] \times [-1,1]$ from a uniform distribution, rejecting the point if $x^2+y^2>1$. Then $\bar{f}=\frac{1}{N}\sum_{i=1}^{N} f_i$ converges to $\pi$ at a rate $ \propto 1/\sqrt{N}$. Going to $(r,\theta)$ coordinates first and performing the $\int d \theta= 2 \pi$ integration, however, we sample only from $r\in [0,1]$ the integrand $f(r)= r$ and obtain much faster convergence.

For a more \emph{apropos} example, consider Yukawa scattering in the first Born approximation. The Yukawa potential $V(r)= \beta \exp(-\mu r)/r$ with $\beta,\mu$ constants representing the strength and $\beta$ and range $1/\mu$ of the Yukawa interaction. After integrating over $\theta$ and $\phi$, the scattering amplitude for a particle with mass $m$ and wave number $\kappa=\sqrt{2mE}/\hbar$ is 

\begin{eqnarray}
f(\theta) & = & - \frac{2 m \beta}{ \hbar^2 \kappa} \int_{0}^{R} dr r \frac{e^{-\mu r}}{r} \sin \kappa r
\end{eqnarray}

\noindent for $R \rightarrow \infty$. The analytical result for the integral is $I(\mu,\kappa)=(\mu^2 + \kappa^2)^{-1}$. For concreteness, consider unit $\mu,\kappa$ so I(1,1)=0.5. See Table \ref{tab:mc} for Python code which implements the Monte Carlo technique for approximating $I(1,1)$. This is an example of \emph{simple} Monte Carlo sampling. Much more sophisticated sampling methods are available.

\begin{table}[t]
\begin{center}
\begin{tabular}{|l|} \hline
\texttt{import math, random, matplotlib.pyplot as plt} \\
\texttt{def function(x):} \\
    \hspace{0.1in} \texttt{return math.exp(-x)*math.sin(x)} \\
\texttt{def monte\_carlo(n,R):} \\
    \hspace{0.2in} \texttt{sample=[R*random.random() for i in range(n)]} \\
    \hspace{0.2in} \texttt{integrands=[function(point) for point in sample]} \\
    \hspace{0.2in} \texttt{return R*sum(integrands)/n} \\
\texttt{n\_values=range(1,10000)} \\
\texttt{approximation=[monte\_carlo(n,100) for n in n\_values]} \\
\texttt{plt.scatter(n\_values,approximation,s=1)} \\
\texttt{plt.show()} \\ \hline
\end{tabular}
\caption{Python code to evaluate the Yukawa scattering amplitude integral $I(1,1)$ discussed in the text by the Monte Carlo technique. The code also plots the approximation \emph{vs.} the number of integrand samplings.}
\label{tab:mc}
\end{center}
\end{table}

Once the integration for the total cross section $\sigma_{tot}$ or decay rate $\Gamma_{tot}$ has been performed with the Monte Carlo technique, a closely related function can be performed. For a fixed initial state, we would like to generate a sample of $N$ events with final states which for large $N$ reproduce the differential cross sections for scattering or decay distributions for decays. For example, for two-body final state four-vectors $p_{\mu}^{1},p_{\mu}^{2}$, generate $N$ events with probabilities consistent with the differential distributions determined by the Feynman rules. We would like the sample of events to reproduce the differential distributions in the limit of large $N$. This is precisely what Monte Carlo \emph{event generators} do. They generate a sample of events with final state four-vectors which correctly reproduce the underlying final state physics.

A straightforward algorithm to correctly reproduce the final state probabilities proceeds as follows. Suppose the integrand $I=\frac{d\sigma}{d\Omega}$ or $I=\frac{d\Gamma}{d\Omega}$ reaches a maximum $I_{max}$ on the integration domain. Then repeat the following steps until $N$ final states are accepted: 

\begin{enumerate}
\item Randomly sample $p_{\mu}^{1},p_{\mu}^{2},...,p_{\mu}^{n}$ from possible final states.
\item Randomly sample the unit interval, $R\in [0,1]$. 
\item If $I(p_{\mu}^{1},p_{\mu}^{2},...,p_{\mu}^{n})/I_{max}<R$ accept the final state, otherwise reject it.
\end{enumerate}

For example, consider the differential cross section for relativistic ($\beta \approx 1$) scattering $e^+ e^- \rightarrow \gamma^{\star} \rightarrow q\bar{q}$. The differential cross section is

\begin{eqnarray}
\frac{d\sigma}{d\Omega} & = & 3 Q_{q}^2 \frac{\alpha^2}{4s} \left( 1+ \cos^{2} \theta \right)
\end{eqnarray}

\noindent so that $I_{max}=3Q_{q}^2 \alpha^2/2s$. If $\theta \in [0,\pi]$ is sampled uniformly and $I(\theta)/I_{max}$ is compared against a random number $R$ sampled from the unit interval, the algorithm above correctly reproduces the $\theta$ distribution of the final state quarks for large $N$.

\subsubsection{Whizard2: $W$, $H$, $Z$}

Polarized electron and positron beams are central to the design of the ILC. Any event generator considered for use in ILC studies must therefore simulate polarized beams. This requirement narrows the field of potential event generators considerably. 
Two events generators which simulate polarized beams are in common use for ILC studies, Whizard2 \cite{Kilian:2007gr} and MG5 aMC@NLO \cite{Alwall:2014hca}.
Whizard2 further distinguishes itself by simulating two additional important initial state effects. 

\emph{Initial state radiation} (ISR) refers to either photon emission from a charged initial state particle or gluon emission from a colored initial state particle. For an $e^+e^-$ collider it refers to photon emission from the electron or positron (or both). 
\emph{Beamstrahlung} refers to bremsstrahlung which occurs when a particle in one collider bunch emits bremsstrahlung due to the field produced by the oncoming bunch, and is therefore sensitive to the detailed structure of the accelerator bunches. 
Whizard2 can simulate both ISR and beamstrahlung. MG5 aMC@NLO simulates neither, although the authors have indicated plans to include ISR in future versions.

\begin{table*}[t]
\begin{center}
\begin{tabular}{|l|l|} \hline
\multicolumn{1}{|c|}{Whizard2} & \multicolumn{1}{|c|}{Madgraph5\_aMC@NLO} \\ \hline
\multicolumn{2}{|c|}{Model, Process and Parameters} \\ \hline
\texttt{model=SM\_CKM} & \texttt{import model sm} \\
\texttt{process zh250pm=E1,e1=>e2,E2,b,bbar} & \texttt{generate e+e-> zh,z> mu+mu-,h> bb\textasciitilde }\\
\hspace{0.1in} \texttt{$\$ \! \!$ restrictions="1+2\textasciitilde{}Z\&\&3+4\textasciitilde{}Z\&\&5+6\textasciitilde{}H"} & \texttt{output zh250pm}\\
\texttt{compile} & \texttt{launch} \\ 
%\multicolumn{2}{|c|}{Model Parameters} \\ \hline
\texttt{mH=125.0 GeV}  & \texttt{set mh 125.0}\\
\texttt{wH=0.004 GeV} & \texttt{set wh 0.004}\\ \hline
\multicolumn{2}{|c|}{Initial State $\sqrt{s}$, ISR and Polarization} \\ \hline
\texttt{beams= E1,e2=>isr,isr} & \texttt{set lpp1 0} \\
\hspace{2.5in} & \texttt{set lpp2 0} \\
\texttt{beams\_pol\_density=@(+1), @(-1)} &  \texttt{set polbeam1 +30}\\
\texttt{beams\_pol\_fraction=30\%, 80\%} & \texttt{set polbeam2 -80} \\
\texttt{sqrts=250 GeV} & \texttt{set ebeam1 125.} \\
\texttt{integrate(zh250pm)} & \texttt{set ebeam2 125.} \\ \hline
\multicolumn{2}{|c|}{Final State QCD Showering and Hadronization} \\ \hline
\texttt{$\$ \! \!$ shower\_method="PYTHIA6"} & \texttt{shower=Pythia8} \\
\texttt{?hadronization\_active=true} & \hspace{2.5in}\\
\texttt{$\$ \! \!$ hadronization\_method="PYTHIA6"} & \\ \hline
\multicolumn{2}{|c|}{Event Generation} \\ \hline
\texttt{n\_events=10000} & \texttt{set nevents 10000}\\
\texttt{seed=12345} & \texttt{set iseed 12345} \\
\texttt{sample\_format=lhef,stdhep} & \texttt{} \\
\texttt{simulate(zh250pm)} & \\ \hline
\end{tabular}
\caption{Whizard2 and Madgraph5\_aMC@NLO scripts to generate $10^5$ Higgstrahlung events at $\sqrt{s}=250$~GeV. In both cases beams are polarized, 30\% positively polarized positrons and 80\% negatively polarized electrons. Here Whizard2 uses Pythia6 for final state showering and hadronization while MG5 aMC@NLO uses Pythia8. Whizard2 simulates ISR while MG5 aMC@NLO does not. }
\label{tab:generators}
\end{center}
\end{table*}

The Whizard2 executable runs with input from a script which specifies the parameters of the event generation. See Table \ref{tab:generators} (left) for a script to calculate the cross section for  Higgstrahlung events with polarized beams and generate $10^5$ events. The first step is to specify a model, which includes all particles, their masses and decay widths, and the interactions. For example, \texttt{model=SM} specifies to use the SM model. This is the default, so if model is not specified then Whizard2 assumes you want the SM with the trivial CKM matrix. Other models implemented in Whizard2 are SM\_CKM (the SM with nontrivial CKM), MSSM\_CKM (the Minimal Supersymmetric SM with nontrivial CKM) and NMSSM\_CKM (the Next to MSSM with nontrivial CKM). Model parameter settings can be displayed and changed: \texttt{show(mH)} displays the default Higgs boson mass, \texttt{mH=125.0 GeV} sets it explicitly.

The next step is to specify a process to simulate and give it a name. For example, \texttt{process zh250pm=E1,e1 => E2,e2,b,bbar} specifies $b$ quark pair and muon pair production from electron positron initial states and names it \texttt{zh250pm} in order to reference it later. Once can impose restrictions on a process when defining it. For example, adding the text \texttt{$\$ \! \!$ restrictions="1+2\textasciitilde{}Z \&\&3+4\textasciitilde{}Z \&\&5+6\textasciitilde{}H" }  to the above requests Whizard2 to couple the $\mu^+ \mu^-$ pair to an internal $Z$ boson and the $b$ quark pair to an internal Higgs boson, thereby requiring the $s$-channel Higgstrahlung. When the \texttt{compile} command is invoked, Whizard2 generates, compiles and loads Fortran code which calculates the amplitudes for the defined process. 

Next one specifies the beam parameters, first defining $\sqrt{s}$ with \texttt{sqrts=250 GeV}, for example, and then specifying beam polarization and ISR, if desired, with the \texttt{beams}, \texttt{beams\_pol\_density} and \texttt{beams\_pol\_fraction} variables. Whizard2 now has all information necessary for calculating a cross section for the defined process, and invoking \texttt{integrate(zh250pm)} integrates the necessary integrals for Higgstrahlung $e^+ e^- \rightarrow ZH$ with $Z \rightarrow \mu^+ \mu^-$ and $H \rightarrow b \bar{b}$ at $\sqrt{s}=250$~GeV with the specified beams.

After specifiying a few more parameters, Whizard2 is ready to generate events which reproduce the correct differential cross sections. The next few commands in Table \ref{tab:generators} instruct Whizard2 to pass the handling of some QCD effects in the final state to Pythia6 (see below). Finally, the desired number of events are specified with \texttt{n\_events=10000}, the desired output formats are specified \texttt{sample\_format=lhef,stdhep}, and the event generation is invoked with \texttt{simulate(zh250pm)}. If the script from Table \ref{tab:generators} is saved in a file \texttt{script.sin} and the binary is \texttt{whizard}, the execution command is simply 

\begin{center}
\texttt{<path>/bin/whizard script.sin}
\end{center}

\noindent The \texttt{lhef} format is the Les Houches Event File (LHEF) format \cite{Alwall:2006yp}, agreed to by a group of generator experts in 2006. The \texttt{stdhep} format is the standardized HEP (StdHep) format \cite{stdhep} which is still in common use. The \texttt{hepmc} format can also be specified for the HEP Monte Carlo format \cite{Dobbs:2001ck}. LHEF reports events prior to final state QCD effects carried out by Pythia6, while StdHep and HepMC formats include such effects.

\subsubsection{MG5 aMC@NLO}

MG5 aMC@NLO is a merging of the older MadGraph program with next leading order techniques, a Monte Carlo at Next Leading Order. Feynman diagrams at arbitrary orders in QED and QCD can be easily specified when defining a process. Table \ref{tab:generators} (right) shows a script for the same process as the Whizard2 script in Table \ref{tab:generators} (left): Higgstrahlung $e^+ e^- \rightarrow ZH$ at $\sqrt{s}=250$~GeV with $Z \rightarrow \mu^+ \mu^-$ and $H \rightarrow b \bar{b}$.

As with Whizard2, the first step is to define a model. Here \texttt{import model sm} specifies the SM. Other models include \texttt{loop\_sm}, \texttt{MSSM\_SLHA2} and \texttt{heft}. The default model is SM. The next step is to define a process within the chosen model, in this case \texttt{generate e+e- > zh, z>mu+mu-, h>bb\~}. Finally invoking \texttt{output zh250pm} names the process, creates a directory for event generation code, and generates Feynman diagrams. Invoking \texttt{launch} then allows the user to set various parameters of the run.

First, the Higgs boson mass and width are set, replacing the default settings. Next the initial state parton density functions are set with \texttt{set lpp1 0} and \texttt{set lpp2 0}, meaning we use trivial parton density functions for electrons. Next the energy and polarizations of each beam are set. Finally, \texttt{shower=Pythia8} instructs MG5 aMC@NLO to pass final state QCD effects to Pythia8 (see below) and after the number of events and a random number seed are set, the cross section calculation and event generation begin. The execution command is 

\begin{center}
\texttt{<path>/bin/mg5\_aMC script.mg5}
\end{center}

Two file formats are saved by default, LHEF and HepMC, in the Events directory. In the bin directory an executable \texttt{generate\_events} is also produced. In the Cards directory one finds the \texttt{run\_card.dat} and \texttt{param\_card.dat}, which specify the parameters of the run and parameters of the model and process, respectively. Finally, \texttt{index.html} includes the cross section and Feynman diagrams.

\subsubsection{Pythia6 and Pythia8 \label{sec:pythia}}

Pythia is frequently used for specialized functions for handling quarks and gluons in the final state. Pythia6 \cite{Sjostrand:2006za} is the last version to use Fortran, the original implementation language. Pythia8 \cite{Sjostrand:2007gs} is the C++ implementation. Both Pythia6 and Pythia8 are in common use (there is no Pythia7).

\emph{Final state radiation} (FSR) refers to either photon emission from a charged final state particle or gluon emission from a colored final state particle. In principle FSR can be included in the Feynman diagram and calculated explicitly, but in practice it is straightforward to use \emph{parton showering} for the $q \rightarrow g q$ and $e \rightarrow \gamma e$ processes after the calculation of the diagram without these vertices. Parton refers to the $q$ or $e$, shower refers to the showerlike cascade of particles from $q \rightarrow g q \rightarrow q\bar{q}q,...$, for example. The Pythia code for parton showering has been extensively tested and tuned against experiment.

\emph{Fragmentation} refers to process of meson and baryon formation from energetic quark pairs under the conditions imposed by QCD confinement. In Pythia, \emph{hadronization} means the formation of hadrons through both fragmentation and the decay of the hadrons. Because QCD confinement is not well understood, fragmentation must be simulated using phenomenological models. Pythia implements the \emph{Lund fragmentation model}, which uses a relativistic massless string connecting the two fragmenting quarks with a linear potential $V=\kappa z$, where $z$ is the distance separating the quarks and $\kappa$ is a constant determined experimentally ($\kappa \approx$~1 GeV/fm). 

In the Lund model, when the initial quarks are separated enough to concentrate a large amount of energy in the string, a new quark pair $q^{\prime} \bar{q}^{\prime}$ appears in the middle of the string and two proto-mesons $q\bar{q}^{\prime}$ and $q^{\prime} \bar{q}$ appear, each pair separated by a new string. This fragmentation repeats until the energy in the proto-meson is close to the mass of a physical meson. Baryons are handled similarly, except that instead of a quark pair separated by a string, a quark and diquark are separated by a string.

In the scripts in Table \ref{tab:generators}, both generators invoke Pythia for showering and hadronization of the $b\bar{b}$ quark pair. Whizard2 employs Pythia6 while MG5 aMC@NLO employs Pythia8 (\texttt{shower=Pythia8} specifies Pythia8 for hadronization as well as showering). In fact both generators can invoke either. For ILC studies, the advantage of using Pythia6 over Pythia8 is that the parameter tunes from LEP2 can be used. In Pythia8 there is no exact correspondence to the Pythia6 tuning parameters. 

\subsection{Simulation of SiD response}

\subsubsection{Geant4: GEometry ANd Tracking}

Geant4 \cite{AGOSTINELLI2003250,Allison:2006ve,ALLISON2016186} is a giant in the world of simulating physical processes in matter, and its applications run well beyond collider physics. Geant4 enables the precise description of detector geometry down to the finest detail, and tracks particles from their origin through the detector, simulating all physical processes which apply to the particles and modifying position, energy and momentum accordingly. GEANT-3, its predecessor, was written in Fortran while Geant4 is written in C++.

Geant4 needs a source of the particles meant to traverse the detector in each event. In the collider detector case this means the files made by generators like Whizard2 and MG5 aMC@NLO containing collider events, but other sources can also be specified. Single particle guns, for example, can be specified, and are useful in evaluating the performance of the detector. Each particle has a definition including all of it properties like charge, mass, spin, necesssary for implementing the processes assigned to apply at each step.

Each event in Geant4 thus contains one or more particles, each particle defined by the particle's physical properties and four-vector. Before processing, the Geant4 event contains only these things, and  after processing the event contains only \emph{hits} and \emph{digitizations}, the energy deposits and electronics responses to those energy deposits in the detector.

In Geant4 detector geometry is described by volumes, the largest of which is the \emph{world} volume. Smaller volumes are placed within the world volume, and each such volume may contain any number of daughter volumes. A \emph{logical} volume, in Geant4, is defined by a shape and the matter which composes it. Shapes may be arbitrarily complex, or simple, as with a box or cylinder. Matter can be defined as an element (atomic mass, atomic number, cross section, etc) or or as a material (density, state, temperature, radiation length, etc). Once a logical volume and is placed physically within its mother volume, it is a \emph{physical} volume. This hierarchy of volumes allows a local, as opposed to global, coordinate systems within each detector volume.

Tracking in Geant4 means applying a list of physical processes to particles in discrete steps, either time steps in the case of decay, or spatial steps in the case of interactions, and altering particle four-vectors accordingly. Of the major categories of physical process in Geant4, most relevant to collider detectors are the electromagnetic, hadronic, decay and transportation processes. For the electromagnetic process the step scale is set by the radiation length $X_0$ of the traversed material, and for the hadronic process by the nuclear interaction length $\lambda$. For the decay process, the step scale is set by the particle lifetime $\tau$.

For each process, there are actions which are applied before, during, and after each step to alter the traversing particle as well as the material around it. As the particle traverses the detector material, each relevant process proposes a numerical value for the step, and the smallest such step is chosen to implement all processes. Geant4 finishes applying processes to particles when the particle decays or the entire detector volume has been traversed.

A minimal Geant4 program to simulate a collider detector might contain the following hierarchy of classes:

\begin{enumerate}

\item \texttt{G4DetectorConstruction}

\item \texttt{G4PhysicsList}

\begin{enumerate}

\item \texttt{G4ElectromagneticProcesses}

\item \texttt{G4HadronicProcesses}

\end{enumerate}

\item \texttt{G4ActionInitialization}

\begin{enumerate}

\item \texttt{G4PrimaryGeneratorAction}

\end{enumerate}

\end{enumerate}

\noindent Within the \texttt{G4DetectorConstruction} class lies the necessary code to construct trackers, calorimeters and any other specialized subdetectors. Among the electromagnetic process classes for an electron in Geant4, for example, are \texttt{G4eIonisation} and \texttt{G4eBremstrahlung}. Within the primary generator class the Whizard2 and MG5 aMC@NLO collider event output can be specified, for example.

\subsubsection{DD4hep: Detector Description for HEP}

While Geant4 is complete and standalone, it is also generic and widely applicable. DD4hep \cite{frank_markus_2018_1464634} was designed as a generic collider detector description toolkit with a more specialized goal: a full, single source detector description suitable for the full lifetime of a collider experiment with full visualization and alignment functionality. 

DD4hep simplifies the use of Geant4 for HEP, in particular for collider detector simulation. It serves as a simplifying interface between the user simulating a collider detector and Geant4. It provides the \emph{compact} XML detector description suitable for early stages of detector design as well as the full detector description suitable for a running experiment. 

Nine XML tags define the detector description in DD4hep. Among them are the \texttt{define}, \texttt{materials}, \texttt{display}, \texttt{readouts} and \texttt{fields} tags. In the \texttt{define} field constants like subdetector component dimensions are defined numerically. In \texttt{materials} all materials and their properties necessary for detector construction are defined. The \texttt{display} tag defines how each subdetector appears visually in a detector display. The \texttt{field} tag defines the magnetic field created by a collider detector solenoid, for example.

The \texttt{readouts} tag defines how each subdetector cell, for example a pixel in the ECal or a strip in the tracker, reports itself. Each cell thus has a unique cell ID identifier string. For example, the string

\begin{center}
\texttt{$\langle$id$\rangle$system:5, barrel:1, layer:2, module:4, sensor:2, side:32:-2, strip:24$\langle$id$/\rangle$}
\end{center}

\noindent defines a cell in the barrel of subdetector 5, layer 2, module 4, sensor 3, side 32:-2, strip 24. Any hit report created by a particle traversing that cell will report this cell ID.

DD4hep also includes tools for detector alignment studies and several utilities useful for debugging detector descriptions: \texttt{geoDisplay} for visualizing the detector with the compact description, \texttt{geoConverter} for converting the DD4hep XML description to other detector representations like the one used by Geant4, \texttt{checkOverlaps.py} for checking if detector volumes intersect, \texttt{checkGeometry.py} for overlap checking and scans for detector boundary crossing, and \texttt{materialScan} for reporting all materials traversed in a specified direction.

\subsubsection{ILCsoft simulation with DD4hep/Geant4}

\begin{table*}[t]
\begin{center}
\begin{tabular}{|l|} \hline
\multicolumn{1}{|c|}{DD4hep SiD ECal XML} \\ \hline
$\langle$ layer repeat="20" $\rangle$  \\
\hspace{0.25in} $\langle$ slice material = "TungstenDens24" thickness = "0.25*cm"  /$\rangle$  \\
\hspace{0.25in} $\langle$ slice material = "Air"     thickness = "0.025*cm"  /$\rangle$  \\
\hspace{0.25in} $\langle$ slice material = "Silicon" thickness = "0.032*cm" sensitive = "yes"/$\rangle$  \\
\hspace{0.25in} $\langle$ slice material = "Copper"  thickness = "0.005*cm" /$\rangle$  \\ 
\hspace{0.25in} $\langle$ slice material = "Kapton"  thickness = "0.030*cm" /$\rangle$  \\
\hspace{0.25in} $\langle$ slice material = "Air"     thickness = "0.033*cm" /$\rangle$  \\
$\langle$ /layer$\rangle$  \\ \hline
\multicolumn{1}{|c|}{Delphes SiD ECal TCL} \\ \hline
module SimpleCalorimeter ECal \{ \\
\hspace{0.25in}	  set ParticleInputArray ParticlePropagator/stableParticles \\ 
\hspace{0.25in}          set TowerOutputArray ecalTowers \\
\hspace{0.25in}	  add EnergyFraction \{0\} \{0.0\} \\
\hspace{0.25in}	  add EnergyFraction \{11\} \{1.0\} \\
\hspace{0.25in}	  add EnergyFraction \{22\} \{1.0\} \\
\hspace{0.25in}	  add EnergyFraction \{111\} \{1.0\} \\
\hspace{0.25in}	  set ResolutionFormula \{sqrt(energy\textasciicircum 2 * 0.01 + energy * 0.17\textasciicircum  2)\} \\
 \} \\ \hline
\end{tabular}
\caption{Full and fast simulation descriptions of the SiD ECal. Above, the DD4hep XML fragment which defines the SiD 20 thin ECal layers. Below, the Delphes TCL fragment which defines the SiD ECal performance.}
\label{tab:fullfast}
\end{center}
\end{table*}

In ILCsoft the executable for invoking Geant4 for detector simulation using a DD4hep compact detector description is \texttt{ddsim}. See Appendix \ref{appendix2} for instructions on installing and testing ILCsoft. 

Assuming a generator file in HepMC format, a DD4hep compact detector description XML file and an output file in LCIO format, the syntax for invoking \texttt{ddsim} is

\begin{center}
\texttt{ddsim --compactFile compact.xml --inputFiles in.hepmc --outputFile out.lcio}
\end{center}

\noindent There are many options for running \texttt{ddsim} not listed here. For concreteness we assumed a HepMC input file, but \texttt{ddsim} also reads generator files in StdHep and LCIO formats. For a full description of the LCIO file format, see ref. \cite{gaede_2017}.

As an example of how a subdetector is configured with DD4hep, consider the XML fragment used to define the 20 thin ECal strips in the SiD compact description. See Table \ref{tab:fullfast} (top), where it is evident how to specify the materials used, their thicknesses, and the number of layers required.

\subsubsection{Delphes fast simulation}

Geant4 detector simulation is an example of \emph{full} simulation, that is, all underlying physical processes are simulated. Full simulation is typically resource intensive, one collider event taking computation time of order 1 to 10 seconds depending on the complexity of the event, complexity of the detector, and the processor speed.

\emph{Fast} simulation bypasses the underlying physics and simply applies parametrized identification efficiencies, fake rates, momentum resolution and energy resolution as measured in data or full simulation. Parametrizations like eq.s \ref{eqn:parametrize1} and \ref{eqn:parametrize2} are applied directly to generator particle four-vectors using random numbers. 

A fast simulation can be considered a map from the domain of generator particle four-vectors (Monte Carlo \emph{truth}) in each event to a range of energy and momentum \emph{smeared} particles which survive a veto to account for identification inefficiencies. The amount of smearing depends on the resolution parametrization as encoded in the fast simulation. Thus there is no detector geometry, material specification or physical process simulation. Such simplification speeds processing time greatly, so that fast simulation can be several orders of magnitude faster than full simulation.

Delphes \cite{Selvaggi:2014mya,Mertens:2015kba} is a powerful and flexible fast simulation C++ program in widespread use. It has been used extensively for fast LHC studies and is now in use for possible future colliders. Delphes uses the TCL scripting language to define detector performance in a detector \emph{card}. The detailed baseline design performance of the SiD detector has been encoded in the Delphes SiD (DSiD) card, available on HepForge \cite{Potter:2016pgp}. See Table \ref{tab:fullfast} (bottom) for the TCL fragment which defines the performance of the SiD ECal in DSiD. Finally see ref. \cite{Potter:2017rlo} for examples of ILC backgrounds generated with Whizard2 and MG5 aMC@NLO and simulated with Delphes using the DSiD card.

Once Delphes is installed, the executable name depends on the format of the generator input file. For StdHep, the executable is \texttt{DelphesSTDHEP}. To run with the DSiD TCL card,

\begin{center}
\texttt{<path>/bin/DelphesSTDHEP delphes\_card\_DSiDi.tcl out.root in.stdhep}
\end{center}

\noindent Delphes can also read HepMC, LHEF and other generator formats. The output file is in Root \cite{Brun:1997pa} format. Root is a C++ data analysis framework, the successor to the Fortran based forerunner PAW. Most recent analysis in HEP is carried out in Root, though Python is becoming more prevalent.

While fast simulation delivers results on a short timescale, the underlying assumptions are typically too optimistic. In Table \ref{tab:fullfast} (bottom), for example, the SiD ECal specifies that 100\% of electrons, photons and neutral pions (since $\pi^0 \rightarrow \gamma \gamma$) energy is contained and reported. This ignores the phenomenon of leakage, in which the electromagnetic shower starts too late for all energy to be contained in the ECal. It also assumes, by default, that all other particles leave no energy in the ECal. But hadrons, for example, will often start hadronic showers in the ECal before entering the HCal, and some hadron showers may be entirely contained in the ECal.

Similar arguments for other subdetectors apply. With tracks, for example, one can specify an identification efficiency, but one cannot easily include fake tracks created by complicated beam scenarios in which particles unrelated to the primary interaction leave hits in the tracker. Because the performance of fast simulation is typically too optimistic, its performance for a particular study is usually compared to performance on a smaller full simulation for validation or, at least, an estimate of systematic uncertainty due to the use of fast simulation.

Fast simulation not only bypasses full simulation of the detector geometry and physical processes, it also bypasses the necessary task of converting hits and digitizations reported in the full simulation into particle candidate four-vectors. We consider this task in the following section.

\subsection{Reconstruction of ILC/SiD events \label{sec:reco}}

\subsubsection{Track finding and fitting}

A charged particle traveling in a constant magnetic field directed in the $z$ direction exhibits helical motion. In the $xy$ plane the particle follows a circle, while in the $z$ direction the particle follows a line. Where the helix intersects the layers of the tracker, the particles most probably leaves a hit detected by the tracker readout. Track \emph{finding} is the project of correctly grouping together the hits left by the particle so that, in track \emph{fitting}, the helix can be explicitly reconstructed.

Suppose a tracker is such that, in any direction, a particle will traverse $m$ layers. Then any stable or quasistable particle traversing the tracker will leave $m$ hits (assuming 100\% hit efficiency). If a collider event contains $n$ charged particles, the task of track finding is to correctly group the $n\times m$ hits into the $n$ tracks left by each such particle. There is a finite number of such groupings,

\begin{eqnarray}
  N_{g} & = & {n m \choose m}  \times  {(n-1)m  \choose m}  \times \cdots  \times {m \choose m} 
\end{eqnarray}

\noindent but $N_g$ gets unreasonably large for even modest track multiplicity $n$. Fortunately there are constraints on each grouping. First, each hit in a nonpathological track must belong to a distinct tracker layer. Second, each track must be consistent with the tracker spatial resolution.

The measure of spatial resolution consistency is $\chi^2 = \frac{1}{n_{hits}-1} \sum_{hits} \vert \vec{r}_{fit}-\vec{r}_{hit} \vert^2/\sigma^2$, where $\vec{r}_{fit}$ is the point of intersection of the helical fit with the tracker layer, $\vec{r}_{hit}$ is the measured position of the associated hit and $\sigma$ is the spatial resolution of the tracker. Thus, of the $N_g$ track groupings, the unique grouping for which for each track all hits lie on distinct layers and the $\chi^2$ is minimal is likely to be the correct one.

Constructing all track groupings for which every track has hits with distinct layers is straightforward using track \emph{seeds}. Two hits underdetermine a circle, four hits overdetermine it. A seed track consists of three hits, each from a distinct layer. The seed track is fitted with a circle in $xy$ and a line in $sz$, where $s$ is arclength, and the fit is extrapolated to the remaining $m-3$ layers not already in the seed. At each such layer the nearest hit is accumulated to the seed. All tracks from such seeds are found and for each the $\chi^2$ is calculated. After requiring $\chi^2 < \chi^{2}_{max}$ for some maximum $\chi^2$, the set of remaining tracks are good candidates for charged particle trajectory reconstruction.

Once the track finding has finished, a final $xy$ circular fit to determine the three circular parameters (center $(x_0,y_0)$ and radius $R$) and $sz$ linear fit to determine two linear parameters (slope $m$ and intercept $b$) are performed for each track. Many fitting procedures are possible: least squares fitting suffices. There are five track fit parameters to extract:

\begin{enumerate}

\item $\Omega=R^{-1}$, the curvature where $R$ is the radius 

\item $d_0$, the transverse impact parameter in $xy$

\item $\phi_0$, the azimuthal angle $\phi$ at closest approach in $xy$

\item $\tan \lambda=\frac{ds}{dz}$, the ratio of arclength $s$ to $z$ traversed
  
\item $z_0$, the azimuthal impact parameter in $z$
  
\end{enumerate}

\noindent The impact parameters $d_0,z_0$ measure how far from the interaction point the track began. The radius of curvature determines the transverse momentum $p_T$ from eq. \ref{eqn:pisqbr}, and $\tan \lambda$ determines the azimuthal momentum $p_z$. For a full definition of these parameters as saved in LCIO files, see ref. \cite{kramer_2006_004}.

This idealized description of track finding and fitting is complicated in practice. Adjustments in track finding must be made for the case of nonzero hit inefficiency. Electrons will frequently emit bremsstrahlung while transiting the tracker, after which the track radius of curvature changes. Frequently there are additional hits in the tracker from events unrelated to the primary event, including noise hits. All of these issues are addressed in track finding. For the case of charged particles leaving less than 3 hits there is no redress.

One example of a track finder and fitter implemented in ILCsoft is Conformal Tracking \cite{Brondolin:2019awm}. This algorithm first maps hits from $(x,y) \mapsto (u,v)=(x/(x^2+y^2),y/(x^2+y^2))$. This map has the property of mapping large radii to small radii and small radii to large radii for circles centered on the origin, and maps circles to lines for circles centered elsewhere. The effect is to make track pattern recognition intuitively clear, though at present no clear performance advantage has been demonstrated.

\subsubsection{Calorimeter cluster finding}

A calorimeter hit is a spatial location, like a tracker hit, together with an energy deposit. A single particle traversing the calorimeter, whether charged or neutral, will leave a hit in each traversed layer if it undergoes sufficient energy loss, and may leave hits in adjacent cells in the same layer.

The group of calorimeter hits left by a single particle is a \emph{cluster}. Cluster \emph{finding} is the project of correctly grouping calorimeter hits in events with multiple particles leaving multiple clusters. Once the hits are grouped into clusters, the cluster energy for each cluster is calculated by summing the cluster hit energies.

One strategy for cluster finding exploits the fact that clusters left by multiple particles are not usually contiguous, they are usually topologically distinct. The algorithm for a \emph{topological} cluster finder begins with seed hits, usually all hits satisfying a minimum energy requirement, and for each seed proceeds by recursively associating hits adjacent to the seed in the same layer and nearby hits in adjacent layers. If two seeds belong to the same cluster, they are naturally merged by the algorithm. The algorithm proceeds until there are no more adjacent hits. Setting the minimum energy for a seed is a tradeoff between the speed of the cluster finding and identification of low energy clusters. If the minimum seed energy is zero, all clusters will be identified but with a time penalty. Conversely, if the seed energy is high, low energy clusters will not be identified but the cluster finding is fast.

Complicating cluster finding are cases in which distinct clusters, by chance, overlap. There may also be noise hits which do not properly belong in any cluster. Clusters may be split between barrel and endcaps, and for hadrons they may be split between ECal and HCal. All such complicating issues are addressed in a cluster finder.

After cluster finding, the concept of \emph{particle flow} enables further classification of clusters. Particle flow relies on the straightforward observation that clusters left by charged particles will have associated tracks left in the tracker, while clusters left by neutral particles will not. A particle flow algorithm will extrapolate tracks into the calorimetry and associate each track to the nearest cluster. Track-associated clusters are considered to be left by charged particles and the momentum of the track, inherently more precise than the energy in the cluster, supersedes the energy of the cluster. Any cluster unassociated to a track is considered to be left by a neutral particle. PandoraPFA \cite{Thomson200925} is an example of a particle flow algorithm which has been implemented in ILCsoft.

Finally, if most of the cluster energy lies in the ECal, the cluster is considered to be left by an electron or photon. If most of the cluster energy lies in the HCal the cluster is considered to be left by a hadron. Thus are built candidate lists of electrons and charged hadrons, which inherit the inherently more precise track momentum, and candidate lists of photons and neutral hadrons, which inherit the cluster energy. Tracks which cannot be associated to any calorimeter cluster but which extrapolate to muon detector hits build candidate lists of muons. Objects identified in this way are called \emph{particle flow objects} (PFOs).

\subsubsection{Jet and vertex finding}

In sect. \ref{sec:pythia} we saw that quarks and gluons produced in a collider event undergo showering and hadronization, that is the gluons split to quark pairs, quarks radiate gluons, confinement generates many mesons and baryons, and those mesons and baryons decay. The project of \emph{jet finding} is to correctly assign the reconstructed stable and quasistable particles in a \emph{jet} and to reconstruct the four-vector of the initiating quark or gluon. The project of \emph{vertex finding} is to identify the location of a hadron decay within a jet from the tracks left by the charged particles in the hadron decay products.

Two types of jet finding algorithms are in common use: cone-based and sequential recombination algorithms. Cone-based algorithms start from a set of seeds, usually objects which satisfy a minimum energy, and group all other objects within a cone of fixed angular radius $R$ together. The objects may be hits, clusters, tracks, or PFOs. Each seed defines a cone axis. For each seed, the energy weighted position $\sum_i E_i r_i/\sum_i E_i$ (the sum is over all objects $i$ in the cone), or \emph{centroid}, defines a new geometric axis and the procedure iterates until the cone axis converges.

Whereas cone algorithms are top down, sequential recombination algorithms are bottom up. They begin by defining a distance measure between all object pairs, $d_{ij}=\min(k_{ti}^{2p},k_{tk}^{2p}) \Delta_{ij}/R^2$ and for single objects $d_{i}=k_{ti}^{2p}$. Here $k_{t}$ is transverse momentum, $\Delta_{ij}$ is the angular distance between objects $i$ and $j$, $R$ is a tuneable parameter and $p=-1,0,+1$ for $k_t$, Cambridge/Aachen, and anti-$k_t$ algorithms. If the minimum over the list of all distance measures is a $d_{ij}$, the objects are merged, the distances are re-calculated, and the procedure repeats. If the minimum is a $d_i$, the object is removed from the list and called a jet.

Jet finders perform differently when faced with radiated low energy gluons. If the gluon is radiated between two jets, it may cause the jets to be incorrectly merged in jet finding. Algorithms which prevent this are called \emph{infrared safe}. Algorithms which prevent problems in jet finding due to gluon emision along the jet axis are called \emph{collinear safe}.

FastJet \cite{Cacciari:2011ma} is a C++ library of jet finders. It has been incorporated into both Delphes and ILCsoft, but it can be installed for standalone jet finding. Of particular interest for $e^+ e^-$ collider events is the Durham algorithm, otherwise known as the $k_t$ algorithm for $e^+ e^-$ events, with distance measure $d_{ij}=2 \min(E_i^2,E_j^2)(1-\cos \theta_{ij})$. All sequential recombinations algorithms are implemented in FastJet, and the various cone algorithms used by collider experiments are implemented as plugins.

Within jets are hadrons produced by fragmentation and decay. Top quarks decay $t \rightarrow bW$ before they can hadronize, but every other quark in the SM hadronizes into mesons and baryons, which then produce a cascade of decays to hadrons composed of bound states of lighter quarks. For a $b$ quark, the cascade is formulated as $b \rightarrow W c \rightarrow W^+ W^- s$, producing $B$ mesons, $D$ mesons, and $K$ mesons respectively, together with the $W$ decay products. The kaons finish the cascade decay to hadrons made from first generation quarks. If two or more charged particles appear in their decays, they form tracks which form a vertex at the location of the decay.

From Table \ref{tab:mesons}, $c\tau$ for $B$ and $D$ range from $0.5$ to $0.1$ mm respectively, and the  decay distance can be significantly increased due to time dilation. Thus we expect a \emph{primary vertex} at the collision point, a \emph{secondary vertex} at the $B$ decay point, and a \emph{tertiary vertex} at the $D$ decay point. The number of vertices and their distance from the primary vertex serve to distinguish events containing $b$ quarks from other events. A \emph{$b$-tag} incorporates such information in making a determination of the jet \emph{flavor}.

In principle the track impact parameters $d_0$ and $z_0$ should be sufficient to determine which tracks form vertices. In practice the uncertainty on those measurements makes this challenging. Nevertheless track impact parameter significances, $d_0/\sigma_{d_0}$ and $z_0/\sigma_{z_0}$ are often used as additional inputs to a $b$-tag since they can be large for tracks from $B$ decay. The number of tracks in an event with large impact parameter significances is also frequently used as an input to a $b$-tag.

In ILCsoft, vertex finding packages have been developed for ILC detectors in the Linear Collider Flavor Identifier (LCFI). LCFIPlus \cite{Suehara:2015ura} combines jet and vertex finding with multivariate techniques for optimal performance. In another approach to vertex finding, one wraps each track with a Gaussian probability tube using the measured track parameter uncertainties. For each track $i$ ($i=1,\dots,n$), the probability function $f_i(r)$ is formed, and for the collision point an ellipsoidal probability function $f_0(r)$  is formed. Then the \emph{vertex function} 

\begin{eqnarray}
V(r) & = & \sum_{i=0}^{n} f_{i}(r) - \frac{\sum_{i=0}^{n} f_{i}^{2}(r)}{\sum_{i=0}^{n} f_{i}(r)}
\end{eqnarray}

\noindent yields maxima at vertices, provided they are resolved, and minima where there is only one track or no track at all. This technique was used successfully at SLD with ZVTOP \cite{JACKSON1997247}.

\subsubsection{ILCsoft reconstruction with Marlin}

In ILCsoft the executable for running the reconstruction chain, including tracking, cluster finding, particle flow, jet finding and vertex finding is \texttt{Marlin}. See Appendix \ref{appendix2} for instructions on installing and testing ILCsoft.

The particular C++ code which is invoked for each reconstruction function is defined in the \texttt{reconstruction.xml} XML file and passed to \texttt{Marlin} as the first argument. The input and output LCIO filenames are the second and third arguments:
\begin{center}
\texttt{Marlin reconstruction.xml --global.LCIOInputFiles=in.lcio / --MyLCIOOutputProcessor.LCIOOutputFile=out.lcio}
\end{center}

\noindent See Table \ref{tab:reco} for an XML fragment within the reconstruction XML which defines the underlying compiled C++ processors to invoke. All such processors are named within the XML \texttt{execute} tag. Algorithm parameters can be passed to the processors from the XML in the \texttt{processor} tags which follow the \texttt{execute} tag.

\begin{table}[t]
\begin{center}
\begin{tabular}{|l|} \hline
\multicolumn{1}{|c|}{Marlin SiD Reconstruction XML} \\ \hline
 $\langle$execute$\rangle$ \\
\hspace{0.2in}    $\langle$processor name="InitDD4hep"/$\rangle$ \\
\hspace{0.2in}    $\langle$processor name="VertexBarrelDigitiser"/$\rangle$       \\       
\hspace{0.2in}    $\langle$processor name="TrackerBarrelPlanarDigiProcessor"/$\rangle$ \\
\hspace{0.2in}    $\langle$processor name="MyConformalTracking"/$\rangle$ \\
\hspace{0.2in}    $\langle$processor name="ECalBarrelDigi"/$\rangle$ \\
\hspace{0.2in}    $\langle$processor name="ECalBarrelReco"/$\rangle$ \\
\hspace{0.2in}    $\langle$processor name="HCalBarrelDigi"/$\rangle$ \\
\hspace{0.2in}    $\langle$processor name="HCalBarrelReco"/$\rangle$ \\
 \hspace{0.2in}   $\langle$processor name="MyDDSimpleMuonDigi"/$\rangle$ \\
\hspace{0.2in}    $\langle$processor name="MyDDMarlinPandora"/$\rangle$ \\
\hspace{0.2in}    $\langle$processor name="MyFastJetProcessor"/$\rangle$ \\
\hspace{0.2in}    $\langle$processor name="MyZVTOP\_ZVRES"/$\rangle$ \\
\hspace{0.2in}    $\langle$processor name="MyLCIOOutputProcessor"/$\rangle$ \\
$\langle$/execute$\rangle$ \\ \hline
\end{tabular}
\caption{Schematic of the processors selected within the \texttt{execute} tag of the SiD reconstruction XML passed to the ILCsoft \texttt{Marlin} executable. Each processor is a compiled C++ program. Algorithm parameters are configured within the \texttt{processor} tags (not shown) and passed to the C++ processors.}
\label{tab:reco}
\end{center}
\end{table}

For the particular reconstruction sequence defined in Table \ref{tab:reco}, the following processors are invoked:

\begin{enumerate}

\item Digitization: default digitization for all subdetectors

\item Track Finding and Fitting: Conformal Tracking \cite{Brondolin:2019awm}

\item Cluster Finding and Particle Flow: PandoraPFA \cite{Thomson200925}

\item Jet Finding: FastJet \cite{Cacciari:2011ma} configured by XML

\item Vertex Finding: ZVTOP \cite{JACKSON1997247} topological vertexing

\end{enumerate}

\noindent The output of each step in the reconstruction chain is the input for the subsequent step: digitization output is input for all following steps, track finding output is input for particle flow, particle flow objects are input for jet finding, jets and tracks are input for vertex finding.

Alternative processors for any of the steps in the reconstruction chain may be specified, as long as they are C++ implemented in ILCsoft. Furthermore each alogorithm processor parameter is configurable by in separate tags in the reconstruction XML. In some cases they are highly configurable, and the user is warned that expert advice is required for optimal configuration. A Python alternative to Marlin reconstruction is described in \cite{Potter:2020ihz}, and a Julia alternative in \cite{Stanitzki:2020bnx}.

\subsubsection{Shortlived particle reconstruction}

Standard reconstruction sequences usually end after tracks, clusters, PFOs, jets and vertices are found, leaving shortlived particle reconstruction to the analysis of individual users. Such analyses are usually carried out in C++ or Python, though in principle any language can be used. See Appendix \ref{appendix2} for instructions on accessing LCIO files with Python.

Particles like the $\pi^0$, $K_S$, $\phi$, $J/\Psi$ and $\Upsilon$ are straightforward to reconstruct using four-vector addition. For $\pi^0 \rightarrow \gamma \gamma$ one simply iterates through all pairs of photons reconstructed in the event, forms four-vectors of each photon in the pair from the measured energy and position of the associated clusters, adds the four-vectors and calculates the invariant mass, which must be near the measured mass of the $\pi^0$. Similarly for $K_S \rightarrow \pi^+ \pi^-$ and $\phi \rightarrow K^+ K^-$ one iterates through pairs of oppositely charged hadrons. For leptonic decays $J/\Psi,\Upsilon \rightarrow \ell^+ \ell^-$ where $\ell=e,\mu$ this is also straightforward.

Gauge bosons can be more challenging. Reconstruction of gluons in $g \rightarrow q\bar{q}$ is straightforward since each quark produces a jet. If the jets are resolved, the gluon is reconstructed from the pair of jets. In many cases the quark pair produces only one jet which is itself the reconstructed gluon. Photons are usually reconstructed directly in the ECal, but can also be reconstructed if they pair produce $\gamma \rightarrow e^+ e^-$.

Electroweak gauge boson reconstruction can be complicated by the presence of one or more neutrinos in the decay products. For a hermetic detector which measures all particles (apart from neutrinos) produced in a collider event, the \emph{missing energy} and \emph{missing momentum} can be calculated by imposing conservation of energy and momentum. For a lepton collider the initial state beam energy and momentum and the final state energy and momentum must be equal, which yields the four-vector sum of the neutrinos in the event. For a hadron collider only the initial state transverse to the beamline is known.

For $Z\rightarrow \nu \bar{\nu}$ there is no recourse unless the signal event topology and missing four-vector provide additional leverage. For $Z \rightarrow \tau^+ \tau^-$ the situation is somewhat improved, but a neutrino pair still make reconstruction challenging. The situation is much easier with $Z \rightarrow \ell^+ \ell^-$ for $\ell=e,\mu$. For $Z \rightarrow q\bar{q}$ one looks for two jets initiated by the quark pair, requiring a jet pair mass consistent with the $Z$ mass within jet energy resolution.

For $W \rightarrow q \bar{q}^{\prime}$ the strategy is the same as for $Z \rightarrow q\bar{q}$. For $W \rightarrow \ell \nu$ where $\ell =e,\mu$ the $W$ may be reconstructed in a signal topology with a single neutrino, whose four-vector may be equated to the missing four-vector. For two or more neutrinos this cannot be done. For $W \rightarrow \tau \nu$ the reconstruction is complicated by the presence of at least one additional neutrino in the $\tau$ decay.

Higgs boson reconstruction occurs in many more final states. For decays to boson pairs $H \rightarrow \gamma \gamma,gg$ the reconstruction is straightforward. For decays to boson pairs $H \rightarrow WW^{\star},ZZ^{\star}$, one boson is virtual so the $W$ or $Z$ mass constraint cannot be applied, but the decay products are nonetheless the same as for on shell decays.

For decays to quark pairs $H \rightarrow b\bar{b},c\bar{c}$ the strategy is the same as for $Z \rightarrow q\bar{q}$, where the $b$-tag may be employed to distinguish the flavor of the quark pairs. For $H \rightarrow \mu^+ \mu^-$ the reconstruction is straightforward, while for $H \rightarrow \tau^+ \tau^-$ the reconstruction is complicated by the presence of a neutrino pair. The \emph{collinear approximation} exploits the signal topology of a massive particle decaying to tau pairs by assuming that the neutrinos are collinear with the visible tau decay products, and can therefore be extracted by projecting the missing momentum onto the visible decay products.

\subsection{Further reading and exercises}

The Particle Data Group reviews on \emph{Monte Carlo Techniques}, \emph{Monte Carlo Event Generators} and the \emph{Monte Carlo Particle Numbering Scheme} \cite{Tanabashi:2018oca} are useful (the last is invaluable). The reprinted paper on Monte Carlo in \emph{Experimental Techniques in High Energy Nuclear and Particle Physics} (Ferbel) \cite{Ferbel:238481} is good. 

For the software discussed here (Whizard2, MG5  aMC@NLO, Pythia6, Pythia8, Geant4, Delphes, DD4hep, FastJet) the technical writeups in journals referenced in the endnotes are, of course, invaluable. So are the user's manuals available, in most cases, on the software webpages. The Pythia6 writeup is a literary classic in technical documentation.

Exercises for this section can be found in sect. \ref{sec:sid} of Appendix \ref{appendix1}.

\section{Conclusion: sensitivity and optimization}

We have given a comprehensive overview of the physics potential for the ILC as well as the software tools available for research and development on SiD, one of two detector proposals detailed in the ILC TDR. While the nominal SiD design is complete, rigorously evaluated, and carefully costed, a final round of costing and optimization involving a larger community of physicists is likely to occur in the event that construction is approved by a host nation and SiD proceeds to the technical design phase. That new round will necessarily involve a new generation of physicists adept at modern software. 

The author hopes that this primer will provide a foundation of scientific knowledge about the ILC and the technical knowledge necessary for exploiting currently available tools explicitly developed for this purpose, as well as pointers to more advanced reading and specialized technical tools. Before concluding this primer, we review the concept of the sensitivity of an experiment and its relation to optimization of the experimental design within cost constraints.

The \emph{sensitivity} of a live experiment refers to the statistical significance, in the case of signal observation, a limit on the maximum possible signal, in the case of signal nonobservation, or the precision with which a physical parameter can be measured. The \emph{expected sensitivity} of a future experiment is evaluated by simulation of the experiment.

If we denote the total number of background events $B$ then the Gaussian statistical \emph{uncertainty} on $B$ is $\sqrt{B}$. For small $B$ a Poisson treatment of uncertainties is necessary, but we assume here we are in the Gaussian regime. If we denote the signal $S$, then the statistical significance of the signal is $S/\sqrt{B}$ and, if we account for uncertainty on the signal as well as the background, $S/\sqrt{S+B}$. Statistical uncertainty is sometimes called somewhat misleadingly \emph{error}. It has become accepted practice in HEP to refer to a signal with significance $S/\sqrt{B} \approx 3$ as signal \emph{evidence} and a significance $S/\sqrt{B} \approx 5$ as signal \emph{observation}. These significances correspond to 68.3\% and 99.7\% of a normally distributed variable. 

The signal $S$ and background $B$ are estimated by an \emph{analysis} of data and simulated data, usually performed with computer code. An analysis can be considered a filter which maximizes signal selection and minimizes background selection in order to maximize the signal significance. The \emph{efficiency} of the analysis for a signal or background is the event count $n$ after applying the filter divided by the event count  $N$ before applying the filter. 

Because efficiency is a proportion, the binomial uncertainty is the appropriate way to evaluate the statistical uncertainty.  For an efficiency $\epsilon=n/N$, the uncertainty is

\begin{eqnarray}
\delta_{\epsilon} & = & z_{ci} \sqrt{\frac{\epsilon (1-\epsilon)}{N}}
\label{eqn:uncertainty}
\end{eqnarray}

\noindent where $z_{ci}=1.00, 1.96,2.58$ for the 68\%, 95\% and 99\% confidence intervals. For the standard $1\sigma$ error, $z_{ci}=1$. As one would expect, an efficiency becomes more precise when a larger number of events $N$ are used to evaluate it.

Another source of uncertainty, qualitatively much different from statistical uncertainty, is \emph{systematic} uncertainty. Systematic uncertainties are attempts to parametrize our ignorance of an experiment. For example, one might identify a systematic uncertainty arising from differences in analysis selection efficiencies by using different event generators to simulate the signal process. Evaluating \emph{systematics} is as much art as it is science, and identifying all possible systematics can be challenging. Systematic and statistical uncertainties are usually reported separately, but when they must be combined they are usually combined in quadrature.

Suppose we would like to optimize the expected ILC sensitivity to Higgstrahlung events. We first write the analysis code to try to maximize the Higgstrahlung event selection and simultaneously minimize the corresponding background selection. Then we evaluate the efficiencies for signal ($\epsilon_s$) and background ($\epsilon_b$) by running the code over simulated signal and background events, taking care to evaluate all possible backgrounds. The expected signal significance for Higgstrahlung at the ILC is overwhelmingly large - this is precisely the argument for building the ILC, a Higgs factory - but claiming this and proving it with full simulation data and a careful consideration of backgrounds are two different things. 

If we assume an integrated luminosity $\int dt \mathcal{L}$ and cross sections for signal Higgstrahlung $\sigma_s$ and backgrounds $\sigma_b$, then 

\begin{eqnarray}
\frac{S}{\sqrt{B}} & = & \frac{\epsilon_s \sigma_s \int dt \mathcal{L}}{\sqrt{\sum_b \epsilon_b \sigma_b \int dt \mathcal{L}}}
\label{eqn:sig}
\end{eqnarray}

\noindent where the sum is over all backgrounds. We emphasize that the luminosity, center of mass energy and beam polarization (which determine the cross section) and their uncertainties depend on the ILC, while the efficiencies and their uncertainties depend on the detector, in our case SiD. 

It should be noted that the precision with which an analysis can measure a property of the Higgs boson, its branching ratios for example, depends on the number of Higgs bosons produced. The numerator $\epsilon_s \sigma_s \int dt \mathcal{L}$ in eq. \ref{eqn:sig}, the signal significance, is also the denominator $N$ in eq. \ref{eqn:uncertainty}, the uncertainty on measuring a proportion like a branching ratio. Clearly, choosing beam parameters to maximize $\sigma_{s}$ and $\int dt \mathcal{L}$, while minimizing $\sigma_{b}$, are part of the optimization. Thus $\sqrt{s}$, beam polarization and luminosity are key ILC parameters. The remaining part of the optimization, maximizing $\epsilon_{s}$, is the job of detector optimization. For recent estimates of the sensitivity of the ILC to a variety of physical parameters, including Higgs boson branching ratios, see \cite{Bambade:2019fyw,Fujii:2019zll}.

The efficiencies $\epsilon_s$ and $\epsilon_{b}$ in eq. \ref{eqn:sig}  with their uncertainties are complex quantities, dependent on many underlying detector performance measures in particle reconstruction, identification and precision. Tracking efficiency and precision, calorimeter cluster finding and precision, jet finding and precision, and vertex finding and precision all play a role in determining $\epsilon_s$, and therefore signal sensitivity, for a given signal process. Ultimately these parameters are all determined by the detector design.

\begin{table}[t]
\begin{center}
\begin{tabular}{|c|c|} \hline
Material & Unit Cost [USD]\\ \hline
ECal Tungsten & ($180 \pm 75$)/kg \\
Silicon Detector & ($6 \pm2$)/cm$^2$ \\
HCal Tungsten & ($105 \pm 45$)/kg \\
HCal Steel & ($4500 \pm 1000$)/ton \\ \hline
\end{tabular}
\caption{Material costs per unit (2008 USD) agreed to by SiD, ILD and CLIC for the ILC TDR. Adapted from the ILC TDR \cite{Behnke:2013lya}.}
\label{tab:material}
\end{center}
\end{table}

Reducing a detector's cost to a minimum is straightforward, but the performance and ultimate physics goals will suffer. Conversely, designing a highly performant detector to maximize $\epsilon_s$ is also straightforward, but the cost may be too high to pay. The right balance between cost and performance must be struck. Typically a detector community targets a performance goal and then, in an effort to minimize the cost necessary to reach that goal, performs a detector \emph{optimization}. If the optimized cost is too high, the performance goals are reduced and the detector is reoptimized. Material cost assumptions are key inputs to this process. For SiD (as well as ILD and CLIC) the estimated material costs of Silicon, Tungsten and Steel assumed in the ILC TDR  are summarized in Table \ref{tab:material}.

In the SiD Loi \cite{Aihara:2009ad} a an optimization of the solenoid field strength $B_z$, the calorimetry inner radius $R$ (equivalently the tracking outer radius) and the HCal depth $n\lambda$ was performed, yielding the nominal configuration $B_z=5$~T, $R=1.25$~m and $n=5$. These parameters together determine the jet energy resolution, a critical factor in determining the precision to which Higgs boson branching ratios may be measured with SiD. The initial LoI optimization yielded the cost estimates detailed in the ILC TDR \cite{Behnke:2013lya} chapter on SiD costs. See Table \ref{tab:cost} for a summary of these costs. 

\begin{table}[t]
\begin{center}
\begin{tabular}{|c|c|c|c|} \hline
Subdetector & Base [MUSD]& Eng. [MY] & Tech. [MY]\\ \hline
Vertex Det. & $2.8 \pm 2.0$ & 8.0 & 13.2 \\
Tracker & $18.5 \pm 7.0$ & 24.0 & 53.2 \\
ECal & $104.8 \pm 47.1$ & 13.0 & 288.0 \\
HCal & $51.2 \pm 23.6$ & 13.0 & 28.1 \\
Solenoid & $115.7 \pm 39.7$ & 28.3 & 11.8 \\
Muon Det. & $8.3 \pm 3.0$ & 5.0 & 22.1 \\ \hline
\end{tabular}
\caption{Baseline material cost (2008 MUSD) and engineering and technical labor (MY) estimated to build SiD subdetectors. Costs not shown are beamline systems, electronics, installation and management. Adapted from the ILC TDR \cite{Behnke:2013lya}.}
\label{tab:cost}
\end{center}
\end{table}

Determining which costs will be borne by the accelerator side and which will be determined by the detector side is a critical component of costing a detector. The TDR detector costing assumes the following costs are borne by the accelerator: detector hall with lighting and electrical power, internet and compressed air utilities, compressed Helium piping, and surface buildings and construction cranes. Another critical component in costing is who bears the cost of gray areas: research and development, detector commissioning, operating costs and physicist salaries. The ILC TDR cost estimate does not include these.

If we consider the HCal and Solenoid optimization to be final, then the most conspicuous cost is for the ECal, requiring a material baseline of $104.8 \pm 47.1$ MUSD and labor costs of 301.0 person years. The nominal ECal design thus requires more than three times the combined vertex detector, tracker, and muon detector cost in material alone. For labor the factor is even larger. A global optimization of ECal design parameters, including the total number of layers and thin and thick Tungsten layer widths, may find that a substantial reduction in cost results with a minimal loss in performance. One preliminary study finds this to be the case \cite{Braun:2020eme}.

\section*{Acknowledgements}

\begin{acknowledgement}

Much of this primer began as lecture notes for the graduate courses \emph{Physics 610: Collider Physics} and \emph{Physics 662: Elementary Particle Phenomenology}, taught at UO in Winter and Spring terms in 2018. Thanks to Jim Brau, who generously allowed the author to fill in as instructor of record for 610 and 662 while Brau attended to ILC matters. Brau first taught the author about measuring Higgs boson branching ratios at a linear collider two decades ago \cite{potter2003}. 

Thanks to doyens Marty Breidenbach and Andy White for sharing their wisdom at the SiD Optimization meetings. Breidenbach, who maintains an incomprehensibly large store of experience, worked on deep inelastic scattering at Stanford, SPEAR and SLD and was an early architect of SiD. Thanks also to Jan Strube and Dan Protopopescu, who have convened the SiD Optimization meetings over the past few years. Protopopescu cowrote and maintains the DD4hep SiD detector description.

The historical material on accelerators and detectors was written while preparing lectures for \emph{Honors College 209H: Discovery of Fundamental Particles and Interactions} taught at UO Winter and Spring 2020. Thanks to Daphne Gallagher, Associate Dean in the Clark Honors College, for encouraging the development of this experimental undergraduate course. 

\end{acknowledgement}

\appendix

\section*{Appendices}

\section{Exercises \label{appendix1}}

\subsection{Higgs factory physics \label{sec:higgs}}

\begin{enumerate}
\item Show that, for scattering from a hard sphere with radius $R$, the impact parameter and scattering angle are related by $b=R\cos \frac{1}{2}\theta$. Calculate the total cross section for $b<R$. What is the luminosity?
\item The Rutherford potential is given by $V(r)=q_1q_2/r^2$, from which it can be shown that $\cot \frac{1}{2}\theta=(2bE/q_1q_2)^2$, where $E$ is the energy of the incident particle.
\begin{enumerate}
\item Calculate the classical differential cross section for Rutherford scattering.
\item Calculate the total cross section for Rutherford scattering. Explain.
\end{enumerate}
\item Show that the $n$th Born Approximation can be written $\Psi_n = \Psi_0 + \sum_{i=1}^{n} \int g^i V^i \Psi_0$ using the recursion relation in eq. \ref{eqn:born}.
\item The Yukawa potential is given by $V(r)=\beta \exp (-\mu r)/r$ where $\beta,\mu$ are constants.
\begin{enumerate}
\item Calculate the Yukawa differential cross section in the first Born approximation.
\item Calculate the Rutherford differential cross section in the first Born approximation.
\end{enumerate}

\item Apply the Euler-Lagrange equation 

\begin{eqnarray*}
\partial_{\mu} (\partial \mathcal{L}/\partial(\partial_{\mu} \phi))-\partial_{\phi} \mathcal{L} & = & 0 
\end{eqnarray*}

 to $\mathcal{L}_0$, $\mathcal{L}_{1/2}$, $\mathcal{L}_1$ to obtain the Klein-Gordon, Dirac and Maxwell equations.
\item Show that $\mathcal{L}_{1/2}+\mathcal{L}_1$ is invariant under $U(1)$ transformation $\psi \rightarrow \exp (-i\alpha(x)) \psi$ if $A \rightarrow A+ \nabla \alpha$.
\item The only leptons which are unstable against decay are the $\mu$ and the $\tau$.
\begin{enumerate}
\item Sketch the tree-level Feynman diagrams for each decay, using vertices from fig. \ref{fig:vertices}.
\item Obtain their lifetimes $\tau_{\mu},\tau_{\tau}$ from the PDG and give an expression for their time-dilated values.  
\end{enumerate}

\item Of all the hadrons (mesons and baryons), only the proton is stable against decay. Sketch the Feynman diagram for neutron decay $n \rightarrow e^- \bar{\nu}_e p$, neutral pion decay $\pi^0 \rightarrow \gamma \gamma$ and charged pion decay $\pi^+ \rightarrow \mu^+ \nu_{\mu}$.
\item The coupling of the Higgs boson to a gauge boson $V$ is $2m_{V}^2/v$, yet the decays $H \rightarrow gg$ and $H \rightarrow \gamma \gamma$ occur despite $m_{g}=m_{\gamma}=0$. Sketch the loop level Feynman diagrams for these decays using vertices from fig. \ref{fig:vertices}.
\item Calculate the decay rate ratios $\Gamma_{H \rightarrow f\bar{f}}/\Gamma_{H \rightarrow b\bar{b}}$ for leptons $f=e,\mu,\tau$ and quarks $f=u,d,s,c$.
\item Using CKM matrix elements from the PDG, estimate the decay rate ratios below.
\begin{enumerate}
\item $\Gamma_{W^+ \rightarrow u\bar{q}_d}/\Gamma_{W^+ \rightarrow u\bar{d}}$ for $q_{d}=s,b$
\item $\Gamma_{W^+ \rightarrow c\bar{d}}/\Gamma_{W^+ \rightarrow c\bar{s}}$
\end{enumerate}
\item Sketch the tree level diagram for $\tau$ decay through a virtual $W$. Based on the decays of the virtual $W$ allowed by energy conservation and the $W$ decay rates in Table \ref{tab:wzbr}, estimate the $\tau$ branching ratios. Compare to branching ratios in the PDG and comment.

\item From the tree level $Z$ and $W$ boson partial  widths given in this section:
\begin{enumerate}
\item Calculate each partial width numerically, and the total width from their sum.
\item Calculate the branching ratios for each decay. Compare your results to the PDG measured values. Explain.
\end{enumerate}
\item From the tree level (and loop level $gg,\gamma\gamma$) $H$ boson partial widths given in this section:
\begin{enumerate}
\item Calculate each partial width numerically, and the total width from their sum.
\item Calculate the branching ratios for each decay. Compare your results to the PDG measured values. Explain.
\end{enumerate}
\item Show that any four-vector contraction $p_{\mu}p^{\mu}$ is a relativistic invariant.
\item Consider the Mandelstam variables $s,t,u$. 
\begin{enumerate}
\item Show that $s-t-u=\sum_i m_{i}^2$.
\item Write $s,t,u$ in terms of $(E_i,\vec{p}_i)$ and the cosine of angles between incident particles.
\item Show that in two-body scattering $\sqrt{s}=E_1+E_2$ if $1$ and $2$ collide at $\theta=\pi$ with $\vec{p}_1=\vec{p}_2$.
\end{enumerate}
\item Define $R(\sqrt{s})\equiv \sum_q \sigma_{e\bar{e} \rightarrow q\bar{q}}/\sigma_{e\bar{e} \rightarrow \mu^+ \mu^-}$, where $N_c=3 \rightarrow N_c=1$ in the cross section for muon pair production. Sketch $R$\emph{vs.}$\sqrt{s}$ for $\sqrt{s} \in [0,20]$~GeV.
\item The Higgs boson decay rate to hadrons dominates all others. Explain why there is no Higgs peak in fig. \ref{fig:eetohad} near $m_{H} \approx125$~GeV.
\item Sketch the Feynman diagrams for the 1f,2f,3f backgrounds with initial states $\gamma e$ and $\gamma \gamma$ discussed in this section.
\item Calculate the number of ILC Higgstrahlung events produced at $\sqrt{s}=250, 350, 500$~GeV for the projected integrated luminosities in Table \ref{tab:beams} using the cross sections in Table \ref{tab:ilcxsec}. Repeat for all backgrounds.
\end{enumerate}

\subsection{ILC: accelerators and detectors \label{sec:ilc}}

\begin{enumerate}
\item The Tevatron accumulated $\mathcal{L}=11$ fb$^{-1}$ during Run2, from 2001 to 2011. The LHC accumulated $\mathcal{L}=23$ fb$^{-1}$ during 2012.
\begin{enumerate}
\item If the Tevatron ($\sqrt{s}=1.96$~TeV) cross section for Higgs boson production is 1pb, how many Higgs bosons are in the Run 2 dataset?
\item If the LHC ($\sqrt{s}=8$~TeV) cross section for for Higgs boson production is 22pb, how many Higgs bosons are in the 2012 dataset?
\end{enumerate}
\item Read the ATLAS and CMS papers announcing discovery of the Higgs boson at the LHC, \cite{Aad:2012tfa} and \cite{Chatrchyan:2012ufa}. Compare and contrast their analyses. How are the detectors different?
\item Calculate the magnetic field $B$ used in Lawrence's 27 in tabletop proton cyclotron if the applied RF frequency was $\nu_0=27$MHz. What was the proton's momentum at maximum radius? What was the relativistic correction factor $\gamma$?
\item Calculate the magnetic field $B$ used in Lawrence's 60 in proton cyclotron if the maximum momentum was 16 MeV. What was the cyclotron frequency $\nu_0=qB/2\pi m$? What was the relativistic correction factor $\gamma$?
\item Plot the orbital frequency \emph{vs.}relativistic energy of a proton in a cyclotron with magnetic field $B$=0.01, 0.1,1~T. At what energies are the frequencies within tolerances of 1,10,100\% of the cyclotron frequency $\nu_0=qB/2\pi m$?
\item  For each collider in Table \ref{tab:circular}, compare the cross sectional area $A$ of the beams to that of SPEAR. Obtain the bunch populations and frequency $n_1,n_2,f$ from the PDG and elsewhere.
\item For each collider in Table \ref{tab:circular}, determine
\begin{enumerate}
\item the beam energy required for a fixed target accelerator to produce the same energy necessary for new particle creation.
\item the magnetic field required to hold the accelerated particle in circular orbit.
\item the power loss of one accelerated particle to synchrotron radiation.
\end{enumerate}
\item Derive an expression for the square of the center of mass energy for head-on collision of particles with energies $E_1$ and $E_2$. Derive the factor by which this is reduced for a crossing angle $\theta$. What is this factor for an ILC crossing angle $\theta=40$ mrad?
\item Derive $a$ with units in eq. \ref{eqn:pisqbr} from the Lorentz law. 
\item Calculate the instantaneous luminosities for each ILC $\sqrt{s}$ in Table \ref{tab:beams}. Then calculate the number of years of continuous running at each instantaneous luminosity required to obtain the projected integrated luminosities for scenario H-20 in the same Table.
\item Consider elements C,Si,Fe,W,U for calorimetry. 
\begin{enumerate}
\item To contain 95\% of electrons and photons, how deep (in m) must an ECal be for each element? 98\%? 99\%?
\item To contain 95\% of hadrons, how deep (in m) must an HCal be for each element? 98\%? 99\%?
\end{enumerate}
\item Modern collider detectors place detectors in order of smallest to largest radius: tracker, ECal, HCal (TEH). For all five other possible configurations (THE, ETH, EHT, HTE, HET), describe the performance for electrons, photons, charged hadrons and neutral hadrons. Why is the muon detector placed at largest radius?
\item Plot the tracking momentum resolution \emph{vs.}$p_T$ at $\theta=\pi/2$ for each detector listed in Table \ref{tab:performance} on the same plot. Comment.
\item Plot the electromagnetic calorimeter energy resolution \emph{vs.}energy for each detector listed in Table \ref{tab:performance} on the same plot. Comment.
\item Plot the hadronic calorimeter energy resolution \emph{vs.}energy for each detector listed in Table \ref{tab:performance} on the same plot. Comment.
\item Calculate the ionization energy loss for a minimum ionizing particle ($\beta^2=0.9$) in the SiD vertex detector and compare it to the energy loss due to electromagnetic showering. Do the same for the SiD tracker.
\item How many nuclear interaction lengths $\lambda$ are in the SiD vertex detector? Tracker? ECal? What fraction of its initial energy will a hadron lose before it enters the HCal?
\item Muon energy loss in SiD.
\begin{enumerate}
\item Calculate the ionization energy loss of a muon traversing the entire SiD detector assuming minimum ionization.
\item Calculate the synchrotron energy loss of a muon traversing the SiD solenoid field. Compare to the ionization energy loss.
\end{enumerate}
\item Calculate the minimum $p_T$ required for a charged particle to reach the inner and outer radii of the SiD
\begin{enumerate}
\item Vertex Detector and Tracker
\item ECal and HCal
\item Muon detector
\end{enumerate}
\item Read the descriptions of SiD and ILD in \cite{Behnke:2013lya}. How does ILD differ from SiD? How is it the same? 
\end{enumerate}

\subsection{SiD: simulation and reconstruction \label{sec:sid}}

\begin{enumerate}
\item For area of the unit circle $\int_{\mathbb{S}^2} dA$,
\begin{enumerate}
\item Write a Python program estimate the integral, sampling from $(x,y) \in [-1,1] \times [-1,1]$.
\item Now go to $(r,\theta)$ coordinates, perform the $\theta$ integration, and sample from $r\in [0,1]$.
\item Plot the approximation \emph{vs.} $N$ for both and compare.
\end{enumerate}
\item Consider the differential cross section for $e^+ e^- \rightarrow \gamma^{\star} \rightarrow q\bar{q}$ with $\beta \approx 1$, 
\begin{eqnarray*}
\frac{d\sigma}{d\Omega} & = & 3 q_{f}^2 \frac{\alpha^2}{3s} \left( 1+ \cos^{2} \theta \right)
\end{eqnarray*}
\begin{enumerate}
\item Integrate analytically to obtain the total cross section.
\item Write a Python program to generate $N$ events which reproduce $d\sigma/d\Omega$ for large $N$.
\end{enumerate}
\item Consider muon decay $\mu \rightarrow e \nu_{e} \nu_{\mu}$ and let $E$ represent the energy of the electron.  The partial width  is given by
\begin{eqnarray*}
\frac{d\Gamma}{dE} & = & \left( \frac{g_W}{m_W} \right)^4 \frac{m_{\mu}^2 E^2}{2 (4 \pi)^3} \left( 1-\frac{4E}{3m_{\mu}} \right)
\end{eqnarray*}
\begin{enumerate}
\item Calculate the total width $\Gamma_{\mu}=\int dE d\Gamma/dE$.
\item Write a Python program to generate $N$ decays which reproduce $d\Gamma/dE$ for large $N$.
\end{enumerate}
\item Install Whizard2 on your local computer. Use the script in Table \ref{tab:generators} (left) to generate Higgstrahlung events at $\sqrt{s}=250$~GeV. What is the reported cross section?
\item Repeate the above exercise, but with ISR turned off. What is the reported cross section?
\item Install MG5 aMC@NLO on your local computer. Use the script in Table \ref{tab:generators} (right) to generate Higgstrahlung events at $\sqrt{s}=250$~GeV. What is the reported cross section?
\item Switch the beam polarizations for the Higgstrahlung events from +30\%,-80\% to -30\%,+80\% for $e^+,e^-$. What are the reported cross sections in Whizard2 and MG5 aMC@NLO?
\item Using either Whizard2 or MG5 aMC@NLO, what is the reported cross section for $e^+e^- \rightarrow W^+ W^-$ with unpolarized beams at $\sqrt{s}=250$~GeV? Repeate with beam polariztions +30\%,-80\% and -30\%,+80\% for $e^+,e^-$. What are the reported cross sections?
\item Radiative return to the $Z$ with Whizard2.
\begin{enumerate}
\item Generate $e^+ e^- \rightarrow q \bar{q}$ events at $\sqrt{s}=250$~GeV with ISR turned off. What is the reported cross section?
\item Now turn ISR on and repeat. What is the reported cross section? Explain why there is such a large difference.
\end{enumerate}
\item Install Geant4 on your local computer.
\begin{enumerate}
\item Build a simple model of the SiD Tracker, ECal and HCal where each is a simple rectangular slab. 
\item Histogram the tracker momentum, ECal energy and HCal energy left by monoenergetic electrons, photons, charged pions, neutral kaons and muons after passing through your simulation
\end{enumerate}
\item Python fast simulation.
\begin{enumerate}
\item Write a Python program which implements a simple fast detector simulation with the SiD Tracker, ECal and HCal. Use the parametrizations in Table \ref{tab:performance}.
\item Histogram the tracker momentum, ECal energy and HCal energy left by monoenergetic electrons, photons, charged pions, neutral kaons and muons after passing through your simulation.
\end{enumerate}
\item Install Delphes on your local computer. Obtain the DSiD detector card from HepForge. Run over the Whizard2 files of Higgstrahlung events at $\sqrt{s}=250$~GeV made in the exercise above. Histogram the kinematic distributions of electrons, photons, charged hadrons and neutral hadrons.
\item Install ILCsoft on your local computer. Run the \texttt{ddsim} executable with the DD4hep compact description of SiD over the Whizard2 files of Higgstrahlung events at $\sqrt{s}=250$~GeV made in the exercise above. Then run the Marlin executable with the nominal SiD reconstruction XML file. Histogram the kinematic distributions of electrons, photons, charged hadrons and neutral hadrons.
\item Repeat the previous exercise, but vary the SiD detector in the DD4hep XML file in the following ways and report on any changes.
\begin{enumerate}
\item Add one (two, three) more layer(s) to the Tracker.
\item Take one (two, three) layer(s) away from the ECal.
\item Use Tungsten for the absorber in the HCal rather than steel.
\item Omit the Vertex Detector.
\end{enumerate}
\item Track parameters.
\begin{enumerate}
\item Derive expressions for the $p_T$ and $p_z$ of a charged particle starting from a subset of the five track parameters.
\item Derive an expression for the transverse impact parameter $d_0$ starting from the circular fit parameters $(x_0,y_0)$ and $R$. 
\end{enumerate}
\item A particle flow algorithm must extrapolate tracker tracks to calorimeter barrels and endcaps. For a track with circular fit parameters $(x_0,y_0)$ and $R$, solve for $(x,y)$ where the track intersects a barrel at radial coordinate $r$.
\item For SiD, calculate the points $(x,y)$ of intersection with tracker, calorimeter and muon detector inner and outer radii for
\begin{enumerate}
\item an electron with $p_T$=20 GeV and a photon with $E$=25 GeV (assume $\phi_0=\pi/4$)
\item a $K_L$ with $p_T$=50 GeV and a charged pion with $p_T$=45 GeV (assume $\phi_0=\pi/2$)
\item a muon with $p_T$=30 GeV (assume $\phi_0=\pi$)
\end{enumerate}
\item Prepare a set of $10^4$ simple events, $e^+ e^- \rightarrow b\bar{b}$ at $\sqrt{s}=250$~GeV, in Whizard2 or MG5 aMC@NLO. Run the nominal ILCsoft SiD simulation and reconstruction on the events. Identify the configurable parameters in the following processors and vary them around the nominal values in the Marlin XML. How do the results compare to the nominal results?

\begin{enumerate}
\item Tracking Processor: minimum hits for track, maximum $\chi^2$, seedfinding
\item Particle Flow Processor: tracks inputs, maximum track/cluster association distance
\item Jet Finding Processor: jet finding algorithm and algorithm parameters
\end{enumerate}
\item Write an analysis in Root C++ to select Higgstrahlung events simulated with Delphes and DSiD. For $\int dt L=1$~ab at $\sqrt{s}=250$~GeV, how many signal events are recovered by your analysis?
\item For the analysis of Higgstrahlung events in the previous problem, run the code over background events simulated with Delphes and DSiD. How many background events survive your analysis?
\end{enumerate}

%\appendix

\section{ILCsoft Installation and Use\label{appendix2}}

It is a truism that technical software documentation is obsolete almost as soon as it is written. Hopefully the instructions below will still be useful at the time the reader tries them. They assume a user in a bash shell on a Linux operating system connected to the Internet.

\subsection{ILCsoft from CVMFS}

ILCsoft version v02-00-02 is current at the time of writing, but make sure to use the most recent version. Presumably the reader is in a directory like /home/potter/ilcsoft/v02-00-02, and will make a new directory under the ilcsoft directory for other versions. In what follows, \texttt{bash>} is the shell prompt. The backslash \texttt{\textbackslash} indicates to continue on the same line \emph{with no space} and \emph{should not be typed}. 

These instructions assume that the CernVM Filesystem is mounted on your local computer. If it is not, your system administrator can mount it with the instructions in sec. \ref{sec:cvmfs}. First put the following two lines in a file called setup\_v02-00-02.sh:

\noindent \dotfill
\begin{verbatim}
source /cvmfs/sft.cern.ch/lcg/releases/gcc/4.8.4/\
       x86_64-slc6/setup.sh
source /cvmfs/ilc.desy.de/sw/x86_64_gcc49_sl6/\
       v02-00-02/init_ilcsoft.sh
\end{verbatim}
\noindent \dotfill

Source these setups, get the lcgeo package from GitHub, and build it. Note that in the cmake command the you need \texttt{\char18 which g++ \char18} and \texttt{\char18 which gcc \char18}. These should be returning paths to the compiler binaries in \texttt{/cvmfs/sft.cern.ch} to the cmake command, and should use the \char18 character (keyboard top left) rather than the ' character (keyboard middle right), despite the typography below:

\begin{verbatim}
bash> source setup_v02-00-02.sh
bash> git clone https://github.com/iLCSoft/lcgeo.git
bash> cd lcgeo
bash> mkdir build
bash> cd build
bash> cmake -DCMAKE_CXX_COMPILER=`which g++` \ 
      -DCMAKE_C_COMPILER=`which gcc` \ 
      -C $ILCSOFT/ILCSoft.cmake ..
bash> make -w -j4 install
\end{verbatim}

\noindent You may get an error indicating that you need to use a more recent version of cmake than you have installed. If so consult your system administrator. Now do some cleanup:

\begin{verbatim}
bash> cd ..
bash> source bin/thislcgeo.sh
bash> rm -rf bin/ddsim lib/python/DDSim
bash> which ddsim 
\end{verbatim}

\noindent The which command should return the executable from \texttt{/cvmfs/ild.desy.de} rather than from your local install. Now  test the simulation executable ddsim:

\begin{verbatim}
bash> cd example
bash> export PYTHONPATH=${LCIO}/src/python:\
      ${ROOTSYS}/lib:$PYTHONPATH
bash> python lcio_particle_gun.py
bash> ddsim --compactFile ../SiD/compact/\
      SiD_o2_v03/SiD_o2_v03.xml \
      --inputFiles mcparticles.slcio -N 10 \
      --outputFile simple_lcio.slcio
bash> anajob simple_lcio.slcio
bash> dumpevent simple_lcio.slcio  1
bash> cd ..
\end{verbatim}

\noindent Check the outputs to make sure they make sense. Now let's install some reconstruction software and run Marlin on the simple simulation file we just generated (simple\_lcio.slcio). First we need to get the SiDPerformance package and update the file SiDPerformance/gear\_sid.xml:

\begin{verbatim}
bash> git clone https://github.com/iLCSoft/\
      SiDPerformance.git
\end{verbatim}

\noindent Make sure to use the right detector description. At the time of writing, this is SiD option 2 version 3, but use the version appropriate to your task. The line in SiDPerformance/gear\_sid.xml should read $\langle$global detectorName="SiD\_o2\_v03"$/\rangle$. for SiD option 2 version 3. Now we run:

\begin{verbatim}
bash> ln -s SiDPerformance/PandoraSettings/
bash> Marlin SiDPerformance/\
      SiDReconstruction_o2_v03_calib1.xml \
      --global.GearXMLFile=\
      SiDPerformance/gear_sid.xml \
      --global.LCIOInputFiles=\
      "example/simple_lcio.slcio" \ 
      --MyLCIOOutputProcessor.LCIOOutputFile=\
      "simple_lcio.reco"
\end{verbatim}

\subsection{LCIO Files With Python (pyLCIO)}

The output LCIO files generated with ddsim and Marlin can be read by a Python program using the pyLCIO package. For Python reconstruction, track and calorimeter hits can be read in and used as input to reconstruction algorithms. For Python analysis of reconstruction objects created with Marlin, one first needs to know what objects are available in the LCIO file.

LCIO objects are stored in collections. The following Python code imports a reader from pyLCIO, uses it to read an LCIO file named 'out.lcio', then cycles through the events in the file and prints the names and types of all collections in the file:

\noindent \dotfill
\begin{verbatim}
from pyLCIO import IOIMPL
reader=IOIMPL.LCFactory.getInstance().\
    createLCReader()
reader.open('infile.lcio')
for event in reader:
    for collectionName, collection in event:
        print collectionName, ' of type ',\
        collection.getTypeName()
reader.close()
\end{verbatim}
\noindent \dotfill

It is frequently useful to use this code to find out the exact names used for collections in an LCIO file since these names are configurable by XML for both ddsim and Marlin and will vary from LCIO file to LCIO file.

We now give an example of code which obtains collections of Monte Carlo truth objects, track objects, and PFO objects and performs some manipulations on the objects:

\noindent \dotfill

\begin{verbatim}
import math
from pyLCIO import IOIMPL
reader=IOIMPL.LCFactory.getInstance().\
    createLCReader()
reader.open('infile.slcio')
Bz=5.
for event in reader:
   mcpCollection=event.getCollection('MCParticle')
   for mcp in mcpCollection:
       if mcp.getGeneratorStatus()==0 and \
       mcp.getEnergy()>1.:
           print 'mcp type', mcp.getPDG(), \
           'energy', mcp.getEnergy()
   trackCollection=event.getCollection('CATracks')
   for track in trackCollection:
       pt=0.0003*math.fabs(Bz/track.getOmega())
       pz=pt*track.getTanLambda()
       print 'track p=', math.sqrt(pt**2+pz**2)
   pfoCollection=event.getCollection\
       ('PandoraPFOsTruthTracks')
   electrons=[pfo for pfo in pfoCollection \
       if abs(pfo.getType())==11]
   photons=[pfo for pfo in pfoCollection \
       if pfo.getType()==22]
   chadrons=[pfo for pfo in pfoCollection \
       if abs(pfo.getType())==211]
   nhadrons=[pfo for pfo in pfoCollection \
       if abs(pfo.getType())==2112]
reader.close()
\end{verbatim}

\noindent \dotfill

For a Monte Carlo truth object the getPDG() method returns the Particle Data Group ID. In the case of a PFO, the getType() method returns not the truth PDG ID, but the PFO hypothesis: 11 for electrons, 22 for photons, 211 for charged hadrons (not just charged pions), 2112 for neutral hadrons (not just neutrons), and 13 for muons.

\begin{figure*}[p]
\begin{center}
\vspace{-1.2in}
\includegraphics[width=0.8\textwidth]{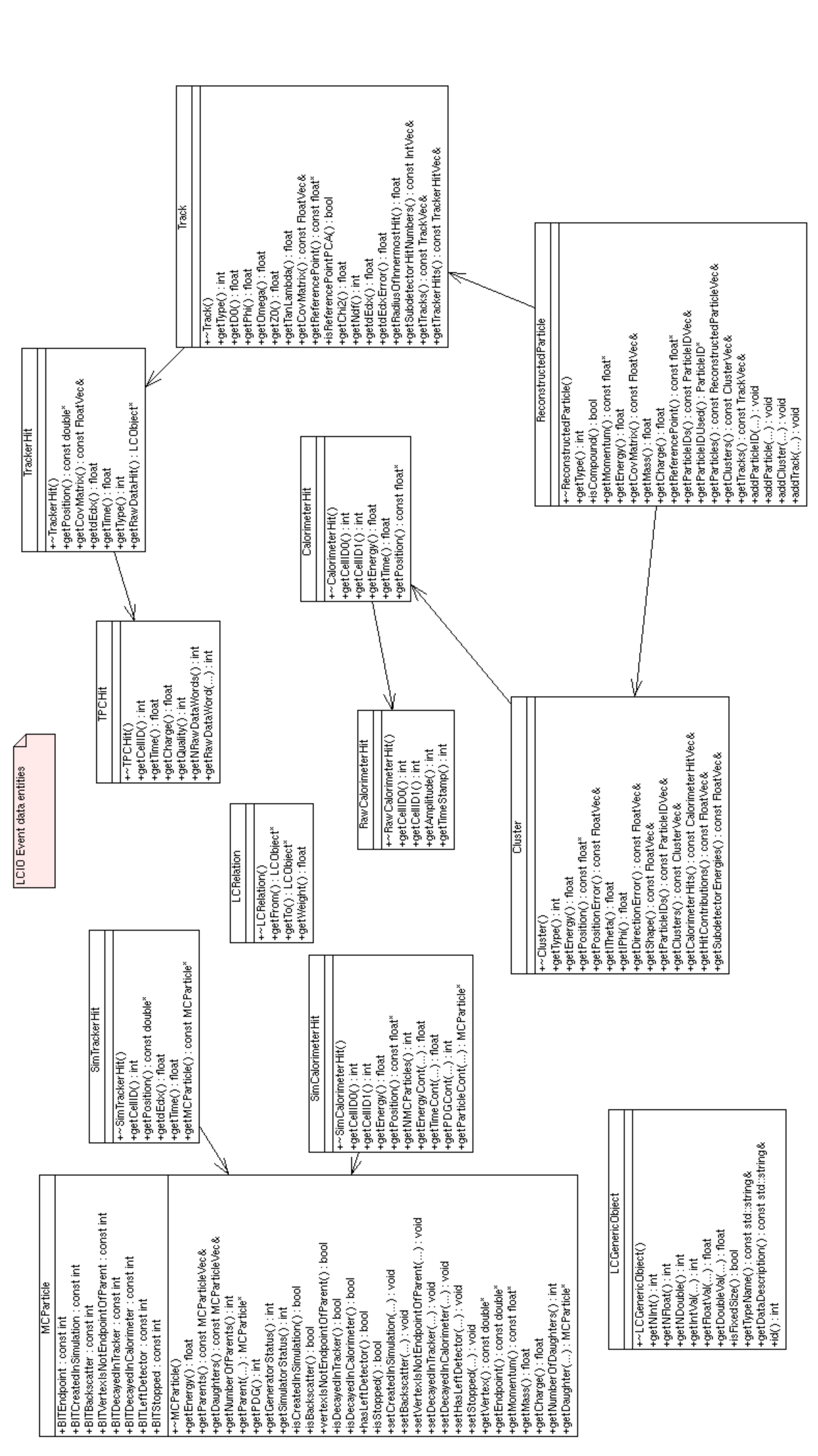}
\caption{Available getters for an event in an LCIO file. \texttt{MCParticle} objects are Monte Carlo truth objects, \emph{ie} truth information provided by the event generator. \texttt{Track} and \texttt{Cluster} objects are reconstructed from detector simulation hit objects. A PFO is an example of a \texttt{ReconstructedParticle}. Credit: LCIO \cite{gaede_2017}.}
\label{fig:lcio}
\end{center}
\end{figure*}

See fig. \ref{fig:lcio} for the available getters in an LCIO event.

\subsection{CernVM Filesystem (CVMFS) \label{sec:cvmfs}}

You will need the root password to install CVMFS. If you do not have it, ask your system administrator. Do the following:

\begin{verbatim}
bash> sudo yum install \
      https://ecsft.cern.ch/dist/cvmfs/\
      cvmfs-release/cvmfs-release-latest.noarch.rpm
bash> sudo yum install cvmfs cvmfs-config-default 
\end{verbatim}

In file /etc/cvmfs/default.local put these lines:

\noindent \dotfill
\begin{verbatim}
CVMFS_REPOSITORIES=sft.cern.ch,ilc.desy.de 
CVMFS_HTTP_PROXY=DIRECT 
CVMFS_SERVER_URL=”http://grid-cvmfs-one.desy.de:\
    8000/cvmfs/@fqrn@” 
\end{verbatim}
\noindent \dotfill

Now obtain the desy.de.pub key 
\begin{verbatim}
bash> wget http://grid.desy.de/etc/cvmfs/keys/\
      desy.de.pub
\end{verbatim}
\noindent and place it in the file /etc/cvmfs/keys/desy.de.pub. In file /etc/cvmfs/domain.d/desy.de.conf put these lines:

\noindent \dotfill
\begin{verbatim}
CVMFS_PUBLIC_KEY=/etc/cvmfs/keys/desy.de.pub
CVMFS_SERVER_URL="http://grid-cvmfs-one.desy.de:\
    8000/cvmfs/@fqrn@"
\end{verbatim}
\noindent \dotfill

Then enter the following commands to setup CVMFS with automounting and check the configuration:

\begin{verbatim}
bash> sudo cvmfs_config setup
bash> sudo service autofs start
bash> sudo chkconfig autofs on
bash> sudo cvmfs_config chksetup
\end{verbatim}

\noindent The necessary mounts from CERN and DESY should be at \texttt{/cvmfs/sft.cern.ch} and  \texttt{/cvmfs/ilc.desy.de}.

\bibliography{paper}

\begin{thebibliography}{10}

\bibitem{Behnke:2013xla}
Ties Behnke, James~E. Brau, Brian Foster, Juan Fuster, Mike Harrison,
  James~McEwan Paterson, Michael Peskin, Marcel Stanitzki, Nicholas Walker, and
  Hitoshi Yamamoto.
\newblock {The International Linear Collider Technical Design Report - Volume
  1: Executive Summary}.
\newblock 2013, arXiv:1306.6327.

\bibitem{Baer:2013cma}
Howard Baer, Tim Barklow, Keisuke Fujii, Yuanning Gao, Andre Hoang, Shinya
  Kanemura, Jenny List, Heather~E. Logan, Andrei Nomerotski, and Maxim
  Perelstein.
\newblock {The International Linear Collider Technical Design Report - Volume
  2: Physics}.
\newblock 2013, arXiv:1306.6352.

\bibitem{Phinney:2007gp}
Gerald Aarons et~al.
\newblock {ILC Reference Design Report Volume 3 - Accelerator}.
\newblock 2007, arXiv:0712.2361.

\bibitem{Behnke:2013lya}
Ties Behnke, James~E. Brau, Philip~N. Burrows, Juan Fuster, Michael Peskin,
  Marcel Stanitzki, Yasuhiro Sugimoto, Sakue Yamada, and Hitoshi Yamamoto.
\newblock {The International Linear Collider Technical Design Report - Volume
  4: Detectors}.
\newblock 2013, arXiv:1306.6329.

\bibitem{Higgs:1964pj}
Peter~W. Higgs.
\newblock {Broken Symmetries and the Masses of Gauge Bosons}.
\newblock {\em Phys. Rev. Lett.}, 13:508--509, 1964.
\newblock [,160(1964)].

\bibitem{Englert:1964et}
F.~Englert and R.~Brout.
\newblock {Broken Symmetry and the Mass of Gauge Vector Mesons}.
\newblock {\em Phys. Rev. Lett.}, 13:321--323, 1964.
\newblock [,157(1964)].

\bibitem{Aad:2012tfa}
Georges Aad et~al.
\newblock {Observation of a new particle in the search for the Standard Model
  Higgs boson with the ATLAS detector at the LHC}.
\newblock {\em Phys.Lett.}, B716:1--29, 2012, arXiv:1207.7214.

\bibitem{Chatrchyan:2012ufa}
Serguei Chatrchyan et~al.
\newblock {Observation of a new boson at a mass of 125 GeV with the CMS
  experiment at the LHC}.
\newblock {\em Phys.Lett.}, B716:30--61, 2012, arXiv:1207.7235.

\bibitem{Aihara:2009ad}
H.~Aihara, P.~Burrows, M.~Oreglia, E.~L. Berger, V.~Guarino, J.~Repond,
  H.~Weerts, L.~Xia, J.~Zhang, Q.~Zhang, et~al.
\newblock {SiD Letter of Intent}.
\newblock 2009, arXiv:0911.0006.

\bibitem{Tanabashi:2018oca}
M.~Tanabashi et~al.
\newblock {Review of Particle Physics}.
\newblock {\em Phys. Rev.}, D98(3):030001, 2018.

\bibitem{Dittmaier:2011ti}
{LHC Higgs Cross Section Working Group}, S.~Dittmaier, C.~Mariotti,
  G.~Passarino, and R.~Tanaka~(Eds.).
\newblock {Handbook of LHC Higgs Cross Sections: 1. Inclusive Observables}.
\newblock {\em CERN-2011-002}, CERN, Geneva, 2011, arXiv:1101.0593.

\bibitem{Dittmaier:2012vm}
{LHC Higgs Cross Section Working Group}, S.~Dittmaier, C.~Mariotti,
  G.~Passarino, and R.~Tanaka~(Eds.).
\newblock {Handbook of LHC Higgs Cross Sections: 2. Differential
  Distributions}.
\newblock {\em CERN-2012-002}, CERN, Geneva, 2012, arXiv:1201.3084.

\bibitem{Heinemeyer:2013tqa}
{LHC Higgs Cross Section Working Group}, S.~Heinemeyer, C.~Mariotti,
  G.~Passarino, and R.~Tanaka~(Eds.).
\newblock {Handbook of LHC Higgs Cross Sections: 3. Higgs Properties}.
\newblock {\em CERN-2013-004}, CERN, Geneva, 2013, arXiv:1307.1347.

\bibitem{Kilian:2007gr}
Wolfgang Kilian, Thorsten Ohl, and Jurgen Reuter.
\newblock {WHIZARD: Simulating Multi-Particle Processes at LHC and ILC}.
\newblock {\em Eur. Phys. J.}, C71:1742, 2011, arXiv:0708.4233.

\bibitem{Murayama:1996ec}
Hitoshi Murayama and Michael~E. Peskin.
\newblock {Physics opportunities of e+ e- linear colliders}.
\newblock {\em Ann. Rev. Nucl. Part. Sci.}, 46:533--608, 1996, hep-ex/9606003.

\bibitem{Potter:2017rlo}
C.~T. Potter.
\newblock {Backgrounds for Fast Simulation $e^+ e^-$ Collider Studies at
  $\sqrt{s}=$91,250,350,500 GeV (1702.04827)}.
\newblock 2017, arXiv:1702.04827.

\bibitem{Alwall:2014hca}
J.~Alwall, R.~Frederix, S.~Frixione, V.~Hirschi, F.~Maltoni, O.~Mattelaer,
  H.~S. Shao, T.~Stelzer, P.~Torrielli, and M.~Zaro.
\newblock {The automated computation of tree-level and next-to-leading order
  differential cross sections, and their matching to parton shower
  simulations}.
\newblock {\em JHEP}, 07:079, 2014, arXiv:1405.0301.

\bibitem{Evans:2017rvt}
Lyn Evans and Shinichiro Michizono.
\newblock {The International Linear Collider Machine Staging Report 2017}.
\newblock 2017, arXiv:1711.00568.

\bibitem{Schulte:1997nga}
Daniel Schulte.
\newblock {\em {Study of Electromagnetic and Hadronic Background in the
  Interaction Region of the TESLA Collider}}.
\newblock PhD thesis, DESY, 1997.
\newblock {\url{http://inspirehep.net/record/888433/files/shulte.pdf}}.

\bibitem{Cottingham:396082}
W~Noel Cottingham and Derek~A Greenwood.
\newblock {\em {An introduction to the standard model of particle physics}}.
\newblock Cambridge Univ. Press, Cambridge, 1998.

\bibitem{Griffiths2004Introduction}
{\em {Introduction to Quantum Mechanics (2nd Edition)}}.
\newblock Pearson Prentice Hall, 2nd edition, April 2004.

\bibitem{Griffiths:2008zz}
David Griffiths.
\newblock {\em {Introduction to elementary particles}}.
\newblock 2008.

\bibitem{Quigg:1983gw}
C.~Quigg.
\newblock {\em {Gauge Theories of the Strong, Weak and Electromagnetic
  Interactions}}, volume~56.
\newblock 1983.

\bibitem{Halzen:1984mc}
F.~Halzen and Alan~D. Martin.
\newblock {\em {Quarks and Leptons: An Introductory Course in Modern Particle
  Physics}}.
\newblock 1 1984.

\bibitem{Barger:1987nn}
Vernon~D. Barger and R.J.N. Phillips.
\newblock {\em {Collider Physics}}.
\newblock 1987.

\bibitem{Peskin:1995ev}
Michael~E. Peskin and Daniel~V. Schroeder.
\newblock {\em {An Introduction to quantum field theory}}.
\newblock Addison-Wesley, Reading, USA, 1995.

\bibitem{nobelpage}
Nobel Prize.
\newblock {The Nobel Prize}.
\newblock \url{https://www.nobelprize.org/}.
\newblock Accessed: January, 2020.

\bibitem{barklow2015ilc}
T.~Barklow, J.~Brau, K.~Fujii, J.~Gao, J.~List, N.~Walker, and K.~Yokoya.
\newblock {ILC Operating Scenarios}, 2015, arXiv:1506.07830.

\bibitem{Rowson:2001cd}
P.~C. Rowson, Dong Su, and Stephane Willocq.
\newblock {Highlights of the SLD physics program at the SLAC linear collider}.
\newblock {\em Ann. Rev. Nucl. Part. Sci.}, 51:345--412, 2001, hep-ph/0110168.

\bibitem{1991275}
{The OPAL detector at LEP}.
\newblock {\em Nuclear Instruments and Methods in Physics Research Section A:
  Accelerators, Spectrometers, Detectors and Associated Equipment}, 305(2):275
  -- 319, 1991.

\bibitem{Collaboration_2008}
The~ATLAS Collaboration.
\newblock The {ATLAS} experiment at the {CERN} large hadron collider.
\newblock {\em Journal of Instrumentation}, 3(08):S08003--S08003, aug 2008.

\bibitem{cahn_goldhaber_2009}
Robert~N. Cahn and Gerson Goldhaber.
\newblock {\em The Experimental Foundations of Particle Physics}.
\newblock Cambridge University Press, 2 edition, 2009.

\bibitem{Pais:1988}
Abraham Pais.
\newblock {\em {Inward Bound: Of Matter and Forces in the Physical World}}.
\newblock Clarendon Press, 1988.

\bibitem{Segrè:100978}
Emilio Segrè.
\newblock {\em {From X-rays to quarks: modern physicists and their
  discoveries}}.
\newblock Freeman, San Francisco, CA, 1980.
\newblock Trans. of : Personaggi e scoperte nella fisica contemporanea. Milano
  : Mondadori, 1976.

\bibitem{Martin:2211679}
Brian~R Martin and Graham Shaw.
\newblock {\em {Particle physics; 4th ed.}}
\newblock Manchester physics series. Wiley, New York, NY, 2017.

\bibitem{Perkins:396126}
Donald~Hill Perkins.
\newblock {\em {Introduction to high energy physics; 4th ed.}}
\newblock Cambridge Univ. Press, Cambridge, 2000.

\bibitem{Edwards:1993qe}
D.A. Edwards and M.J. Syphers.
\newblock {\em {An Introduction to the Physics of High-Energy Accelerators}}.
\newblock Wiley Series in Beam Physics and Accelerator Technology. Wiley, New
  York, 1992.

\bibitem{Wangler:368392}
Thomas~P Wangler.
\newblock {\em {Principles of RF linear accelerators}}.
\newblock Wiley Beam Phys. Accel. Technol. Wiley, New York, NY, 1998.

\bibitem{Ferbel:238481}
Thomas Ferbel.
\newblock {\em {Experimental techniques in high-energy nuclear and particle
  physics; 2nd ed.}}
\newblock World Scientific, Singapore, 1991.

\bibitem{Bambade:2019fyw}
Philip Bambade et~al.
\newblock {The International Linear Collider: A Global Project}.
\newblock 2019, arXiv:1903.01629.

\bibitem{jan_strube_2020_3766518}
Jan Strube and saveliev77.
\newblock linearcollider/detectorliaisonreport 2012.1.3, April 2020.

\bibitem{Alwall:2006yp}
Johan Alwall et~al.
\newblock {A Standard format for Les Houches event files}.
\newblock {\em Comput. Phys. Commun.}, 176:300--304, 2007, hep-ph/0609017.

\bibitem{stdhep}
L.~Garren.
\newblock {StdHep 5.06.01 Monte Carlo Standardization at FNAL Fortran and C
  Implementation}.
\newblock
  \url{http://cd-docdb.fnal.gov/0009/000903/015/stdhep_50601_manual.ps}.
\newblock Accessed: March 14, 2016.

\bibitem{Dobbs:2001ck}
Matt Dobbs and Jorgen~Beck Hansen.
\newblock {The HepMC C++ Monte Carlo event record for High Energy Physics}.
\newblock {\em Comput. Phys. Commun.}, 134:41--46, 2001.

\bibitem{Sjostrand:2006za}
Torbjorn Sjostrand, Stephen Mrenna, and Peter~Z. Skands.
\newblock {PYTHIA 6.4 Physics and Manual}.
\newblock {\em JHEP}, 0605:026, 2006, hep-ph/0603175.

\bibitem{Sjostrand:2007gs}
Torbjorn Sjostrand, Stephen Mrenna, and Peter~Z. Skands.
\newblock {A Brief Introduction to PYTHIA 8.1}.
\newblock {\em Comput.Phys.Commun.}, 178:852--867, 2008, arXiv:0710.3820.

\bibitem{AGOSTINELLI2003250}
S.~Agostinelli et~al.
\newblock {Geant4 — a simulation toolkit}.
\newblock {\em Nuclear Instruments and Methods in Physics Research Section A:
  Accelerators, Spectrometers, Detectors and Associated Equipment}, 506(3):250
  -- 303, 2003.

\bibitem{Allison:2006ve}
John Allison et~al.
\newblock {Geant4 developments and applications}.
\newblock {\em IEEE Trans. Nucl. Sci.}, 53:270, 2006.

\bibitem{ALLISON2016186}
J.~Allison et~al.
\newblock {Recent developments in Geant4}.
\newblock {\em Nuclear Instruments and Methods in Physics Research Section A:
  Accelerators, Spectrometers, Detectors and Associated Equipment}, 835:186 --
  225, 2016.

\bibitem{frank_markus_2018_1464634}
Markus Frank, Frank Gaede, Marko Petric, and Andre Sailer.
\newblock {AIDASoft/DD4hep}, 2018.
\newblock \url{http://dd4hep.cern.ch/}.

\bibitem{gaede_2017}
F.~Gaede and H.~Vogt.
\newblock {LCIO - Users Manual}, 2017.
\newblock \url{http://lcio.desy.de/v02-09/doc/manual.pdf}.

\bibitem{Selvaggi:2014mya}
Michele Selvaggi.
\newblock {DELPHES 3: A modular framework for fast-simulation of generic
  collider experiments}.
\newblock {\em J. Phys. Conf. Ser.}, 523:012033, 2014.

\bibitem{Mertens:2015kba}
Alexandre Mertens.
\newblock {New features in Delphes 3}.
\newblock {\em J. Phys. Conf. Ser.}, 608(1):012045, 2015.

\bibitem{Potter:2016pgp}
C.~T. Potter.
\newblock {DSiD: a Delphes Detector for ILC Physics Studies}.
\newblock In {\em {Proceedings, International Workshop on Future Linear
  Colliders (LCWS15): Whistler, B.C., Canada, November 02-06, 2015}}, 2016,
  arXiv:1602.07748.

\bibitem{Brun:1997pa}
R.~Brun and F.~Rademakers.
\newblock {ROOT: An object oriented data analysis framework}.
\newblock {\em Nucl. Instrum. Meth.}, A389:81--86, 1997.

\bibitem{kramer_2006_004}
Thomas Kramer.
\newblock {Track Parameters in LCIO}, 2006.
\newblock LC-DET-2006-004.
  \url{http://flc.desy.de/lcnotes/noteslist/index_eng.html}.

\bibitem{Brondolin:2019awm}
Erica Brondolin, Frank Gaede, Daniel Hynds, Emilia Leogrande, Marko Petrič,
  André Sailer, and Rosa Simoniello.
\newblock {Conformal Tracking for all-silicon trackers at future
  electron-positron colliders}.
\newblock {\em Nucl. Instrum. Meth.}, A956:163304, 2020, arXiv:1908.00256.

\bibitem{Thomson200925}
M.A. Thomson.
\newblock {Particle flow calorimetry and the PandoraPFA algorithm}.
\newblock {\em Nuclear Instruments and Methods in Physics Research Section A:
  Accelerators, Spectrometers, Detectors and Associated Equipment}, 611(1):25
  -- 40, 2009.

\bibitem{Cacciari:2011ma}
Matteo Cacciari, Gavin~P. Salam, and Gregory Soyez.
\newblock {FastJet User Manual}.
\newblock {\em Eur. Phys. J.}, C72:1896, 2012, arXiv:1111.6097.

\bibitem{Suehara:2015ura}
Taikan Suehara and Tomohiko Tanabe.
\newblock {LCFIPlus: A Framework for Jet Analysis in Linear Collider Studies}.
\newblock {\em Nucl. Instrum. Meth. A}, 808:109--116, 2016, 1506.08371.

\bibitem{JACKSON1997247}
David~J Jackson.
\newblock {A topological vertex reconstruction algorithm for hadronic jets}.
\newblock {\em Nuclear Instruments and Methods in Physics Research Section A:
  Accelerators, Spectrometers, Detectors and Associated Equipment}, 388(1):247
  -- 253, 1997.

\bibitem{Potter:2020ihz}
C.~T. Potter.
\newblock {pySiDR: Python Event Reconstruction for SiD}.
\newblock In {\em {International Workshop on Future Linear Colliders (LCWS
  2019) Sendai, Miyagi, Japan, October 28-November 1, 2019}}, 2020,
  arXiv:2002.05804.

\bibitem{Stanitzki:2020bnx}
Marcel Stanitzki and Jan Strube.
\newblock {Performance of Julia for High Energy Physics Analyses}.
\newblock 2020, arXiv:2003.11952.

\bibitem{Fujii:2019zll}
Keisuke Fujii et~al.
\newblock {Tests of the Standard Model at the International Linear Collider}.
\newblock 2019, arXiv:1908.11299.

\bibitem{Braun:2020eme}
L.~Braun, D.~Austin, J.~Barkeloo, J.~Brau, and C.~T. Potter.
\newblock {Correcting for Leakage Energy in the SiD Silicon-Tungsten ECal}.
\newblock In {\em {International Workshop on Future Linear Colliders (LCWS
  2019) Sendai, Miyagi, Japan, October 28-November 1, 2019}}, 2020,
  arXiv:2002.04100.

\bibitem{potter2003}
C.T. Potter, J.E. Brau, and Nikolai Sinev.
\newblock {A CCD vertex detector for measuring Higgs boson branching ratios at
  a linear collider}.
\newblock {\em Nuclear Instruments and Methods in Physics Research Section A:
  Accelerators, Spectrometers, Detectors and Associated Equipment},
  511:225--228, 09 2003.

\end{thebibliography}

\end{document}